\documentclass[11pt,a4paper]{article}
\usepackage{amsmath,amssymb,amsfonts,a4wide,graphicx,bm,times}
\usepackage{mcite}%\usepackage{mciteplus}
\usepackage{tikz}
\usepackage{hyperref}
\usepackage{amsthm} %%%%% proof environment
\usepackage{braket}
\usepackage{mathtools}
\usepackage[mathscr]{euscript}

\makeatletter \let\old@startsection=\@startsection
\renewcommand{\@startsection}[6]
{\old@startsection{#1}{#2}{#3}{#4}{#5}{#6\mathversion{bold}}}
\makeatother

\makeatletter

\@addtoreset{equation}{section}
\makeatother
\let\refOld\ref
\renewcommand{\ref}[1]{(\refOld{#1})}

 \newcommand{\mat}[4]{\begin{smallmatrix}
 #1 & #2\\
 #3 & #4
 \end{smallmatrix}}
 \newcommand{\mr}[1]{\mathring{#1}}
 \newcommand{\dket}[1]{\ket{#1}\!\rangle}
 \newcommand{\dbra}[1]{\langle\!\bra{#1}}
 \newcommand{\superp}[2]{\genfrac{}{}{0pt}{}{#1}{#2}}
 \newcommand{\superpsim}[1]{\genfrac{}{}{0pt}{3}{\sim}{#1}}
 
 \def\Abox{\tikz[scale=0.007cm] \draw (0,0) rectangle (1,1);}
 \def\sAbox{\tikz[scale=0.004cm] \draw (0,0) rectangle (1,1);}
 \def\mAbox{\tikz[scale=0.0055cm] \draw (0,0) rectangle (1,1);}
 
 \def\sAboxF{\tikz[scale=0.004cm] \draw [fill=black] (0,0) rectangle (1,1);}
 \def\mAboxF{\tikz[scale=0.0055cm] \draw [fill=black] (0,0) rectangle (1,1);}

 \def\d{\delta}

 \def\p{\partial}
 
 \def\a{\alpha}
 \def\b{\beta}
 \def\g{\gamma}
 \def\d{\delta}
 
 \def\e{\varepsilon}
 \def\th{\theta}
 
 \def\k{\kappa}
 \def\l{\lambda}
 
 \def\n{\nu}

 \def\s{\sigma}
 \def\t{\tau}
 \def\th{\theta}
 \def\z{\zeta }
 
 \def\G{\Gamma}
 \def\D{\Delta}

 \def\L{\Lambda}
 \def\O{\Omega}
 \def\o{\omega }
 
\def\CA{{\mathcal{A}}}

\def\CG{{\mathcal{G}}}

\def\CM{{\mathcal{M}}}
\def\CN{{\mathcal{N}}}
\def\CO{{\mathcal{O}}}
\def\CP{{\mathcal{P}}}

\def\CS{{\mathcal{S}}}
\def\CT{{\mathcal{T}}}

\def\CY{{\mathcal{Y}}}

\def\la{\left\langle}
\def\ra{\right\rangle}

\def\hf{\dfrac{1}{2}}

\def\bc{{\bar{c}}}

\def\implies{\quad\Rightarrow\quad}

\def\vphi{\varphi}
\def\sign{\text{sign }}

\def\tPhi{\tilde{\Phi}}

\def\Zv{\mathcal{Z}_\text{vect.}}
\def\Zbf{\mathcal{Z}_\text{bfd.}}

\def\Zinst{\mathcal{Z}_\text{inst.}}
\def\ZCS{\mathcal{Z}_\text{CS}}

\def\tg{\tilde{g}}
\def\res{\mathop{\text{Res}}}

\def\qf{\mathfrak{q}}

\def\vac{\varnothing}

\def\rag{\right\rangle_\text{gauge}}

\def\bd{{\bar d}}
\def\bc{{\bar c}}

\def\bi{{\bar i}}

\def\bl{{\boldsymbol{\lambda}}}
\def\blc{{\boldsymbol{\lambda}^{\boldsymbol{c}}}}
\def\blinf{{\boldsymbol{\lambda}^\infty}}
\def\bv{{\boldsymbol{v}}}
\def\CYY{\mathcal{Y}_{\omega}^{[\boldsymbol{\lambda}]}}
\def\fYY{f_{\omega}^{[\boldsymbol{\lambda}]}}
\def\PsiY{\Psi_{\omega}^{[\boldsymbol{\lambda}]}}

\def\gbY{\b_{\omega}^{[\boldsymbol{\lambda}]}}

\def\bo{{\bar\omega}}
\def\tx{\tilde{x}}
\def\tpsi{\tilde{\psi}}
\def\qf{\mathfrak{q}}
\def\Zv{\mathcal{Z}_\text{vect.}}
\def\Zbf{\mathcal{Z}_\text{bfd.}}
\def\td{\tilde{d}}
\def\tp{\tilde{p}}
\def\tq{\tilde{q}}

\def\ta{\tilde{a}}
\def\bz{{\bar z}}

\def\bc{{\bar c}}
\def\mZ{\mathbb{Z}}
\def\mC{\mathbb{C}}
\def\gf{\mathfrak{g}}
\def\Zp{\mathbb{Z}_p}
\def\glp{\mathfrak{gl}(p)}
\def\oY{_\omega^{[\bl]}}
\def\boY{_{\bo}^{[\bl]}}
\def\bi{\bar{i}}
\def\bj{\bar{j}}
\def\kO{\Omega}
\def\tD{\tilde{\Delta}}

\begin{document}

\begin{titlepage}
\vspace*{-2cm}
\begin{flushright}
KIAS - Q19005  \\YITP-SB-19-18
\end{flushright}	
\vskip 22mm

\begin{center}
{\huge\bf New quantum toroidal algebras \\from 5D $\CN=1$ instantons on orbifolds}
\end{center}
\vskip 30mm
\begin{center}
{\Large Jean-Emile Bourgine$^\dagger$, Saebyeok Jeong$^\ast$}
\\[.4cm]
{\em {}$^\dagger$Korea Institute for Advanced Study (KIAS)}\\
{\em Quantum Universe Center (QUC)}\\
{\em 85 Hoegiro, Dongdaemun-gu, Seoul, South Korea}\\
\texttt{bourgine@kias.re.kr}
\\[.4cm]
{\em {}$^\ast$C.N. Yang Institute for Theoretical Physics}\\
{\em Stony Brook University}\\
{\em Stony Brook, NY 11794-3840, USA}\\
\texttt{saebyeok.jeong@gmail.com}
\\[.4cm]
\end{center}
\vfill
\begin{abstract}
Quantum toroidal algebras are obtained from quantum affine algebras by a further affinization, and, like the latter, can be used to construct integrable systems. These algebras also describe the symmetries of instanton partition functions for 5D $\mathcal{N}=1$ supersymmetric quiver gauge theories. We consider here the gauge theories defined on an orbifold $S^1\times\mathbb{C}^2/\mathbb{Z}_p$ where the action of $\mathbb{Z}_p$  is determined by two integer parameters $(\nu_1,\nu_2)$. The corresponding quantum toroidal algebra is introduced as a deformation of the quantum toroidal algebra of $\mathfrak{gl}(p)$. We show that it has the structure of a Hopf algebra, and present two representations, called \textit{vertical} and \textit{horizontal}, obtained by deforming respectively the Fock representation and Saito's vertex representations of the quantum toroidal algebra of $\glp$. We construct the vertex operator intertwining between these two types of representations. This object is identified with a $(\nu_1,\nu_2)$-deformation of the refined topological vertex, allowing us to reconstruct the Nekrasov partition function and the $qq$-characters of the quiver gauge theories.
\end{abstract}
\vfill
\end{titlepage}

% \vspace*{1cm}
% 
% \begin{center}
% {\huge {\bf   New quantum toroidal algebras from a correspondence with 5D $\CN=1$ instantons on orbifolds}}
% % {\huge {\bf   Algebraic construction of $\CN=1$ BPS-observables for quiver gauge theories on orbifolded spacetimes $S_R^1\times(\mathbb{C}_{\e_1}\times\mathbb{C}_{\e_2})/\mathbb{Z}_p$}}
% \end{center}

\tableofcontents

\section{Introduction}
Non-perturbative dynamics of supersymmetric gauge theories is a prolific research subject in theoretical physics. Since the innovative work \cite{Nekrasov2002}, direct microscopic studies on four-dimensional gauge theories with $\mathcal{N}=2$ supersymmetry became largely accessible, from exact computations of their partition functions on the (non-commutative) $\mathbb{C}^2$. The divergence in the partition function coming from the non-compactness of $\mathbb{C}^2$ is properly regularized by introducing the $\Omega$-background \cite{Nekrasov2002,Nekrasov2003}, effectively localizing the four-dimensional theory to the origin. In turn, the path integral reduces to an equivariant integration on the finite dimensional framed moduli space of non-commutative instantons, for which equivariant localization can be applied for its exact evaluation. The Nekrasov partition function has been a powerful tool to investigate the correspondences of four-dimensional $\mathcal{N}=2$ supersymmetric quiver gauge theories with other objects in mathematical physics, i.e., quantum integrable systems \cite{NS2009,NS2009',NS2009''}, two-dimensional CFTs \cite{Alday2010,Wyllard2009,Nekrasov_BPS1,Nekrasov_BPS2, Nekrasov_BPS3,Nekrasov_BPS4,Nekrasov_BPS5}, flat connections on Riemann surfaces \cite{NRS2011,Jeong2018}, and isomonodromic deformations of Fuchsian systems \cite{LLNZ2013, GIL2012, ILT2014}.

Very rich algebraic structures lie at the heart of these correspondences \cite{NPS}. For instance, the AGT correspondence \cite{Alday2010,Wyllard2009} between Nekrasov partition functions and conformal blocks of Liouville/Toda 2D CFTs can be understood algebraically as the action of W-algebras on the cohomology of instantons moduli space \cite{Nakajima1997,MO2012,Schiffmann2012}. In this context, the W-algebra currents are coupled to an infinite Heisenberg algebra, and the total action is formulated in terms of a quantum algebra, namely the Spherical Hecke central algebra \cite{Schiffmann2012} (isomorphic to the affine Yangian of $\mathfrak{gl}(1)$ \cite{Tsymbaliuk2014,Prochazka2015}). The coupling to an Heisenberg algebra is essential for the definition of a coalgebraic structure, thus emphasizing the underlying quantum integrability since the coproduct provides the R-matrix satisfying the celebrated quantum Yang-Baxter equation.

A closely related but different connection with quantum algebras arises from the type IIB strings theory realization of the five-dimensional uplifts of 4D $\CN=2$ gauge theories, that is the 5D $\CN=1$ quiver gauge theories compactified on $S^1$. In this construction, $\CN=1$ gauge theories emerge as the low-energy description of the dynamics of 5-branes webs \cite{Aharony1997,Aharony1997a}. Here, each brane carries the charges $(p,q)$, generalizing D5-branes (charge $(1,0)$) and NS5-branes (charge $(0,1)$). Their world-volume include the five-dimensional gauge theory spacetime, together with an extra line segment in the 56-plane. Individual branes' line segments are joined by trivalent vertices, and form the $(p,q)$-branes web. Alternatively, the $(p,q)$-brane web can be seen as the toric diagram of a Calabi-Yau threefold on which topological strings can be compactified \cite{Leung1998}. The trivalent vertices are then identified with the (refined) topological vertex, thereby leading to a very efficient method of computing 5D Nekrasov partition functions as topological strings amplitudes \cite{Aganagic2005,Iqbal2007}. 

Awata, Feigin and Shiraishi observed in \cite{Awata2011'} that a specific representation of the quantum toroidal $\mathfrak{gl}(1)$ algebra (or Ding-Iohara-Miki algebra \cite{Ding1997,Miki2007}) can be associated to each edge of the $(p,q)$-branes web. The charges $(p,q)$ are identified with the values of the two central charges while the brane position define the weight of the representation. As such, the D5-branes correspond to the so-called \textit{vertical representation} while an \textit{horizontal representation} is associated to NS5-branes (possibly dressed by extra D5-branes).\footnote{In fact, the vertical representation is simply the q-deformation of the affine Yangian action mentioned previously, it is expected to describe a quantum toroidal action on the K-equivariant cohomology of the quiver variety describing the instanton moduli space. The equivalent of the horizontal representation can also be defined for 4D $\CN=2$ theories, thus extending the whole algebraic construction of the Nekrasov partition functions \cite{Bourgine2018}. However, for this purpose, it is necessary to consider the central extension of the Drinfeld double of the affine Yangian following from the construction given in \cite{Khoroshkin1996}.} The refined topological vertex is then identified with an intertwiner between vertical and horizontal representations, that is in fact the toroidal version of the vertex operator introduced in \cite{Davies1992} for the quantum group $U_q(\widehat{\mathfrak{sl}(2)})$. In this way, the Nekrasov partition function is written as a purely algebraic object using the quantum toroidal algebra, just like conformal blocks with W-algebras \cite{Awata2016a,Mironov2016}. This algebraic construction turns out to be useful in probing various properties of the partition function, e.g. in addressing the (q-deformed) AGT correspondence \cite{Awata2009,Awata2011}, or in studying strings' S-duality \cite{Awata2009a,Bourgine2018a}.

In \cite{Nekrasov_BPS1}, an important class of half-BPS observables, called \textit{$qq$-characters}, were defined, whose characteristic property is the regularity of their gauge theory expectation values. This regularity property encodes efficiently an infinite set of constraints on the partition function called \textit{non-perturbative Dyson-Schwinger equations} \cite{Nekrasov_BPS1}. The algebraic nature of these constraints was observed in \cite{Bourgine2015c,Bourgine2016}. Actually, the constraints take an even more elegant form in the algebraic construction described above as they express the invariance of an operator $\CT$ under the adjoint action of the quantum toroidal algebra \cite{Bourgine2017b}. This operator is obtained by gluing intertwiners along the edges of the $(p,q)$-branes web, and its vacuum expectation value reproduces the 5D Nekrasov instanton partition.

A natural question is how to generalize the algebraic construction to gauge theories on more complicated manifolds. Among other manifolds, the $\mathbb{Z}_p$-orbifolded $\mathbb{C}^2$ are of a particular interest, since the partition functions on these spaces can be computed by simply projecting out the contributions which are not invariant under the $\mathbb{Z}_p$-action \cite{Kanno2011,Bonelli2012,Nekrasov_BPS3}. The generalization of the algebraic construction is not entirely straightforward since it is necessary to introduce the information of the  
\textit{coloring} corresponding to the $\Zp$-action of the orbifolding. In this scope, deformations of the quantum toroidal $\mathfrak{gl}(1)$ algebra must be considered. A special case of the $\mathbb{Z}_p$-orbifolded $\mathbb{C}^2$ is the (un-resolved) $A_p$-type ALE spaces. The ALE instantons were introduced by Kronheimer in \cite{Kronheimer1989}, and the ALE instanton moduli spaces were constructed as quiver varieties in \cite{Kronheimer1990, Nakajima1994}. The algebraic construction of the corresponding 5D Nekrasov partition functions has been realized recently using an underlying quantum toroidal $\mathfrak{gl}(p)$ algebra \cite{Awata2017}. There, the index carried by the Drinfeld currents renders the $\Zp$-coloring due to the orbifolding. Incidentally, the vertical representation of this quantum toroidal algebra should coincide with the q-deformation of the affine Yangian of $\mathfrak{gl}(p)$ acting on the cohomology of the moduli space of ALE instantons, extending by further affinization the algebraic actions discovered in \cite{Nakajima1994,Nakajima1999}.

In this work, we extend the algebraic construction of 5D Nekrasov partition functions to a more general $\mathbb{Z}_p$-orbifolding depending on two integer parameters $(\nu_1,\nu_2)$. We propose an extended quantum toroidal algebra relevant to the construction, and prove its Hopf algebra structure. We define both horizontal and vertical representations, and derive the vertex operator which intertwines between these representations. Finally, using these ingredients, we give an algebraic construction of Nekrasov partition functions and $qq$-characters. The orbifolds considered in this work incorporate the case of codimension-two defect insertion, whose applications to BPS/CFT correspondence, Bethe/gauge correspondence, and Nekrasov-Rosly-Shatashvili correspondence have been largely investigated \cite{Nekrasov_BPS4,Nekrasov_BPS5,Jeong2017',Jeong2017,Jeong2018}. 

This paper is written in such a way that mathematicians interested only in the formulation of the extended algebra can focus on the reading of section three, together with the appendices \refOld{AppA} (quantum toroidal $\glp$), \refOld{AppC} (representations) and \refOld{AppD} (automorphisms and gradings) for more details. Instead, the section two provides a brief description of the physical context in which the algebra emerges, i.e. instantons of 5D $\CN=1$ gauge theories on the spacetime $\mC^2/\Zp$. Finally, the section four is dedicated to the algebraic construction of gauge theories observables, giving the expression of the $(\nu_1,\nu_2)$-colored refined topological vertex and a few examples of application.

\section{Instantons on orbifolds}
We consider the moduli space of solutions to the anti-self-duality equation $F = - \star F$ defined on the orbifold $\mathbb{C}^2/\mathbb{Z}_p$. The gauge field is allowed to be singular along the orbifold singularity, and its singular behavior determines how the gauge group is broken on the singularity and how the gauge coupling fractionalizes. For the purpose of computing the instanton partition function, it is convenient to encode these additional data directly in the ADHM construction, following \cite{Kanno2011, Bonelli2012} for example.

\subsection{Action of the abelian group $\mathbb{Z}_p$ on the ADHM data}
In order to derive the group action on the instanton moduli space, we focus first on the case of a pure $U(m)$ gauge theory. In this case, the ADHM construction of the moduli space \cite{Atiyah1978} involves only two vector spaces $M$ and $K$ of dimension $m$ and $k$ respectively, where $k$ is the instanton number. Introducing the four matrices $B_1,B_2: K\to K$ $I:M\to K$ and $J: K\to M$, the instanton moduli space is identified with the quiver variety (see, for instance, \cite{Nakajima2003})
\begin{equation}
\CM_k=\{B_1,B_2,I,J\diagup[B_1,B_2]+IJ=0,\mathbb{C}[B_1;B_2]I(M)=K\}\diagup\text{GL}(K).
\end{equation} 
The complexified global symmetry group $\text{GL}(M)\times\text{SL}(2,\mathbb{C})^2$ acts on the ADHM matrices, preserving the quiver variety $\CM_k$. It contains an $(m+2)$-dimensional torus that acts follows,
\begin{equation}\label{torus_action}
(B_1,B_2,I,J)\to (t_1B_1,t_2B_2,It,t^{-1}Jt_1t_2),\quad (t,t_1,t_2)\in (\mathbb{C}^\times)^m\times (\mathbb{C}^\times)^2.
\end{equation} 
The fixed points of this action parameterize the configurations of instantons with total charge $k$. They are in one-to-one correspondence with the $m$-tuple partitions $\bl=(\l^{(1)},\cdots\l^{(m)})$ of the integer $k$, here identified with the $m$-tuple Young diagrams with $|\bl|=k$ boxes. At the fixed point, the vector space $K$ is decomposed into
\begin{equation}\label{basis_K}
K=\bigoplus_{\a=1}^m\bigoplus_{(i,j)\in\l^{(\a)}}B_1^{i-1}B_2^{j-1}I(M_\a),
\end{equation} 
where $M_\a$ denotes the one-dimensional vector spaces generated by the basis vectors of $M$. Thus, each box $\Abox=(\a,i,j)$ of the $m$-tuple partition $\bl$ with coordinate $(i,j)\in\bl^{(\a)}$ corresponds to a one-dimensional vector space $B_1^{i-1}B_2^{j-1}I(M_\a)$. We further associate to the box $\Abox$ the complex variable $\phi_{\sAbox}=a_\a+(i-1)\e_1+(j-1)\e_2$ called \textit{instanton position} or, sometimes, the \textit{box content of} $\Abox$. The parameters $a_1,\cdots,a_m$ are the \textit{Coulomb branch vevs} of the gauge theory. We also define the exponentiated quantities $v_\a=e^{R a_\a}$, $(q_1,q_2)=(e^{R\e_1},e^{R\e_2})$ and $\chi_{\sAbox}=e^{R\phi_{\sAbox}}=v_\a q_1^{i-1}q_2^{j-1}$.

In this paper, gauge theories are considered on the 5D orbifolded omega-background $S_R^1\times(\mathbb{C}_{\e_1}\times\mathbb{C}_{\e_2})/\mathbb{Z}_p$ where $\Zp=\mZ/p\mZ$ is a subgroup of the torus ${U}(1)^2 \subset {SO}(4)$. The action of the group $\Zp$ on the spacetime is parameterized by two integers $\nu_1,\nu_2$,
\begin{equation}\label{Zp_action}
(\th,z_1,z_2)\in S_R^1\times\mathbb{C}_{\e_1}\times\mathbb{C}_{\e_2} \to(\th,e^{2i\pi\nu_1/p}z_1,e^{2i\pi\nu_2/p}z_2),\quad\text{with}\quad(\nu_1,\nu_2)\in\mathbb{Z}_p\times\mathbb{Z}_p.
\end{equation}
Furthermore, it is possible to combine it with a global gauge transformation in the subgroup $U(1)^m\subset U(m)$. As a result, we obtain an action of $\mZ_p$ on the ADHM data by specialization of the $(m+2)$-torus action \ref{torus_action}, taking
\begin{equation}
t=\text{diag}(e^{2i\pi c_\a/p})_{\a =1, \cdots, m},\quad t_1=e^{2i\pi\nu_1/p},\quad t_2=e^{2i\pi\nu_2/p}.
\end{equation}
This action of the abelian group $\mZ_p$ is parameterized by the $m+2$ integers $(c_\a,\nu_1,\nu_2)$ considered modulo $p$. The transformation of the vector spaces in the decomposition \ref{basis_K} of $K$ leads to associate to each box $\Abox=(\a,i,j)\in\bl$, in addition to the complex variables $\phi_{\sAbox}$ and $\chi_{\sAbox}$, the integer $c(\Abox)$ such that
\begin{equation}
B_1^{i-1}B_2^{j-1}I(M_\a)\to e^{2i\pi c(\sAbox)/p}B_1^{i-1}B_2^{j-1}I(M_\a),\quad\text{with}\quad c(\Abox)=c_\a+(i-1)\nu_1+(j-1)\nu_2\in\mZ_p.
\end{equation}
We call \textit{color} any integer parameter defined modulo $p$. For short, we also say that $c_\a$ and $\nu_1,\nu_2$ are respectively  color of the Coulomb branch vevs, and of the parameters $q_1,q_2$. The map $c:\bl\to\mZ_p$ defines a coloring of the $m$-tuple partitions $\bl$, and $K$ has a natural decomposition into sectors of a given color $c(\Abox)=\o$,
\begin{equation}
K=\bigoplus_{\o\in\mZ_p}K_\o(\bl).
\end{equation} 

\paragraph{Notations} We denote $C_\o(m)$ the subset of $[\![1,m]\!]$ such that the Coulomb branch vevs $a_\a$ (or $v_\a$) with $\a\in C_\o(m)$ have color $c_\a=\o$ (or, equivalently, that the box $(1,1)\in\l^{(\a)}$ with $\a\in C_\o(m)$ is of color $c(\a,1,1)=c_\a=\o$). Similarly, $K_\o(\bl)$ denotes the set of boxes $\Abox\in\bl$ of the $m$-tuple colored partition $\bl$ that carry the color $c(\Abox)=\o$. Besides, in the generic case $\nu_1+\nu_2\neq0$, the shift of color indices $\o$ by the quantity $\nu_1+\nu_2$ appears in many formulas. To simplify these expressions, we introduce the notation $\bar\o=\o+\nu_1+\nu_2$ for the shifted indices, along with the map $\bar c(\Abox)=c(\Abox)+\nu_1+\nu_2$. Finally, we also introduce the extra variables $q_3$ and $\nu_3$ such that $q_1q_2q_3=1$ and $\nu_1+\nu_2+\nu_3=0$. Due to the fact that the $\Zp$-action coincides with a subgroup of the torus action, in all formulas the shift of color indices $\o+\nu_i$ coincide with a factor $q_i$ multiplying the parameters associated to instanton positions in the moduli space.

\paragraph{McKay subgroups in $SO(4)$} Although we are considering here a different problem, it is interesting to make a short parallel with the action of $SU(2)_L\times SU(2)_R\subset SO(4)$ on the omega-background (see, for instance, \cite{Johnson1996}). This action takes a simpler form if we employ the quaternionic coordinates
\begin{equation}
Z=\left(\mat{z_1}{-\bz_2}{z_2}{\bz_1}\right),\quad (z_1,z_2)\in\mathbb{C}_{\e_1}\times\mathbb{C}_{\e_2}.
\end{equation} 
Then the $2\times2$ matrices $(G_L,G_R)\in SU(2)_L\times SU(2)_R$ act on the quaternions as $Z\to G_LZG_R$. The McKay subgroups of $SU(2)$ possess an ADE-classification. For instance, the $A_{p-1}$ series corresponds to the action of $\mathbb{Z}_p$, it is generated (multiplicatively) by the diagonal matrices
\begin{equation}
G=\left(\mat{e^{2i\pi/p}}{0}{0}{e^{-2i\pi/p}}\right).
\end{equation} 
Considering only the action of the $A_{p-1}$ subgroup on the left, the background coordinates transform as $(z_1,z_2)\to(e^{2i\pi/p}z_1,e^{-2i\pi/p}z_2)$. This transformation can be recovered from the action \ref{Zp_action} of $\mathbb{Z}_p$ by choosing $\nu_1=-\nu_2=1$. The orbifold of the spacetime under this action of $\mathbb{Z}_p$ reproduces the ALE space constructed in \cite{Kronheimer1989}. Instantons of $\CN=1$ gauge theories defined on ALE spacetimes have been extensively studied \cite{Kronheimer1989,Kronheimer1990,Nakajima1994,Nakajima1999}. In \cite{Awata2017}, their contributions to the gauge theory partition functions have been reproduced using algebraic techniques based on the quantum toroidal algebra of $\glp $. The generalization to DE-type McKay subgroups with only left action is expected to involve quantum toroidal algebras based on either $\mathfrak{so}(p)$ or $\mathfrak{sp}(p)$ Lie algebras \cite{Nakajima1999}.

It is also possible to consider simultaneously the action of two McKay subgroups $A_{p_1-1}$ and $A_{p_2-1}$, with one acting on the left, the other on the right. As a result, coordinates now transform as
\begin{equation}
(z_1,z_2)\to (e^{2i\pi(p_1+p_2)/(p_1p_2)}z_1,e^{2i\pi(p_1-p_2)/(p_1p_2)}z_2).
\end{equation}
We recognize here another particular case of the $\mathbb{Z}_p$-action defined in \ref{Zp_action}, albeit more general than before. It is simply obtained by the specialization $\nu_1=p_1+p_2$, $\nu_2=p_1-p_2$ and $p=p_1p_2$. Thus, the action \ref{Zp_action} leads to a particularly rich context. Moreover, taking $\nu_1=0$, the first coordinate $z_1$ is invariant and the orbifolded spacetime can be reinterpreted as the insertion of a codimension-two defect in a 5D omega background with no orbifold \cite{Nekrasov_BPS4,Kanno2011}. We build here a general algebraic framework to address this kind of problems. It may be possible to further generalize our approach to the action of DE-type McKay subgroups with both left and right actions, but this is beyond the scope of this paper.

\subsection{Instantons partition function}
The computation of the Nekrasov instanton partition function on such $\Zp$-orbifolds has been performed in \cite{Kanno2011,Bonelli2012,Nekrasov_BPS3}.\footnote{Strictly speaking, in \cite{Bonelli2012}, the authors consider the instantons on the minimal resolution of the orbifold. Instead, here, following \cite{Kanno2011,Nekrasov_BPS3}, we simply consider the $\Zp$-invariant part of the instanton moduli space $\CM_k$. Both approaches should provide the same result \cite{Belavin2011a}.} For simplicity, we do not introduce fundamental matter multiplets, those being obtained in the limit $\qf\to0$ of the gauge coupling parameters. Furthermore, we only discuss the case of linear quiver gauge theories $A_r$, with $U(m^{(i)})$ gauge groups at each node $i=1\cdots r$. Thus, the node $i$ carries the following parameters:
\begin{itemize}
\item a set of colored exponentiated gauge couplings $\qf_{\o,i}$,
\item a $p$-vector of colored Chern-Simons levels $\boldsymbol{\k}^{(i)}=(\k_\o^{(i)})_{\o\in\Zp}$,
\item an $m^{(i)}$-vector of Coulomb branch vevs $\boldsymbol{a}^{(i)}=(a_\a^{(i)})_{\a=1}^{m^{(i)}}$ defining the exponentiated parameters $\bv^{(i)}=(v_\a^{(i)})_{\a=1}^{m^{(i)}}$ with $v_\a^{(i)}=e^{R a_\a^{(i)}}$,
\item an associated vector of colors $\boldsymbol{c}^{(i)}=(c_\a^{(i)})_{\a=1}^{m^{(i)}}$.
\end{itemize}
In addition, each link $i\to j$ between two nodes $i$ and $j$ represent a chiral multiplet of matter fields in the bifundamental representation of the gauge group $U(m^{(i)})\times U(m^{(j)})$, with mass $\mu_{ij}\in\mathbb{C}$. For linear quivers, all bifundamental masses can be set to $q_3^{-1}$ by a rescaling of the Coulomb branch vevs. 

The instantons contribution to the gauge theories partition function is expressed as a sum over the content of $r$ $m^{(i)}$-tuple Young diagrams $\bl^{(i)}$ describing the configuration of instantons at the $i$th node. Each term can be further decomposed into the contributions of vector (gauge) multiplets, bifundamental chiral (matter) multiplets, and Chern-Simons factors:
\begin{align}\label{def_Zv_Zbf}
\begin{split}
&\Zinst=\sum_{\bl^{(i)}}\prod_{i=1}^{r}\left(\prod_{\o\in\Zp}\qf_{\o,i}^{|K_\o(\bl^{(i)})|}\Zv(\bv^{(i)},\bl^{(i)})\ZCS(\boldsymbol{\k}^{(i)},\bl^{(i)})\right)\prod_{i\to j}\Zbf(\bv^{(i)},\bl^{(i)},\bv^{(j)},\bl^{(j)}|\mu_{ij}),\\
&\text{with}\quad \Zv(\bv,\bl)=N(\bv,\bl|\bv,\bl)^{-1},\quad \Zbf(\bv,\bl,\bv',\bl'|\mu)=N(\bv,\bl|\mu\bv',\bl'),\quad \ZCS(\boldsymbol{\k},\bl)=\prod_{\mAbox\in\bl}\chi_{\sAbox}^{\k_{c(\sAbox)}}.
\end{split}
\end{align}
Vector and bifundamental contributions are written in terms of the Nekrasov factor $N(\bv,\bl|\mu\bv',\bl')$. For a better readability, we drop the node indices in the following, and simply distinguish the two nodes involved in the definition of the Nekrasov factor with a prime. In order to write down the expression of $N(\bv,\bl|\mu\bv',\bl')$ given in \cite{Nekrasov_BPS3}, we need to introduce the equivariant character $M_\bv$ and $K_\bl$ of the vector spaces $M$ and $K$ associated to each node,
\begin{equation}\label{equiv_char}
M_{\bv}=\sum_{\a=1}^m e^{Ra_\a},\quad K_\bl=\sum_{\mAbox\in\bl}e^{R\phi_{\sAbox}},
\end{equation}
A linear involutive operation $\ast$ acts on such characters by flipping the sign of $R$: $(e^{R a_\a})^\ast=e^{-Ra_\a}$, $(q_1^\ast,q_2^\ast)=(q_1^{-1},q_2^{-1})$ and thus $(e^{R\phi_{\sAbox}})^\ast=e^{-R\phi_{\sAbox}}$ (see \cite{Nekrasov_BPS1,Nekrasov_BPS2,Nekrasov_BPS3} for more details on these notations). Introducing $S_\bl=M-P_{12}K_\bl$ with $P_{12}=(1-q_1)(1-q_2)$, the Nekrasov factor writes
\begin{equation}\label{expr_N}
N(\bv,\bl|\bv',\bl')=\mathbb{I} \left[\dfrac{M_{\bv}M_{\bv'}^\ast-S_{\bl}S_{\bl'}^\ast}{P_{12}^\ast}\right]^{\mathbb{Z}_p} = \mathbb{I} \left[ M_{\bv} K_{\bl'} ^* + q_{3} ^{-1} M_{\bv'} ^* K_{\bl} -P_{12} K_{\bl} K_{\bl'}^* \right]^{\mathbb{Z}_p},
\end{equation}
where the $\mathbb{I}$-symbol is the equivariant index functor,
\begin{equation}
\mathbb{I}\left[\sum_{i\in I_+}e^{R w_i}-\sum_{i\in I_-}e^{R w_i}\right]=\dfrac{\prod_{i\in I_+} 1-e^{R w_i}}{\prod_{i\in I_-} 1-e^{R w_i}},
\end{equation}
and $\left[ \cdots \right]^{\mathbb{Z}_p}$ denotes the operation of keeping only the $\mathbb{Z}_p$-invariant parts. In particular, the RHS of \eqref{expr_N} involves a coloring function $c:\mZ[a_\a,\e_1,\e_2]\to\Zp$ defined on weights $w_i$ as the linear map taking the values $c(a_\a)=c_\a$, $c(\e_1)=\nu_1$ and $c(\e_2)=\nu_2$ so that $c(\phi_{\sAbox})=c(\Abox)$ (justifying our slight abuse of notations). The $[\cdots]^{\mathbb{Z}_p}$ projects on $\Zp$-invariant factors.

Replacing the equivariant characters by their expressions \ref{equiv_char}, the Nekrasov factor can be written in a more explicit form,
\begin{equation}\label{Nekrasov}
N(\bv,\bl|\bv',\bl')=\prod_{\superp{\mAbox\in\bl}{\mAboxF\in\bl'}}S_{c(\sAbox)c(\sAboxF)}(\chi_{\sAbox}/\chi_{\sAboxF})\times\prod_{\mAbox\in\bl}\prod_{\a\in C_{\bar c(\sAbox)}(m')}\left(1-\dfrac{\chi_{\sAbox}}{q_3v_{\a}'}\right)\times \prod_{\mAbox\in\bl'}\prod_{\a\in C_{c(\sAbox)}(m)}\left(1-\dfrac{v_\a}{\chi_{\sAbox}}\right).
\end{equation} 
The function $S_{\o\o'}(z)$ is sometimes called the \textit{scattering function}, it carries two color indices $\o,\o'$:
\begin{equation}\label{def_S}
S_{\o\o'}(z)=\dfrac{(1-q_1z)^{\d_{\o,\o'-\nu_1}}(1-q_2z)^{\d_{\o,\o'-\nu_2}}}{(1-z)^{\d_{\o,\o'}}(1-q_1q_2z)^{\d_{\o,\o'-\nu_1-\nu_2}}}.
\end{equation}
In this expression, the non-zero matrix elements have been expressed in a compact way using the delta function $\d_{\o,\o'}$ defined modulo $p$ (i.e. $\d_{\o,\o'}=1$ iff $\o=\o'$ modulo $p$, zero otherwise). In fact, $S_{\o\o'}(z)$, and more generally all the matrices of size $p\times p$ with indices $\o,\o'$ appearing in this paper, are circulant matrices: their matrix elements only depend on the difference $\o-\o'$ of row and column indices. In particular, $S_{\o+\nu\ \o'}(z)=S_{\o\ \o'-\nu}(z)$ for all $\nu\in\mathbb{Z}_p$. Finally, the function $S_{\o\o'}(z)$ satisfies a sort of \textit{crossing symmetry},
\begin{equation}\label{crossing}
S_{\o\bar\o'}(q_3/z)=f_{\o\o'}(z)S_{\o'\o}(z),
\end{equation} 
with the function $f_{\o\o'}(z)=F_{\o\o'}z^{\b_{\o\o'}}$ defined by\footnote{The function $f_{\o\o'}(z)$ also controls the asymptotics of the scattering function since $S_{\o\o'}(z)\superpsim{0}1$ and $S_{\o\o'}(z)\superpsim{\infty}f_{\o'\o}(z)^{-1}$. It obeys an important reflection symmetry $f_{\bo\o'}(q_3/z)=f_{\o'\o}(z)^{-1}$ coming from $F_{\o\o'}F_{\bar\o'\o}=q_3^{-\b_{\o\o'}}$ and $\b_{\bar\o'\o}=\b_{\o\o'}$.}
\begin{equation}\label{def_beta}
\b_{\o\o'}=\d_{\o\o'}+\d_{\o\o'+\nu_1+\nu_2}-\d_{\o\o'+\nu_1}-\d_{\o\o'+\nu_2},\quad F_{\o\o'}=(-1)^{\d_{\o\o'}}(-q_3)^{-\d_{\o,\o'-\nu_3}}
(-q_1)^{-\d_{\o\o'+\nu_1}}(-q_2)^{-\d_{\o\o'+\nu_2}}.
\end{equation}

\subsection{$\CY$-observables}
A new class of BPS-observables for supersymmetric gauge theories was introduced in \cite{Nekrasov_BPS1}, they are called \textit{$qq$-characters}. As the name suggests, they correspond to a natural deformation of the $q$-characters of Frenkel-Reshetikhin \cite{Frenkel1998} from the gauge theory point of view \cite{NPS}. They were defined in \cite{Nekrasov_BPS1} as particular combinations of chiral ring observables in such a way that their expectation values exhibit an important regularity property \cite{Nekrasov_BPS1,Nekrasov_BPS2}. This regularity property encodes an infinite set of constraints called \textit{non-perturbative Dyson-Schwinger equations}. From a different viewpoint, $qq$-characters in 5D gauge theories can also be defined in terms of Wilson loops \cite{Kim2016} (see also \cite{Kimura2017} for a string theory perspective).\footnote{In \cite{Nekrasov_BPS4}, the $qq$-characters of 4D $\CN=2$ gauge theories with the insertion of surface defects were considered. In this case, the non-perturbative Dyson-Schwinger equations produce either Knizhnik - Zamolodchikov equations or BPZ equations that are satisfied by the surface defect partition functions \cite{Nekrasov_BPS5,Jeong2017}. These surface defect partition functions were investigated in the context of Bethe/gauge correspondence in \cite{Jeong2017'}, and in their relation to the oper submanifold of the moduli space of flat connections on Riemann surfaces in \cite{Jeong2018}.}

The $qq$-characters are half-BPS observables written as combinations of \textit{$\CY$-observables}. In the case of a $\Zp$-orbifold, it is natural to introduce two inequivalent $\CY$-observables $\CYY(z)=\mathbb{I}\left[e^{-R\z}S_\bl \right]^{\mathbb{Z}_p}$ and $\CY_\o^{[\bl]\ast}(z)=\mathbb{I}\left[e^{R\z}S_\bl^\ast \right]^{\mathbb{Z}_p}$ where $z=e^{R\z}$ and $c(\z)=\o$. These two observables encode the recursion relations satisfied by Nekrasov factors, 
\begin{equation}\label{rec_N}
\dfrac{N(\bv,\bl|\bv',\bl'+\Abox)}{N(\bv,\bl|\bv',\bl')}=\CY_{c(\sAbox)}^{[\bl]}(\chi_{\sAbox})\quad\dfrac{N(\bv,\bl+\Abox|\bv',\bl')}{N(\bv,\bl|\bv',\bl')}=\CY_{\bc(\sAbox)}^{[\bl']\ast}(q_3^{-1}\chi_{\sAbox}),
\end{equation}
Replacing the equivariant characters with the expressions \ref{equiv_char}, we find the explicit formulas
\begin{equation}\label{def_Y}
\CYY(z)=\prod_{\a\in C_\o(m)}(1-v_\a/z)\times\prod_{\mAbox\in\bl}S_{c(\sAbox)\o}(\chi_{\sAbox}/z),\quad \CY_\o^{[\bl]\ast}(z)=\prod_{\a\in C_\o(m)}(1-z/v_\a)\prod_{\mAbox\in\bl}S_{\o\bc(\sAbox)}(q_3z/\chi_{\sAbox}).
\end{equation}
Due to the crossing symmetry \ref{crossing}, these two $\CY$-observables satisfy the relation
\begin{align} \label{eq:yandy}
\CY_\o^{[\bl]\ast}(z)=f\oY(z)\CY\oY(z),
\end{align}
with
\begin{equation} \label{eq:f}
\fYY(z)=\prod_{\a\in C_\o(m)}(-z/v_\a)\prod_{\mAbox\in\bl}f_{\o c(\sAbox)}(\chi_{\sAbox}/z).
\end{equation} 
It will allow us to express all the equations below in terms of $\CYY(z)$ only.\footnote{The presence of the function $\fYY(z)$ can be interpreted as
follows. Note that $\mathbb{I}(X^\ast)=(-1)^{\text{rk}X^\ast}\det
X^\ast\ \mathbb{I}(X)$, for $X=\sum_{i\in I_+}e^{R w_i}-\sum_{i\in I_-
}e^{Rw_i}$, $\text{rk} X=|I_+|-|I_-|$, and $\det X=\prod_{i\in
I_+}e^{Rw_i}/\prod_{i\in I_-}e^{Rw_i}$. Applying this reflection
relation to $X=\left[e^{-R\z}S_\bl \right]^{\mathbb{Z}_p}$, we recover the relation \eqref{eq:yandy} with $\fYY(z)$
given in \eqref{eq:f} identified with $(-1)^{\text{rk}X^\ast}\det
X^\ast$.} Furthermore, the $\CY$-observables possess an alternative expression following from the \textit{shell formula} derived in appendix \refOld{AppB0},
\begin{equation}\label{shell_Y}
\CYY(z)=\dfrac{\prod_{\mAbox\in A_\o(\bl)}(1-\chi_{\sAbox}/z)}{\prod_{\mAbox\in R_{\o-\nu_1-\nu_2}(\bl)}(1-\chi_{\sAbox}/(q_3z))},\quad \fYY(z)=\dfrac{\prod_{\mAbox\in R_{\o-\nu_1-\nu_2}(\bl)}(-\chi_{\sAbox}/(q_3z))}{\prod_{\mAbox\in A_\o(\bl)}(-\chi_{\sAbox}/z)}.
\end{equation}
Here, the sets $A_\o(\bl)$ and $R_\o(\bl)$ denote respectively the set of boxes of color $\o$ that can be added to or removed from the $m$-tuple Young diagram $\bl$. This expression arises from the cancellations of contributions by neighboring boxes, it plays an essential role in the definition of the vertical representation of the algebra.

\section{New quantum toroidal algebras}
In order to reconstruct the instanton partition functions on the general orbifold \ref{Zp_action}, the definition of a new quantum toroidal algebra is necessary. In addition to the complex parameters $q_1,q_2$ and the rank $p\in\mZ^{>0}$, this algebra will depend on the integers $(\nu_1,\nu_2)$ modulo $\Zp$. Taking $\nu_1=-\nu_2=1$, the $\Zp$-action \ref{Zp_action} reduces to the standard action defining ALE spaces. Thus, in this limit the $(\nu_1,\nu_2)$-deformed algebra should reduce to the quantum toroidal algebra of $\glp$. In fact, this is true only up to a twist in the definition of the Drinfeld currents (see the appendix \ref{AppE}). A brief reminder on the quantum toroidal algebra of $\glp$ is given in appendix \refOld{AppA}, it includes its two main representations called, in the gauge theory context, \textit{vertical} and \textit{horizontal} representations.

The key ingredient to define the deformation of the quantum toroidal algebra of $\glp$ is the scattering function $S_{\o\o'}(z)$ defined in \ref{def_S}. Indeed, this function plays an essential role in the two elementary representations involved in the algebraic engineering of partition functions and $qq$-characters. In the vertical representation, it enters through the definition \ref{def_Y} of the $\CY$-observables that describe the recursion relations among Nekrasov factors. Instead, in the horizontal representation, it expresses the normal-ordering relations between vertex operators. Thus, from the physics perspective, the scattering function is the natural object to consider for the deformation of the algebra. Moreover, through the crossing symmetry relation \ref{crossing}, this function defines the $p\times p$ matrix $\b_{\o\o'}$ that could be identified with the underlying Cartan matrix of the deformed quantum toroidal algebra (see the subsection \refOld{AppA4}). Note that the matrix $\b_{\o\o'}$ naturally reduces to the generalized Cartan matrix of the Kac-Moody algebra $\widehat{\glp}$ when $\nu_1=-\nu_2=1$. In general, it is non-symmetrizable, yet, like in the case of $\widehat{\glp}$, it is a circulant matrix. Its eigenvectors $v_j=(1,\O_j,\O_j^2,\cdots,\O_j^{p-1})$ are written in terms of the $p$th root of unity $\O_j=e^{2i\pi j/p}$, and the corresponding eigenvalues read
\begin{equation}
\l_j=-4e^{i\pi\nu_3 j/p}\sin(\pi\nu_1 j/p)\sin(\pi\nu_2 j/p).
\end{equation}
In particular, the eigenvector $v_0=(1,1,\cdots,1)$ has the eigenvalue zero which relates $\b_{\o\o'}$ to the Cartan matrix of affine Lie algebras, and thus justifies the designation \textit{toroidal} of the deformed algebra.

\subsection{Definition of the algebra}
Like in the case of $\glp$, the $(\nu_1,\nu_2)$-deformed quantum toroidal algebra is defined in terms of a central element $c$ and $4p$ Drinfeld currents, denoted $x_\o^\pm(z)$ and $\psi_\o^\pm(z)$, with $\o\in\Zp$. The currents $\psi_\o^\pm(z)$ (together with $c$) generate the Cartan subalgebra, while the currents $x_\o^\pm(z)$ deform the notion of Chevalley generators $e_\o,f_\o$. The algebraic relations obeyed by the currents resemble those defining the quantum toroidal algebra of $\glp$ in \ref{algebra_glp}, the main difference being the presence of shifts in the indices $\o$ by the product $\nu_3 c$:\footnote{Comparing with the standard definition of quantum toroidal algebras, the Drinfeld currents have been redefined as follows: $x^\pm(z)\to x^\pm(q_3^{\pm c/4}z)$, $\psi_\o^+(z)\to\psi_\o^+(z)$ and $\psi_\o^-(z)\to\psi_\o^-(q_3^{-c/2}z)$. This redefinition makes the coincidence between shifts of indices $\o\pm\nu_3c$ and spectral parameters $zq_3^{\pm c}$ manifest. In fact, this asymmetric form of the algebraic relations appears naturally in the construction of a central extension of the Yangian double \cite{Khoroshkin1996}.}\footnote{We note that the algebra is well-defined only for representations with integer level $c$ because of the shifts $-\nu_3c$ in the indices of the matrix $g_{\o\o'}(z)$. More rigorously, one should introduce the structure functions $g(z,\o,\o')$ that reduce to the expression $g_{\o\o'}(z)$ given in \ref{def_g} for $\o,\o'\in\mZ$.}
\begin{align}\label{algebra}
\begin{split}
&x_\o^\pm(z)x_{\o'}^\pm(w)=g_{\o\o'}(z/w)^{\pm1}x_{\o'}^\pm(w)x_\o^\pm(z),\quad \psi^+_\o(z)x^\pm_{\o'}(w)=g_{\o\o'}(z/w)^{\pm1}x^\pm_{\o'}(w)\psi^+_\o(z),\\
&\psi^-_\o(z)x^+_{\o'}(w)=g_{\o-\nu_3c\ \o'}(q_3^{-c}z/w)x^+_{\o'}(w)\psi^-_\o(z),\quad \psi^-_\o(z)x^-_{\o'}(w)=g_{\o\o'}(z/w)^{-1}x^-_{\o'}(w)\psi^-_\o(z),\\
&\psi_\o^+(z)\psi_{\o'}^-(w)=\dfrac{g_{\o\o'-\nu_3c}(q_3^{c} z/w)}{g_{\o\o'}(z/w)}\psi_{\o'}^-(w)\psi_\o^+(z),\quad [\psi_\o^\pm(z),\psi_{\o'}^\pm(w)]=0,\\
&[x_\o^+(z),x_{\o'}^-(w)]=\kO\left[\d_{\o,\o'}\d(z/w)\psi_\o^+(z)-\d_{\o,\o'-\nu_3c}\d(q_3^{c} z/w)\psi^-_{\o+\nu_3c}(q_3^{c}z)\right].
\end{split}
\end{align}
In the last relation $\d(z)=\sum_{k\in\mZ}z^k$ denotes the multiplicative Dirac delta function and we introduced the complex parameter
\begin{equation}
\kO=\dfrac{(1-q_1)^{\d_{\nu_1,0}}(1-q_2)^{\d_{\nu_2,0}}}{(1-q_1q_2)^{\d_{\nu_1+\nu_2,0}}}F^{1/2},\quad F=F_{\o\o}=-\prod_i(-q_i)^{-\d_{\nu_i,0}}.
\end{equation}
The other relations in \ref{algebra} involve the \textit{structure function} $g_{\o\o'}(z)$ defined as a ratio of two scattering functions. This function depends on the variables $(q_1,q_2)\in\mathbb{C}^\times\times\mathbb{C}^\times$ and the integers $(\nu_1,\nu_2)\in\Zp\times\Zp$:
\begin{equation}\label{def_g}
g_{\o\o'}(z)=\dfrac{S_{\o\o'}(z)}{S_{\o'\o}(z^{-1})}=f_{\o\o'}(z^{-1})\prod_{i=1,2,3}\dfrac{(1-q_iz)^{\d_{\o,\o'-\nu_i}}}{(1-q_i^{-1}z)^{\d_{\o,\o'+\nu_i}}},
\end{equation}
where the extra variables $q_3$ and $\nu_3$ obey $q_1q_2q_3=1$ and $\nu_1+\nu_2+\nu_3=0$. Note that the invariance under the $S_3$-permutation of indices $(\nu_i,q_i)$ is broken to $S_2$ corresponding to exchange $(\nu_1,q_1)$ and $(\nu_2,q_2)$. The structure function satisfies the property $g_{\o\o'}(z)g_{\o'\o}(z^{-1})=1$ necessary for the definiteness of the algebraic relations. 

The algebraic relations \ref{algebra} are expected to include additional Serre relations. However, the Drinfeld currents employed here are a twisted version of those used in the formulation of the quantum toroidal algebra of $\glp$. This explains why the function $g_{\o\o'}(z)$ defined in \ref{def_g} does not quite reproduce the $\glp$ structure function \ref{def_g_glp} as we set $\nu_1=-\nu_2=1$. Even in the case of $\glp$, the twist of the currents make the derivation of Serre relations difficult. We hope to come back to this question in the near future.

Due to the non-trivial power of $z$ in the asymptotics of the functions $g_{\o\o'}(z)$, namely
\begin{equation}
g_{\o\o'}(z)\superpsim{0}f_{\o\o'}(z^{-1}),\quad g_{\o\o'}(z)\superpsim{\infty}f_{\o'\o}(z)^{-1},
\end{equation}
the Cartan currents $\psi_\o^\pm(z)$ cannot be expanded in powers of $z^{\mp k}$ with $k>0$ as it is usually the case for quantum groups. Instead, it is necessary to introduce a \textit{zero modes} part using extra operators $a_{\o,0}^\pm$:
\begin{equation}
x_\o^\pm(z)=\sum_{k\in\mathbb{Z}}z^{-k}x_{\o,k}^\pm,\quad \psi_\o^\pm(z)=\psi_{\o,0}^\pm z^{\mp a_{\o,0}^\pm}\exp\left(\pm\sum_{k>0}z^{\mp k}a_{\o,\pm k}\right).
\end{equation} 
In the appendix \ref{AppD}, the operators $\psi_{\o,0}z^{\mp a_{\o,0}^\pm}$ are constructed as a specific combination of grading operators. The Cartan zero modes $\psi_{\o,0}^\pm$ are invertible, they can be used to define another central element $\bc$ setting
\begin{equation}\label{def_bc}
q_3^{-\bc}=\left(\prod_{\o\in\Zp}\psi_{\o,0}^+\right)\left(\prod_{\o\in\Zp}\psi_{\o,0}^-\right)^{-1}=\prod_{\superp{\o,\o'=0}{\o\leq\o'}}^{p-1}\dfrac{F_{\o'\o}}{F_{\o'\o+\nu_3c}}\prod_{\o\in\Zp}\psi_{\o,0}^+(\psi_{\o,0}^-)^{-1}.
\end{equation} 
Note that the ordering of the zero modes is important since they do not commute. It is chosen here such that the expression of the coproduct defined below simplifies.

\paragraph{Coalgebraic structure} A Hopf algebra $\CA$ over the field $\mC$ is a $\mC$-module equipped with a unit $1_\CA$, a product $\nabla$, a counit $\e$, a coproduct $\D$ and an antipode $S$ satisfying the following properties \cite{Chari1995}.
\begin{itemize}
\item $\CA$ is both an algebra and a coalgebra. This implies the property $\nabla(1\otimes\e)\D=\nabla(\e\otimes1)\D=1$ and the coassociativity of the coproduct $(1\otimes\D)\D=(\D\otimes1)\D$.
\item The counit $\e:\CA\to\mC$ and the coproduct $\D:\CA\to\CA\otimes\CA$ are homomorphisms of algebras. The compatibility with the scalar multiplication and the addition are trivially satisfied. On the other hand, the compatibility with the product requires to verify $\e(ee')=\e(e)\e(e')$ and $\D(e)\D(e')=\D(ee')$ for any two elements $e,e'\in\CA$. %The preservation of units further requires $\e(1)=1$ and $\D(1)=1\otimes 1$.
\item The unit $1_\CA:\mC\to\CA$ and product $\nabla:\CA\otimes\CA\to\CA$ are homomorphisms of coalgebras. This means $\D(1_\CA)=1_\CA\otimes 1_\CA$, $\e(1_\CA)=1$, and, once again, $\D(e)\D(e')=\D(ee')$.
\item The antipode $S:\CA\to\CA$ is a bijective $\mC$-module map satisfying $\nabla(S\otimes1)\D=\e=\nabla(1\otimes S)\D$.
\end{itemize}
The algebra \ref{algebra} is a Hopf algebra with the coproduct, counit and antipode given by
\begin{align}
\begin{split}\label{Drinfeld_coproduct}
&\D(x_\o^+(z))=x_\o^+(z)\otimes 1+\psi_{\o+\nu_3 c_{(1)}}^-(q_3^{c_{(1)}}z)\otimes x_\o^+(z),\\
&\D(x_\o^-(z))=x_\o^-(z)\otimes \psi_{\o-\nu_3c_{(1)}}^+(q_3^{-c_{(1)}}z)+1\otimes x_{\o-\nu_3c_{(1)}}^-(q_3^{-c_{(1)}}z),\\
&\D(\psi_\o^+(z))=\psi_\o^+(z)\otimes\psi_{\o-\nu_3 c_{(1)}}^+(q_3^{-c_{(1)}}z),\quad \D(\psi_\o^-(z))=\psi_{\o-\nu_3c_{(2)}}^-(q_3^{-c_{(2)}}z)\otimes\psi_{\o-\nu_3c_{(1)}}^-(q_3^{-c_{(1)}}z),\\
&S(x_\o^+(z))=-\psi_{\o+\nu_3c}^-(q_3^cz)^{-1}x_\o^+(z),\quad S(x_\o^-(z))=-x_{\o+\nu_3c}^-(q_3^cz)\psi_{\o+\nu_3c}^+(q_3^cz)^{-1},\quad \e(x_\o^\pm(z))=0,\\
&S(\psi_\o^+(z))=\psi_{\o+\nu_3c}^+(q_3^{c}z)^{-1},\quad S(\psi_\o^-(z))=\psi_{\o+2\nu_3c}^-(q_3^{2c}z)^{-1},\quad \e(\psi_\o^\pm(z))=1,
\end{split}
\end{align}
with the standard notation $c_{(1)}=c\otimes1$, $c_{(2)}=1\otimes c$. The central element $c$ obeys $\D(c)=c_{(1)}+c_{(2)}$, $S(c)=-c$ and $\e(c)=0$. The proof is a tedious but straightforward calculation that the axioms defining a Hopf algebra hold for any pair of currents. The antipode is an anti-homomorphism of algebra, it satisfies $S^2=(-1)^{1+\e}\text{Id}$. Using the coproduct of the Cartan zero modes $\psi_{\o,0}^\pm$, it is possible to compute the coproduct of the central charge $\bc$ defined in \ref{def_bc}, we find\footnote{The extra factor in the RHS comes from the shifts of the currents' arguments in the coproduct that brings
\begin{equation}
\D(\psi_{\o,0}^+)=\psi_{\o,0}^+\otimes q_3^{2c_{(1)}a_{\o-\nu_3c_{(1)}}^+}\psi_{\o-\nu_3c_{(1)},0}^+,\quad \D(\psi_{\o,0}^-)=\psi_{\o-\nu_3c_{(2)},0}^-q_3^{-2c_{(2)}a_{\o-\nu_3c_{(2)}}^-}\otimes \psi_{\o-\nu_3c_{(1)},0}^-q_3^{-2c_{(1)}a_{\o-\nu_3c_{(1)}}^-}.
\end{equation}}
\begin{equation}\label{coprod_bc}
\D(q_3^{-\bc})=(q_3^{-\bc}\otimes q_3^{-\bc})\left(q_3^{c_{(2)}\sum_{\o\in\Zp} a_{\o,0}^-}\otimes q_3^{c_{(1)}\sum_{\o\in\Zp} a_{\o,0}^+}q_3^{c_{(1)}\sum_{\o\in\Zp} a_{\o,0}^-}\right).
\end{equation} 

In order to reconstruct the instanton partition functions, we need to introduce two types of representations: a vertical representation $\rho^{(V)}$ with level $c=0$ and a horizontal representation $\rho^{(H)}$ with level $c=1$. Such representations are already known in the case of quantum toroidal algebras of $\glp$ (see \cite{Feigin2012,Saito1996}, or the brief summary presented in appendix \refOld{AppA}), but also for the quantum toroidal $\mathfrak{gl}(1)$ algebra (or Ding-Iohara-Miki algebra \cite{Ding1997,Miki2007}) \cite{feigin2011quantum,Feigin2009a}. In fact, there are two different point of view concerning these representations. In the mathematics literature \cite{feigin2011quantum,Feigin2009a,Feigin2012}, one often considers a single module, the \textit{Fock module}, and present the action of two subalgebras called horizontal and vertical. Miki's automorphism $\CS$ \cite{Miki1999,Miki2007} exchanges the two subalgebras, allowing us to define (for instance) $\rho^{(H)}=\rho^{(V)}\circ\CS$. On the opposite, physicists usually introduce two different types of modules referred as vertical and horizontal modules, somehow fixing the choice of subalgebra. Of course, the modules are isomorphic thanks to Miki's automorphism and the two point of views are equivalent \cite{Bourgine2018a}. However, no analogue of Miki's automorphism is known yet for the $(\nu_1,\nu_2)$-deformed algebra. Thus, at this stage, we have no choice but to follow the second approach and define two distinct representations. This will be done in the next two subsections.

\subsection{Vertical representation}\label{sec_vert}
The vertical representation presented here is a deformation of the Fock representation for the quantum toroidal algebra of $\glp$ \cite{Feigin2012} (see appendix \refOld{AppA3}). This representation is similar to the usual finite dimensional representations of quantum groups. Indeed, the Cartan currents $\psi_\o^\pm(z)$ are diagonal on a set of weight vectors. The currents $x_\o^-(z)$ annihilates the highest weight (or vacuum) $\dket{\vac}$, and $x_\o^+(z)$ creates excitations. However, the weight vectors are labeled here by the box configurations of an $m$-tuple Young diagrams $\bl$. Thus, this representation is infinite dimensional, yet it is graded by the total number of boxes $|\bl|$.

From the gauge theory perspective, the vertical representation of the algebra \ref{algebra} describes the relation between sectors of different instanton numbers. Thus, vertical modules are characterized by a basis of states $\dket{\bl}$ labeled by instanton configurations. Accordingly, the representation depend on a set of $m$ (highest) weights $\bv=(v_\a)_{\a=1\cdots m}$ and a choice of color $c_\a$ for each weight. This coloring defines the integers $m_\o=|C_\o(m)|$ corresponding to the number of weights $v_\a$ of color $\o$. The integers $m_\o$ provide the levels of the vertical representation: $\rho^{(V)}(c)=0$ and $\rho^{(V)}(\bc)=m$ with $m=\sum_{\o\in\Zp}m_\o$. As mentioned previously, the Cartan currents $\psi_\o^\pm(z)$ are diagonal on the basis $\dket{\bl}$. On the other hand, the operators $x_\o^\pm(z)$ relate the sectors of instanton charge $|\bl|$ and $|\bl|\pm1$ by adding/removing a box to the $m$-tuple Young diagram $\bl$. Their action encodes the recursion relation \ref{rec_N} obeyed by Nekrasov factors \cite{Kanno2013}. The action of the Drinfeld currents on the states $\dket{\bl}$ is derived in appendix \refOld{AppB1}, it reads\footnote{The definition of the vertical representation is not unique due, for instance, to the following invariance of Drinfeld currents at $c=0$:
\begin{equation}
\psi_\o^\pm(z)\to C_\o z^{\a_\o}\psi_\o^\pm(z),\quad x_\o^+(z)\to x_\o^+(z),\quad x_\o^-(z)\to C_\o z^{\a_\o}x_\o^-(z).
\end{equation} 
Here a particular choice is made to simplify the derivation of intertwiners in section 4 below.}
\begin{align}
\begin{split}\label{vert_x_psi_mr}
&\rho^{(V)}(x_\omega^+(z))\dket{\bl}=F^{1/2}\sum_{\mAbox\in A_\omega(\boldsymbol\l)}\d(z/\chi_{\sAbox}) \res_{z=\chi_{\sAbox}}z^{-1}\CYY(z)^{-1}  \dket{\bl+\Abox},\\
&\rho^{(V)}(x_\omega^-(z))\dket{\bl}= \mr{f}_{\bo}^{[\bl]}(q_3^{-1}z)\sum_{\mAbox\in R_\omega(\boldsymbol\l)}\d(z/\chi_{\sAbox})\res_{z=\chi_{\sAbox}}z^{-1}\CY_{\bo}^{[\bl]}(q_3^{-1}z) \dket{\bl-\Abox},\\
&\rho^{(V)}(\psi_\omega^\pm(z))\dket{\bl}=\left[\PsiY(z)\right]_\pm\dket{\bl}.
\end{split}
\end{align}
In the first two lines, $A_\o(\bl)$ and $R_\o(\bl)$ correspond respectively to the set of boxes of color $\o$ that can be added to or removed from $\bl$. In the last line, the subscript $\pm$ denotes the expansion of the function $\PsiY(z)$ for $|z|^{\pm1}\to\infty$. This function is written as a ratio of the $\CY$-observables defined in \ref{def_Y},
\begin{equation}\label{def_Psi}
\PsiY(z)=\mr{f}_{\bo}^{[\bl]}(q_3^{-1}z)\dfrac{\CY_{\bar\o}^{[\bl]}(q_3^{-1}z)}{\CYY(z)},\quad\text{with}\quad \mr{f}\oY(z)=f_{\o}^{[\bl]}(z)\prod_{\a\in C_\o(m)}(-v_\a/z)=\prod_{\mAbox\in\bl}f_{\o c(\sAbox)}(\chi_{\sAbox}/z).
\end{equation}

We notice that the highest weights are still encoded in the form of a Drinfeld polynomial $p_\o(z)$:
\begin{equation}
\Psi_\o^{[\vac]}(z)=z^{m_\o-m_\bo}\dfrac{\prod_{\a\in C_{\bo}(m)}(-q_3 v_\a)}{\prod_{\a\in C_{\o}(m)}(-v_\a)}\dfrac{p_\bo(q_3^{-1/2}z)}{p_\o(q_3^{1/2}z)},\quad\text{with}\quad p_\o(z)=\prod_{\a\in C_\o(m)}(1-q_3^{-1/2}z/v_\a).
\end{equation} 
When $\nu_3=0$, we have $\bo=\o$ and the prefactor reduces to the usual expression $q_3^{m_\o}$ where $m_\o=\deg p_\o(z)$.

The functions $\fYY(z)$ and $\mr{f}\oY(z)$ control the asymptotics of the functions $\CYY(z)$ and $\PsiY(z)$,
\begin{equation}\label{asympt_PsiY_g}
\CYY(z)\superpsim{\infty}1,\quad \CYY(z)\superpsim{0}\fYY(z)^{-1}\implies \PsiY(z)\superpsim{0}\fYY(z)\dfrac{\mr{f}_{\bo}^{[\bl]}(q_3^{-1}z)}{f_{\bo}^{[\bl]}(q_3^{-1}z)},\quad \PsiY(z)\superpsim{\infty}\mr{f}_{\bo}^{[\bl]}(q_3^{-1}z).
\end{equation}
As a result, the action of the zero-modes of the Cartan currents read
\begin{align}
\begin{split}
&\rho^{(V)}(\psi_{\o,0}^+)\dket{\bl}=\mr{f}\boY(q_3^{-1})\dket{\bl},\quad \rho^{(V)}(\psi_{\o,0}^-)\dket{\bl}=\mr{f}\oY(1)\dfrac{\prod_{\a\in C_\bo(m)}(-q_3v_\a)}{\prod_{\a\in C_\o(m)}(-v_\a)}\dket{\bl},\\
&\rho^{(V)}(a_{\o,0}^+)\dket{\bl}=\left(\sum_{\mAbox\in\bl}\b_{\bo c(\sAbox)}\right)\dket{\bl},\quad \rho^{(V)}(a_{\o,0}^-)\dket{\bl}=\left(m_\o-m_\bo-\sum_{\mAbox\in\bl}\b_{\o c(\sAbox)}\right)\dket{\bl}
\end{split}
\end{align}
The value of the second central charge is obtained by taking the product over $\o$, we recover $\rho^{(V)}(\bc)=m$.

\paragraph{Contragredient representation} The definition of intertwiners in the next section requires the introduction of the dual basis $\dbra{\bl}$. The algebra \ref{algebra} acts on the dual basis with the contragredient representation $\rho^{(V)\ast}$, defined such that
\begin{equation}
\dbra{\bl}\left(\rho^{(V)}(e)\dket{\bl'}\right)=\left(\dbra{\bl}\rho^{(V)\ast}(e)\right)\dket{\bl'},
\end{equation} 
for any element $e$ of the algebra. Thus, the action of the contragredient representation depends on the choice of a scalar product for the vertical states. It turns out that the analysis of intertwining relations simplifies for a particular choice of scalar product for which states are orthogonal but not orthonormal,
\begin{equation}
\dbra{\bl}\bl'\rangle\!\rangle=a_{\bl}(\bv)^{-1}\d_{\bl,\bl'}.
\end{equation}
The norms $a_{\bl}(\bv)^{-1}$ are chosen so that the contragredient representation of $x^\pm_\o(z)$ acts on $\dbra{\bl}$ in the same way as the original representation $\rho^{(V)}(x^\mp_\o(z))$ acts on $\dket{\bl}$ (note that $x^\pm_\o$ becomes $x^\mp_\o$). As a result, the norms have to obey the two following recursion relations for a box $x$ of color $c(\Abox)=\o$:
\begin{align}
\begin{split}
&\dfrac{a_{\bl-\mAbox}(\bv)}{a_\bl(\bv)}=\kO^{-1}f_{\bo}^{[\bl]}(q_3^{-1}\chi_{\sAbox})\res_{z=\chi_{\sAbox}}z^{-1}\CY_\o^{[\bl]}(z)\CY_\bo^{[\bl]}(q_3^{-1}z),\\
&\dfrac{a_{\bl+\mAbox}(\bv)}{a_\bl(\bv)}=-\kO^{-1}Ff_{\bo}^{[\bl]}(q_3^{-1}\chi_{\sAbox})^{-1}\res_{z=\chi_{\sAbox}}z^{-1}\CY_\o^{[\bl]}(z)^{-1}\CY_\bo^{[\bl]}(q_3^{-1}z)^{-1}.
\end{split}
\end{align}
The solution is expressed in terms of the vector contribution $\Zv(\bv,\bl)$ defined in \ref{def_Zv_Zbf},
\begin{equation}
a_{\bl}(\bv)=(-F^{1/2})^{|\bl|}\Zv(\bv,\bl)\prod_{\mAbox\in\bl}\prod_{\a\in C_{\bc(\sAbox)}(m)}(-\chi_{\sAbox}/(q_3v_\a)).
\end{equation} 

\subsection{Horizontal representation}
The horizontal representation of the algebra \ref{algebra} is the equivalent of the vertex representations constructed by Saito in \cite{Saito1996} for quantum toroidal algebras of $\glp$. It has level $\rho^{(H)}(c)=1$ and depends on $p$ weights $u_\o\in\mathbb{C}^\times$ and $p$ integers $n_\o\in\mZ$. In this representation, Drinfeld currents are constructed as a direct product of two (commuting) algebras. The first algebra is called here the \textit{zero modes} factor, it is defined in terms the two operators $Q_\o(z),P_\o(z)$ satisfying the exchange relation
\begin{equation}\label{rel_PQ}
P_\o(z)Q_{\o'}(w)=f_{\o\o'}(w/z)Q_{\o'}(w)P_\o(z).
\end{equation} 
In appendix \refOld{AppB2}, these operators are constructed explicitly in terms of $2p$ Heisenberg algebras. As a result, the operator $P_\o(z)$ acts on the vacuum state $\ket{\vac}$ as $P_\o(z)\ket{\vac}=\ket{\vac}$, and $Q_\o(z)$ acts on the dual vacuum $\bra{\vac}$ as $\bra{\vac}Q_\o(z)=\bra{\vac}$. Accordingly, we define the normal ordering of these operators by writing the $Q_\o(z)$-dependence on the left.

The second algebra involved in the horizontal representation is defined upon the modes $\a_{\o,k}$ of $p$ coupled free bosons ($\o\in\mathbb{Z}_p$ and $k\in\mathbb{Z}^\times$) satisfying the commutation relations,\footnote{The RHS of these commutation relations involves the coefficients $\s_{\o\o'}^{(k)}=-\s_{\bo'\o}^{(-k)}=kq_3^{k/2}\left[\d_{\o\o'}+q_3^{-k}\d_{\o\bo'}-q_1^k\d_{\o\ \o'+\nu_1}-q_2^k\d_{\o\ \o'+\nu_2}\right]$ appearing in the expansion of the scattering function \ref{def_S},
\begin{equation}\label{def_sigma}
\left[S_{\o\o'}(z)\right]_-=\exp\left(\sum_{k>0}\dfrac{z^k}{k^2}q_3^{-k/2}\s_{\o'\o}^{(k)}\right),\quad \left[S_{\o\o'}(z)\right]_+=f_{\o'\o}(z)^{-1}\exp\left(-\sum_{k>0}\dfrac{z^{-k}}{k^2}q_3^{k/2}\s_{\o'\o}^{(-k)}\right).
\end{equation}}
\begin{equation}\label{com_rel}
[\a_{\o,k},\a_{\o',l}]=k\d_{k+l}q_3^{k/2}\left[\d_{\o\o'}+q_3^{-k}\d_{\o\bo'}-q_1^k\d_{\o\ \o'+\nu_1}-q_2^k\d_{\o\ \o'+\nu_2}\right],\quad (k>0).
\end{equation} 
As usual, the vacuum state $\ket{\vac}$ is annihilated by the positive modes ($k>0$), while negative modes create excitations. The dual state $\bra{\vac}$ is annihilated by negative modes. Thus, these modes are normal ordered by moving the positive modes to the right. The representation of the Drinfeld currents $x_\o^\pm$ and $\psi_\o^\pm$ is given in terms of the vertex operators
\begin{align}
\begin{split}\label{def_eta_vphi}
&\eta_\o^+(z)=\exp\left(\sum_{k>0}\dfrac{z^k}{k}\a_{\o,-k}\right)\exp\left(-\sum_{k>0}\dfrac{z^{-k}}{k}q_3^{-k/2}\a_{\o,k}\right),\quad \eta_\o^-(z)=\exp\left(-\sum_{k>0}\dfrac{z^k}{k}\a_{\o,-k}\right)\exp\left(\sum_{k>0}\dfrac{z^{-k}}{k}q_3^{k/2}\a_{\bo,k}\right),\\
&\vphi_\o^+(z)=\exp\left(-\sum_{k>0}\dfrac{z^{-k}}{k}(q_3^{-k/2}\a_{\o,k}-q_3^{k/2}\a_{\bo,k})\right),\quad \vphi_\o^-(z)=\exp\left(\sum_{k>0}\dfrac{z^{k}}{k}(q_3^{-k}\a_{\bo,-k}-\a_{\o,-k})\right).
\end{split}
\end{align}

Combining the zero modes and vertex operators, the horizontal representation writes
\begin{align}
\begin{split}\label{rep_H}
&\rho^{(H)}(x^+_\o(z))=u_\o z^{-n_\o}Q_\o(z)\eta_\o^+(z),\quad \rho_{u}^{(H)}(x^-_\o(z))=u_\o^{-1}z^{n_\o}Q_\o(z)^{-1}P_{\bo}(q_3^{-1}z)\eta_\o^-(z),\\
&\rho^{(H)}(\psi^+_\o(z))=F^{-1/2}P_{\bo}(q_3^{-1}z)\vphi_\o^+(z),\\
&\rho^{(H)}(\psi_\o^-(z))=F^{1/2}\dfrac{u_{\bo}}{u_\o}q_3^{n_{\bo}}z^{n_\o-n_{\bo}}\dfrac{Q_{\bo}(q_3^{-1}z)}{Q_{\o}(z)}P_{\bo}(q_3^{-1}z)\vphi_\o^-(z).
\end{split}
\end{align}
It is shown in appendix \refOld{AppB2} that the expressions in the RHS obey the algebraic relations \ref{algebra} at the levels $\rho^{(H)}(c)=1$ and $\rho^{(H)}(\bc)=n+p$ if $\nu_1+\nu_2<p$ and $\rho^{(H)}(\bc)=n$ otherwise, where $n=\sum_{\o\in\Zp} n_\o$. Note that even in the ALE case $\nu_1=-\nu_2=1$, the horizontal representation given here is slightly more general than the one proposed in \cite{Zenkevich2018}. Indeed, in the latter the $\Zp$-symmetry is broken by a choice of color $\o_0$, setting $u_\o=u\d_{\o,\o_0}$ and $n_\o=n\d_{\o,\o_0}$. Instead, in our construction of the gauge theory partition functions, it is necessary to keep $u_\o$ and $n_\o$ arbitrary in order to be able to assign a different gauge coupling $\qf_\o$ and Chern-Simons level $\k_\o$ for each color $\o$.

\section{Algebraic engineering}
The algebraic engineering of 5D $\CN=1$ quiver gauge theories on $\mC_{\e_1}\times\mC_{\e_2}\times S_R^1$ (without orbifold) follows from their correspondence with topological string theories in which the Nekrasov instanton partition function is obtained as a topological strings amplitude \cite{Aganagic2005}. Indeed, these amplitudes are computed using the (refined) topological vertex \cite{Iqbal2007,Awata2005,Taki2007} that was identified in \cite{Awata2011} with an intertwiner between certain modules of the Ding-Iohara-Miki algebra \cite{Ding1997,Miki2007}, also known as the quantum toroidal algebra of $\mathfrak{gl}(1)$. This intertwiner is in fact the toroidal analogue of the vertex operators introduced in \cite{Davies1992} to compute the form factors of the XXZ Heisenberg spin chain. As result, the powerful topological strings computational methods for supersymmetric gauge theories can be reformulated in the language of quantum integrability.

The correspondence between 5D $\CN=1$ gauge theories and quantum toroidal algebras is better formulated using the $(p,q)$-brane realization of the gauge theories in type IIB string theory \cite{Aharony1997,Aharony1997a}. In this realization, quiver gauge theories are reproduced by the low energy dynamics of a network of 5-branes with charges $(p,q)$. These branes generalize both NS5-branes $(0,1)$ and D5-branes $(1,0)$. They wrap the 5-dimensional spacetime, and define a line segment in the 56-plane of the ten dimensional strings spacetime. These segments meet at trivalent vertices and form a web called the \textit{$(p,q)$-branes web}. For instance, in the case of linear quivers, a set of $m$-D5 branes is associated to each node bearing a $U(m)$ gauge group. These D5-branes are suspended between dressed NS5-branes (i.e. branes of charge $(n,1)$). In this context, the relevant quantum toroidal algebra is determined by the spacetime of the gauge theory. Then, each brane of the $(p,q)$-branes web is associated to a representation of the algebra, identifying the levels with the charges $\rho(c)=q$, $\rho(\bc)=p$ and the weights with the (exponentiated) position of the branes \cite{Awata2017,Mironov2016,Awata2016a}. Thus, to a D5-brane corresponds a vertical representation with $m=1$, while horizontal representations are associated to dressed NS-branes of charge $(n,1)$. It was further noticed in \cite{Bourgine2017b} that the set of $m$ D5-branes of a single node (with a $U(m)$ gauge group) can be directly described by a vertical representation with $\rho^{(V)}(\bc)=m$. Following the identification of the $(p,q)$-branes web with the toric diagram of the Calabi-Yau in topological strings \cite{Leung1998}, the trivalent junctions of branes coincide with the vertex operator of the algebra acting on the modules determined by the branes charge. Finally, the automorphisms of the algebra renders the various geometrical operations (translations, rotations) applied to the branes web \cite{Bourgine2018a}.

For each $(p,q)$-branes web it is possible to write down an operator $\CT$ constructed by `gluing' the vertex operators of nodes connected by an edge. The gluing procedure is done by a product of operators in horizontal representations (NS5), and a scalar product in vertical ones (D5). The $\CT$-operator obtained in this way acts on the tensor product of representations corresponding to the external branes of the web (i.e. the semi-infinite line segments). These representations are in fact horizontal modules, and the vacuum expectation value of the $\CT$-operator reproduces the instantons partition function. The $qq$-characters are further obtained by introducing algebra elements (in the proper representation) within the vacuum expectation value \cite{Bourgine2017b}. We will give several examples below.

This algebraic construction of gauge theories BPS-observables has been generalized in several directions: D-type quivers \cite{Bourgine2017c}, 6D spacetime and elliptic algebras \cite{Foda2018}, 4D $\CN=2$ gauge theories and the affine Yangian of $\mathfrak{gl}(1)$ \cite{Bourgine2018}, 5D $\CN=1$ gauge theories on ALE spaces \cite{Awata2017}, and 3D $\CN=2^\ast$ gauge theories \cite{Zenkevich2018}. In this section, we present yet another generalization corresponding to deformed ALE spaces with the $\Zp$-action described in section two. However, we do not wish to reproduce the whole construction here as it is a straightforward application of the methods developed earlier \cite{Awata2017,Mironov2016,Awata2016a,Bourgine2017b}. Instead, we will only provide the main ingredient, namely the expression of the vertex operators, and a few selected examples to illustrate the construction.

\subsection{Vertex operators}\label{sec_Intw}
We consider two types of vertex operators, denoted $\Phi$ and $\Phi^\ast$, and obtained, up to a normalization factor, by solving the following equations
\begin{equation}\label{intw}
\rho^{(H')}(e)\Phi=\Phi\ \left(\rho^{(V)}\otimes\rho^{(H)}\ \D(e)\right),\quad \left(\rho^{(V)}\otimes\rho^{(H)}\ \D'(e)\right)\ \Phi^\ast=\Phi^\ast\rho^{(H')}(e),
\end{equation} 
where $e$ is any of the currents $x_\o^\pm(z),\psi_\o^\pm(z)$ or the central charge $c$.\footnote{In fact, these relations are also satisfied for the grading operator $\xi_\o(z)$ (see appendix \refOld{AppD}).} Here $\D'$ denotes the opposite coproduct obtained by permutation $\D'=\CP\D\CP$. In order to distinguish the two horizontal representations, we denoted them $\rho^{(H)}$ and $\rho^{(H')}$, they depend on the parameters $u_\o,n_\o$ and $u_\o', n_\o'$ respectively. Thus, the vertex operator $\Phi$ (and also $\Phi^\ast)$ depends on the set of weights $u_\o,u_\o',v_\o$ and integers $n_\o,n'_\o,m_\o$. A solution to the equations \ref{intw} is found only if these parameters satisfy the two constraints
\begin{equation}\label{rel_u_n}
u'_\o=u_\o\prod_{\a\in C_\bo}(-q_3v_\a),\quad n'_\o=n_\o+m_\bo.
\end{equation} 
The first relation expresses a constraint among the position of the branes in the 56-planes. The second equation is the charge conservation at the vertex. Due to the spacetime orbifold, the branes charges $p$ in $(p,q)$ degenerates into charges $p_\o$ with $\o\in\Zp$ identified with the integers $n_\o$ and $m_\bo$ of horizontal/vertical representations.\footnote{There is an unfortunate conflict of notations here since the integer $p$ labeling the $\Zp$-orbifold is unrelated to the charge $p=\sum_\o p_\o$ of the branes.} Summing over $\o$, these constraints reproduce the conservation of the levels $n'=n+m$ that follows from the application of the intertwining relations \ref{intw} to the element $e=\bc$ with the coproduct \ref{coprod_bc}. Due to the presence of an algebra automorphism exchanging $c$ and $\bc$ in the $\glp$-case \cite{Miki1999}, we expect a similar degeneration of the charge $q$ into $q_\o$. It is not observed here because only a single charge $q=1$ flow through the topological vertex.

By definition, the vertex operator $\Phi^\ast$ is a vector in the vertical module while $\Phi$ is a dual vector,
\begin{equation}\label{Phi_vert}
\Phi=\sum_{\bl}\Phi_{\bl}\ \dbra{\bl},\quad \Phi^\ast=\sum_{\bl}\Phi_{\bl}^{\ast}\ \dket{\bl}.
\end{equation}
Each vertical component $\Phi_\bl$ (or $\Phi_\bl^\ast$) is a Fock vertex operator acting on the horizontal module,
\begin{align}\label{def_Phi}
\begin{split}
&\Phi_{\bl}=t_{\bl}:\Phi_\vac\prod_{\mAbox\in\bl}\eta_{c(\sAbox)}^+(\chi_{\sAbox}):,\quad \Phi_{\bl}^\ast=t_{\bl}^\ast:\Phi_\vac^\ast\prod_{\mAbox\in\bl}\eta_{c(\sAbox)-\nu_1-\nu_2}^-(q_3\chi_{\sAbox}):,\\
&t_{\bl}=F^{-|\bl|/2}\prod_{\mAbox\in\bl}u_{c(\sAbox)}'\chi_{\sAbox}^{-n_{c(\sAbox)}'}\prod_{\mAbox\in\bl}Q_{c(\sAbox)}(\chi_{\sAbox}),\\
&t_{\bl}^\ast=F^{-|\bl|/2}\prod_{\mAbox\in\bl}(-u_{c(\sAbox)-\nu_1-\nu_2})^{-1}(q_3\chi_{\sAbox})^{n_{c(\sAbox)-\nu_1-\nu_2}}\prod_{\mAbox\in\bl}:Q_{c(\sAbox)-\nu_1-\nu_2}(q_3\chi_{\sAbox})^{-1}P_{c(\sAbox)}(\chi_{\sAbox}):.
\end{split}
\end{align}
A sketch of the derivation can be found in the appendix \refOld{AppF}, together with the (rather lengthy) expressions of the vacuum components $\Phi_\vac$ and $\Phi_\vac^\ast$. The vertex operators $\Phi$ and $\Phi^\ast$ given here are a generalization of the colored refined topological vertex derived in \cite{Awata2017,Chaimanowong2018} with extra parameters $(\nu_1,\nu_2)$. The relation with the vertex operators of the quantum toroidal $\glp$ algebra is investigated in appendix \ref{AppE}. It is seen that an extra zero-modes factor is necessary due to the twist of the coproduct. This factor leads to a different identification between the algebraic parameters (levels, weights) and the physical quantities, i.e. the branes charges (or Chern-Simons levels) and positions (or instanton counting parameters and Coulomb branch vevs).

The vertical components \ref{def_Phi} of the vertex operators obey important normal ordering relations, from which we recover the vector and bifundamental contributions to the partition functions \cite{Bourgine2017b},\footnote{To simplify the notations, we have omitted the dependence of the Nekrasov factors in the vectors of colors $\boldsymbol{c}=(c_\a)_{\a=1}^m$ and $\boldsymbol{c}'$. The shortcut notation $q_3^{-1}\bv'$ in $N(\bv,\bl|q_3^{-1}\bv',\bl')$ should be understood as a shift of the weights $q_3^{-1}v_\a'$ together with the corresponding shift of indices $c_\a'-\nu_3=\bc_\a'$. Thus, we have the important relation
\begin{align}
\begin{split}
&N(\bv,\bl|q_3^{-1}\bv',\bl')=N(\bv',\bl'|\bv,\bl)f(\bv,\bl|\bv',\bl'),\quad \text{with:}\\
&f(\bv,\bl|\bv',\bl')=\prod_{\superp{\sAbox\in\bl}{\sAboxF\in\bl'}}f_{c(\sAbox)c(\sAboxF)}(\chi_{\sAboxF}/\chi_{\sAbox})\times\prod_{\sAbox\in\bl}\prod_{\a\in C_{c(\sAbox)}(m')}\left(-\dfrac{\chi_{\sAbox}}{v_{\a}'}\right)\times \prod_{\sAbox\in\bl'}\prod_{\a\in C_{\bc(\sAbox)}(m)}\left(-\dfrac{v_\a}{q_3\chi_{\sAbox}}\right).
\end{split}
\end{align}}
\begin{align}\label{ordering_Phi}
\begin{split}
&\Phi_{\bl}\Phi_{\bl'}=\CG(\bv'|\bv)^{-1}N(\bv',\bl'|\bv,\bl)^{-1}:\Phi_{\bl}\Phi_{\bl'}:\\
&\Phi_{\bl}\Phi_{\bl'}^\ast=\CG(\bv'|q_3^{-1}\bv)N(\bv',\bl'|q_3^{-1}\bv,\bl):\Phi_{\bl}\Phi_{\bl'}^\ast:\\
&\Phi_{\bl}^\ast\Phi_{\bl'}=\CG(\bv'|\bv)N(\bv',\bl'|\bv,\bl):\Phi_{\bl}^\ast\Phi_{\bl'}:\\
&\Phi_{\bl}^\ast\Phi_{\bl'}^\ast=\CG(\bv'|q_3^{-1}\bv)^{-1}N(\bv',\bl'|q_3^{-1}\bv,\bl)^{-1}:\Phi_{\bl}^\ast\Phi_{\bl'}^\ast:,
\end{split}
\end{align}
The expression of the one-loop factors $\CG(\bv|\bv')$ can be found in appendix \refOld{AppF}, formula \ref{def_CG}. Note also that, following the method presented in \cite{Awata2016a,Awata2017}, it is a priori possible to show that $\Phi_{\bl}$ and $\Phi_{\bl}^\ast$ are solutions of the double deformed Knizhnik - Zamolodchikov (or $(q,t)$-KZ) equations.

\subsection{Partition functions and $qq$-characters}
The simplest example of algebraic engineering is given by the pure $U(m)$ gauge theory with quiver $A_1$. In this case, the $(p,q)$-brane web can be described roughly as a set of $m$ D5-branes suspended between two (dressed) NS5-branes. The corresponding $\CT$-operator is obtained as a product of vertex operators $\Phi$ and $\Phi^\ast$ in the vertical channel \cite{Bourgine2017b}, it acts on the tensor product of two horizontal modules,
\begin{align}
\CT[U(m)] = \Phi\cdot \Phi^\ast = \sum_{\boldsymbol\lambda} a_{\bl}(\bv)\ \Phi_\bl\otimes\Phi_\bl^\ast\  :\ H\otimes H'_\ast\to H'\otimes H_\ast.
\end{align}
In order to distinguish the horizontal modules, we added the subscript $\ast$ to the ones on which $\Phi^\ast$ act. Accordingly, we denote the parameters of these representations $(n_\o^\ast,u_\o^\ast)$ and $(n_\o^{\ast\prime},u_\o^{\ast\prime})$. Evaluating the vacuum expectation value of this operator, we recover the instanton partition function of the underlying gauge theory:
\begin{equation}
\mathcal{Z}_{\text{inst}}= \bra{\vac}\otimes\bra{\vac}\ \mathcal{T}[U(m)] \ket{\vac}\otimes\ket{\vac} = \sum_{\boldsymbol\lambda} \prod_{\omega \in \mathbb{Z}_p}  \mathfrak{q}_\omega ^{K_\omega (\boldsymbol\lambda)} \mathcal{Z}_{\text{vect}} (\bv, \bl) \mathcal{Z}_{\text{CS}} (\boldsymbol{\k},\bl)
\end{equation} 
where we have identified the colored gauge coupling $\qf_\o$ and Chern-Simons level $\k_\o$ with
\begin{align}\label{id_q_k}
\qf_\o = F^{-1/2} \dfrac{u_\o}{u_{\o+\nu_3}^\ast}q_3 ^{n _{\o+\nu_3}^\ast} , \quad \k_\o=n_{\o+\nu_3}^\ast-n_\o.
\end{align}

By construction, the operator $\CT[U(m)]$ commutes with the action of the algebra defined by the opposite coproduct $\D'$ \cite{Bourgine2017b}, namely,
\begin{align} \label{eq:commute}
\left(\rho^{(H')}\otimes\rho^{(H_\ast)}\right) \D'(e ) \, \mathcal{T}[U(m)] = \mathcal{T}[U(m)] \, \left( \rho^{(H_\ast)} \otimes \rho^{(H')} \right) \D'(e), \quad e \in \mathcal{A}.
\end{align}
For this reason, $\CT[U(m)]$ plays the role of the \textit{screening operator} in \cite{Awata2016a}. The gauge theory expectation value of the fundamental $qq$-characters is obtained by insertion of $\D' ( x^- _{\o +\nu_3} (q_3 z))$ in the horizontal vacuum expectation value,
\begin{equation}\label{qq_character}
\la\mathcal{X}_\bo^{[\bl]\ast} (q_3^{-1}z) \rag=u_\o^{-1}z^{n_\o} \frac{ \bra{\vac}\otimes\bra{\vac}\ \left(\rho^{(H')}\otimes\rho^{(H_\ast)}\ \D'(x_{\o+\nu_3}^-(q_3z))\right)\mathcal{T}[U(m)] \ket{\vac}\otimes\ket{\vac} } {\bra{\vac}\otimes\bra{\vac}\ \mathcal{T}[U(m)] \ket{\vac}\otimes\ket{\vac}}
\end{equation}
where the gauge averaging of a chiral ring observable $\CO^{[\bl]}$ is performed over the instanton configurations weighted by the vector (and Chern-Simons) contributions to the partition function,
\begin{equation}
\la \CO^{[\bl]}\rag=\dfrac1{\Zinst}\sum_{\boldsymbol\lambda} \prod_{\omega \in \mathbb{Z}_p}  \mathfrak{q}_\omega ^{K_\omega (\boldsymbol\lambda)} \mathcal{Z}_{\text{vect}} (\bv, \bl) \mathcal{Z}_{\text{CS}} (\boldsymbol{\k},\bl) \CO^{[\bl]},
\end{equation} 
and the $qq$-character writes 
\begin{equation}
\mathcal{X}_\o^{[\bl]\ast} (z)=\CY_\o^{[\bl]\ast}(z)+\qf_{\o+\nu_3} \dfrac{(q_3z)^{\k_{\o+\nu_3}}}{\CY_{\o+\nu_3}^{[\bl]}(q_3z)}.
\end{equation} 
Note that the first term involves the $\CY$-observable $\CY_\o^{[\bl]\ast}(z)=f\oY(z)\CY\oY(z)$. As shown in \cite{Bourgine2017b}, it follows from the commutation relations \eqref{eq:commute} that the quantity $\la\mathcal{X}_\bo^{[\bl]} (q_3^{-1}z) \rag$ is a finite Laurent series in $z$ (i.e. a polynomial upon multiplication by a positive power of $z$). This is in fact due to the radial ordering of operators in the horizontal Fock spaces. Indeed, when $x_\o^-(z)$ is inserted on the left of $\CT$, the correlator as a well-defined expansion around $z=\infty$. On the other hand, when $x_\o^-(z)$ in inserted on the right, the expansion around $z=0$ is now well-defined. The non-trivial equality between the two expansions \eqref{eq:commute} implies that both series are finite, and thus that the correlator is a finite Laurent series in $z$. Asymptotically, the $\CY$-observables behave as $\CYY\superpsim{0}z^{-\gbY}$, $\CY_\o^{[\bl]\ast}\superpsim{0}1$ and $\CYY\superpsim{\infty}1$, $\CY_\o^{[\bl]\ast}\superpsim{\infty}z^{\gbY}$ with $\gbY=|A_\o(\bl)|-|R_{\o+\nu_3}(\bl)|$. When $\nu_3=0$, the exponent $\gbY$ becomes independent of $\bl$, $\gbY=m_\o$. As a result, the gauge average of the $qq$-character $\mathcal{X}_{\o}(z)$ is a polynomial of degree $m_\o$ when $|\k_\o|<m_\o$. Unfortunately, when $\nu_3\neq0$ not much can be said.

Another fundamental $qq$-character can be obtained using the generator $x_\o^+(z)$ instead,
\begin{align} \label{eq:antifund}
\begin{split}
&\la\mathcal{X}_\bo^{[\bl]} (q_3^{-1}z)\rag=(u_\o^\ast)^{-1}z^{n_\o^\ast} \frac{ \bra{\vac}\otimes\bra{\vac}\ \left(\rho^{(H')}\otimes\rho^{(H_\ast)}\ \D'(x_\o^+(z))\right)\mathcal{T}[U(m)] \ket{\vac}\otimes\ket{\vac} } {\bra{\vac}\otimes\bra{\vac}\ \mathcal{T}[U(m)] \ket{\vac}\otimes\ket{\vac}},\\
&\text{with}\quad \mathcal{X}_\o^{[\bl]} (z)=\CY\oY(z)+\qf_{\o+\nu_3} F \dfrac{(q_3z)^{\k_{\o+\nu_3}}}{f\oY(z)\CY_{\o+\nu_3}^{[\bl]}(q_3z)}.
\end{split}
\end{align}
The presence of two different fundamental $qq$-characters is a specificity of 5D $\CN=1$ gauge theories on orbifolds: when $p=1$, the two $qq$-characters are equivalent (they only differ by multiplication of a constant times a power of $z$). Further, as we shall see below, in the 4D limit $R\to0$, the $qq$-characters $\mathcal{X}_\o^{[\bl]} (z)$ and $\mathcal{X}_\o^{[\bl]\ast}(z)$ reduce to the same expression. The gauge averages \eqref{qq_character} and \eqref{eq:antifund} for the $qq$-characters have been computed at the first few orders in the gauge couplings $\qf_\o$ for the gauge groups $U(1)$ and $U(2)$ and various orbifold parameters. In all cases, it has been observed that these quantities are indeed finite Laurent series in the argument $z$. Finally, it is worth mentioning that higher $qq$-characters can be obtained by multiple insertions of the coproducts $\D'(x_\o^\pm(z))$. We refer to \cite{Bourgine2017b} for more details on the computation of $qq$-characters, and to the appendix \refOld{AppG} for explicit formulas at the first few orders in the instanton counting parameters.

\paragraph{4D limit}  When the radius $R$ of the background circle $S_R^1$ is sent to zero, the gauge theory reduces to a 4D $\CN=2$ gauge theory. This limit can be performed directly on the partition functions and $qq$-characters, re-introducing the radius dependences in the parameters $(q_1,q_2)=(e^{R\e_1},e^{R\e_2})$, $v_\a=e^{R a_\a}$, $\chi_{\sAbox}=e^{R\phi_{\sAbox}}$,... Sending $R\to0$ in the expression \ref{def_Zv_Zbf} of the instanton partition function, we observe that the Chern-Simons contribution is subdominant while, after setting the spectral variable to $z=e^{R\z}$, the scattering function \ref{def_S} becomes
\begin{equation}
S_{\o\o'}^{(\text{4D})}(\z)=\dfrac{(\z+\e_1)^{\d_{\o,\o'-\nu_1}}(\z+\e_2)^{\d_{\o,\o'-\nu_2}}}{\z^{\d_{\o,\o'}}(\z+\e_1+\e_2)^{\d_{\o,\o'-\nu_1-\nu_2}}}.
\end{equation}
This function satisfies a simpler crossing symmetry $S_{\o\bo'}^{(\text{4D})}(-\z-\e_1-\e_2)=f_{\o\o'}^{(\text{4D})}S_{\o'\o}^{(\text{4D})}(\z)$ where $f_{\o\o'}^{(\text{4D})}=(-1)^{\b_{\o\o'}}$ in now independent of the spectral variable $\z$. As a result, the function $f\oY(z)$ reduces to a sign. When $\nu_3=0$, this sign is simply $(-1)^{m_\o}$, it can be absorbed in the definition of $\qf_\o$. In this way, both $\mathcal{X}_\o^{[\bl]} (z)$ and $\mathcal{X}_\o^{[\bl]\ast}(z)$ reproduce the expression of the 4D fundamental $qq$-character given in \cite{Nekrasov_BPS3,Nekrasov_BPS4}.

\paragraph{$A_2$ quiver} Linear quiver gauge theories can be treated along the same lines. For instance, the $A_2$ quiver gauge theory with gauge group $U(m_1)\times U(m_2)$ is obtained by considering two sets of $m_1$ and $m_2$ D5-branes suspended between three dressed NS5-branes. The $\CT$-operator is simply the product of the single nodes operators $\CT[U(m_1)]$ and $\CT[U(m_2)]$ in a common horizontal representation,
\begin{equation}
\CT[U(m_1)\times U(m_2)] = \Phi_1\cdot \Phi_2\Phi_1^\ast\cdot \Phi_2^\ast = \sum_{\bl^{(1)},\bl^{(2)}} a_{\bl^{(1)}}(\bv^{(1)})a_{\bl^{(2)}}(\bv^{(2)})\ \Phi_{\bl^{(1)}}^{(1)}\otimes\Phi_{\bl^{(2)}}^{(2)}\Phi_{\bl^{(1)}}^{(1)\ast}\otimes\Phi_{\bl^{(2)}}^{(2)\ast}.
\end{equation} 
The vacuum expectation value is computed using the normal ordering relation \ref{ordering_Phi} for the product $\Phi_{\bl^{(2)}}^{(2)}\Phi_{\bl^{(1)}}^{(1)\ast}$,
\begin{align}
\begin{split}
\Zinst&= \dfrac1{\CG(\bv^{(1)}|q_3^{-1}\bv^{(2)})}\bra{\vac}\otimes\bra{\vac}\otimes\bra{\vac}\ \mathcal{T}[U(m_1)\times U(m_2)]\ket{\vac}\otimes\ket{\vac} \otimes\ket{\vac}.%\\
% &= \sum_{\bl^{(1)},\bl^{(2)}}\prod_{\omega \in \mathbb{Z}_p}  (\qf^{(1)}_\omega)^{K_\omega (\bl^{(1)})}(\qf^{(2)}_\omega)^{K_\omega (\bl^{(2)})} \Zv(\bv^{(1)},\bl^{(1)})\ZCS(\bl^{(1)})\Zv(\bv^{(2)},\bl^{(1)})\ZCS(\bl^{(2)})\\
% &\quad\quad\times N(\bv^{(1)},\bl^{(1)}|q_3^{-1}\bv^{(2)},\bl^{(2)})
\end{split}
\end{align}
It reproduces the instanton partition function \ref{def_Zv_Zbf} for the $A_2$ quiver gauge theory, with the identification \ref{id_q_k} of the parameters at each node $i=1,2$. The $qq$-characters can also be constructed along the lines of \cite{Bourgine2017b}.

\section{Concluding remarks}
In this paper, we have reconstructed algebraically the instanton partition functions for $\CN=1$ linear quiver gauge theories with unitary gauge groups on the five dimensional background $S_R^1\times(\mathbb{C}_{\e_1}\times\mathbb{C}_{\e_2})/\mathbb{Z}_p$. The action of the abelian group considered here is a generalization by two integers $(\nu_1,\nu_2)$ of the standard action defining ALE spaces. These extra parameters led us to introduce a deformation of the quantum toroidal algebra of $\glp$. Conjecturally, this new algebra should coincide with the quantum toroidal algebra built upon a Kac-Moody algebra with the (non-symmetrizable) generalized Cartan matrix $\b_{\o\o'}$ given in \ref{def_beta}. We have shown that this deformed algebra still possesses the structure of a Hopf algebra with the coproduct given in \ref{Drinfeld_coproduct}. We have also presented two different representations, called \textit{vertical} and \textit{horizontal}, that are respectively the deformation of the Fock module \cite{Feigin2012} and the vertex representation \cite{Saito1996} of the quantum toroidal algebra of $\glp$. Other types of representations should exist, like the Macmahon representation obtained for $\glp$ as a tensor products of Fock modules in \cite{Feigin2012}. Although the definition of this new algebra may appear intricate, the physical context in which it emerges is very natural, and its representations are simple generalizations of the usual ones.

Quantum toroidal algebras extend the definition of quantum affine algebras (or quantum groups) by an extra affinization. In fact, the quantum toroidal algebra of $\glp$ is generated by two orthogonal quantum affine subalgebra $U_q(\widehat{\mathfrak{sl}(p)})$ \cite{Ginzburg1995,Saito1997}. Then, one may wonder if the $(\nu_1,\nu_2)$-deformed algebra possesses a similar property. Of course, it is assuming that a quantum affine algebra built upon the Cartan matrix $\b_{\o\o'}$ can be defined properly. In fact, we expect that this is indeed the case, and that such quantum affine algebra retains a quasitriangular Hopf algebra structure, making it suitable for the construction of new quantum integrable systems.

On the gauge theory side, several generalizations of our approach could be implemented. For instance, the abelian group $\Zp$ could be replaced by a Mckay subgroup of $SU(2)$ of type DE, with either left, right, or both left-right action. As shown by Nakajima in \cite{Nakajima1994,Nakajima1999}, in the first two cases a quantum affine algebra of type $\mathfrak{so}/\mathfrak{sp}$ acts on the cohomology of the instanton moduli space. This action is expected to be lifted to a quantum toroidal algebra in K-theory. Accordingly, the algebraic engineering should involve the quantum toroidal $\mathfrak{so}/\mathfrak{sp}$ algebras. However, the effective construction requires some new developments in the representation theory of these algebras.

When $\nu_2=0$, the orbifold can be interpreted as the presence of a surface defect \cite{Nekrasov_BPS4}. In this case, the Cartan matrix $\b_{\o\o'}$ appears to vanish but the algebra remains non-trivial,
\begin{align}
\begin{split}
&S_{\o\o'}(z)=\left(\dfrac{1-q_2z}{1-z}\right)^{\d_{\o,\o'}}\left(\dfrac{1-q_1z}{1-q_1q_2z}\right)^{\d_{\o,\o'-\nu_1}},\\
&g_{\o\o'}(z)=\left(q_2^{-1}\dfrac{1-q_2z}{1-q_2^{-1}z}\right)^{\d_{\o\o'}}\left(\dfrac{1-q_1z}{1-q_3^{-1}z}\right)^{\d_{\o,\o'-\nu_1}}\left(q_2\dfrac{1-q_3z}{1-q_1^{-1}z}\right)^{\d_{\o,\o'+\nu_1}}.
\end{split}
\end{align}
When $\nu_1=1$, the structure function $g_{\o\o'}(z)$ reproduces the one that defines the quantum toroidal algebra of $\glp$ with $q_2$ and  $q_3$ exchanged (up to a factor $q_3^{m_{\o\o'}/2}$). However, the function $S_{\o\o'}(z)$ is different from the one appearing in \ref{S_glp_II}, and thus  horizontal and vertical representations of the $(\nu_1,\nu_2)$-deformed algebra degenerate into new representations for the quantum toroidal algebra of $\glp$. We hope to come back to the study of this problem in a future publication.

Finally, an important question was left behind in our study, namely the correspondence with (q-deformed) W-algebras. This type of correspondences is now well-understood in the case of quantum toroidal $\mathfrak{gl}(1)$. There, the q-W-algebras appearing in horizontal or vertical representations play different roles. In the horizontal case, a representation of level $c=m$ can be built by tensoring $m$ level one representations. It is thus expressed in terms of $m$ sets of bosonic modes that are coupled through their commutation relations. Diagonalizing these relations, the Drinfeld currents can be expressed in terms of q-$W_m$ currents coupled to an infinite Heisenberg algebra. This dual q-W-algebra corresponds to the quiver W-algebra of Kimura and Pestun \cite{Kimura2015}. Using Miki's automorphism \cite{Miki1999,Miki2007}, vertical representations of level $\bc=m$ can be mapped on horizontal ones, and thus expressed in terms of q-$W_m$ currents coupled to the Heisenberg algebra. In the vertical case, the dual W-algebra is responsible for the AGT-like correspondence with q-deformed conformal blocks \cite{Awata2009}. Alternatively, the AGT correspondence can also be seen directly in the degenerate limit $R\to0$ in which the vertical representation of the toroidal algebra reduces to a representation of the affine Yangian of $\mathfrak{gl}(1)$ that is known to contain the action of $W_m$-currents \cite{Schiffmann2012,Tsymbaliuk2014,Prochazka2015}.

A similar type of duality is believed to hold between the degenerate limit of the quantum toroidal algebra of $\glp$ and the coset\footnote{The parameter $\a$ is determined by the omega-background parameters $(\e_1,\e_2)$.} $\widehat{\mathfrak{gl}(\a)_m}/\widehat{\mathfrak{gl}(\a-p)_m}$, leading to an AGT correspondence between instantons on ALE spaces and parafermionic conformal field theories \cite{Belavin2011, Bonelli2011a, Bonelli2011b, Belavin2011a,Pedrini2014}. This conjecture has been verified for small values of $p$ and $m$ by comparing the conformal blocks of the coset theory with the gauge theories instanton partition functions \cite{Belavin2011,Belavin2011b,Wyllard2011,Nishioka2011,Ito2011,Alfimov2011}, or the limit $R\to0$ of 5D topological strings amplitudes 
\cite{Chaimanowong2018}. There are two main strategies to extend this duality to the $(\nu_1,\nu_2)$-deformed algebra. One possibility is again to compare instanton partition functions with conformal blocks. This approach was taken in \cite{Bonelli2012} where the gauge theory calculations led to conjectural expressions for these conformal blocks. But, unfortunately, the corresponding conformal field theory appears to be unknown. Another possible approach consists in identifying directly the (q-deformed) coset algebra generators acting on the vertical modules of the quantum toroidal algebra. For this purpose, one could diagonalize the commutation relations for the modes $\a_{\o,k}$ in the horizontal representations, and then define the analogue of Miki's automorphism to map the horizontal representations to the vertical ones. From the strings theory perspective, the latter is expected to exist since it should describe the fiber-base duality of the topological strings (or, the S-duality in Type IIB string theory) \cite{Awata2009a,Bourgine2018a}. This approach appears very promising and we hope to be able to report soon on this problem.

\section*{Acknowledgments} 
JEB would like to thank Omar Foda for discussions and generous support during his visit of the university of Melbourne. SJ is greatly indebted to Nikita Nekrasov for numerous discussions and supports. SJ is also grateful to Korea Institute for Advanced Study for providing support during his visit. The work of SJ was supported in part by the NSF grant PHY 1404446 and also by the generous support of the Simons Center for Geometry and Physics.

\appendix

\section{Quantum toroidal algebra of $\glp$}\label{AppA}
In this appendix, we remind the definition of the quantum toroidal algebra of $\glp$, give its vertical and horizontal representations, and comment on the reduction $\nu_1=-\nu_2=1$ of the algebraic relations \ref{algebra}.

\subsection{Definition}
Quantum toroidal algebras were introduced by V. Ginzburg and M. Kapranov and E. Vasserot in \cite{Ginzburg1995}. In general, they can be  built over an affine Kac-Moody algebra $\hat{\gf}$, but we will focus in this appendix on the case of an algebra of type $A_{p-1}^{(1)}$, also called quantum toroidal $\gf=\glp $ algebra. This algebra is formulated in terms of the Drinfeld currents
\begin{equation}
x_\o^\pm(z)=\sum_{k\in\mathbb{Z}}z^{-k}x_{\o,k}^\pm,\quad \psi_\o^\pm(z)=\sum_{k\geq0}z^{\mp k}\psi_{\o,\pm k}^\pm.
\end{equation}
Like the Chevalley generators, the operators $x_\o^\pm(z)$ are associated to the simple roots $\a_\o$ of $\hat{\gf}$. On the other hand, the operators $\psi_{\o}^\pm(z)$ describe the Cartan sector of the algebra, they are naturally associated to the coroots $\a_\o^\vee$. We denote the Cartan matrix $\b_{\o\o'}=\la \a_\o^\vee,\a_{\o'}\ra$, in the case of $\glp $, we have $\b_{\o\o'}=2\d_{\o,\o'}-\d_{\o,\o'+1}-\d_{\o,\o'-1}$ (here $\d_{\o,\o'}$ denotes the Kronecker delta with indices taken modulo p). In this case, the original relations can be deformed by an extra central parameter $\k$, using the antisymmetric matrix $m_{\o\o'}=\d_{\o,\o'-1}-\d_{\o,\o'+1}$ \cite{Saito1996}:\footnote{The generators of this algebra are sometimes denoted $x_\o^+(z)\to E_i(z)$, $x_\o^-(z)\to F_i(z)$, $\psi_\o^\pm(z)\to K_i^\pm(z)$. More rigorously the $x^\pm-x^\pm$ exchange relation should be written
\begin{equation}
(\k^{m_{\o\o'}}z-q^{\pm \b_{\o\o'}}w)x_\o^\pm(z)x_{\o'}^\pm(w)=(\k^{m_{\o\o'}}q^{\pm \b_{\o\o'}}z-w)x_{\o'}^\pm(w)x_\o^\pm(z).
\end{equation} 
This subtlety only affects the colliding points $z=q^{\pm \b_{\o\o'}}\k^{-m_{\o\o'}}w$. The parameter $\k$ here bears no connection with the Chern-Simons levels $\k_\o$ of the gauge theory.}
\begin{align}\label{algebra_glp}
\begin{split}
&[\psi_\o^\pm(z),\psi_{\o'}^\pm(w)]=0,\quad \psi_\o^+(z)\psi_{\o'}^-(w)=\dfrac{g_{\o\o'}(q^c z/w)}{g_{\o\o'}(q^{-c}z/w)}\psi_{\o'}^-(w)\psi_\o^+(z),\quad x_\o^\pm(z)x_{\o'}^\pm(w)=g_{\o\o'}(z/w)^{\pm1}x_{\o'}^\pm(w)x_\o^\pm(z),\\
&\psi_{\o}^+(z)x_{\o'}^\pm(w)=g_{\o\o'}(q^{\pm c/2}z/w)^{\pm1}x_{\o'}^\pm(w)\psi_{\o}^+(z),\quad \psi_{\o}^-(z)x_{\o'}^\pm(w)=g_{\o\o'}(q^{\mp c/2}z/w)^{\pm1}x_{\o'}^\pm(w)\psi_{\o}^-(z)\\
&[x_\o^+(z),x_{\o'}^-(w)]=\dfrac{\d_{\o,\o'}}{q-q^{-1}}\left[\d(q^{-c}z/w)\psi_\o^+(q^{-c/2}z)-\d(q^c z/w)\psi^-_{\o}(q^{c/2} z)\right]\\
&\sum_{\s\in S_2}\left[x_\o^\pm(z_{\s(1)})x_\o^\pm(z_{\s(2)})x_{\o\pm1}^\pm(w)-(q+q^{-1})x_\o^\pm(z_{\s(1)})x_{\o\pm1}^\pm(w)x_\o^\pm(z_{\s(2)})+x_{\o\pm1}^\pm(w)x_\o^\pm(z_{\s(1)})x_\o^\pm(z_{\s(2)})\right]=0,
\end{split}
\end{align}
and $\psi_{\o,0}^+\psi_{\o,0}^-=\psi_{\o,0}^-\psi_{\o,0}^+=1$. In these relations, $q\in\mathbb{C}^\times$, $c$ is a central element, and the matrix $g_{\o\o'}(z)$ writes\footnote{This matrix is sometimes also written
\begin{equation}
g_{\o\o'}(z)=\th_{\b_{\o\o'}}(\k^{m_{\o\o'}}z),\quad \th_m(z)=q^{-m}\dfrac{1-q^mz}{1-q^{-m}z},\quad \th_m(z^{-1})=\th_m(z)^{-1}=\th_{-m}(z).
\end{equation}}
\begin{equation}
g_{\o\o'}(z)=q^{-\b_{\o\o'}}\dfrac{1-q^{\b_{\o\o'}}\k^{m_{\o\o'}}z}{1-q^{-\b_{\o\o'}}\k^{m_{\o\o'}}z},\quad g_{\o\o'}(z^{-1})=g_{\o'\o}(z)^{-1}=g_{\o\o'}(\k^{-2m_{\o\o'}}z)^{-1}.
\end{equation}
In order to compare with the gauge theory quantities, we should set $q=q_3^{1/2}$, $\k=(q_1/q_2)^{1/2}$, then
\begin{equation}\label{def_g_glp}
g_{\o\o'}(z)=\left(q_3^{-1}\dfrac{1-q_3z}{1-q_3^{-1}z}\right)^{\d_{\o,\o'}}\left(q_3^{1/2}\dfrac{1-q_1 z}{1-q_2^{-1} z}\right)^{\d_{\o,\o'-1}}\left(q_3^{1/2}\dfrac{1-q_2 z}{1-q_1^{-1} z}\right)^{\d_{\o,\o'+1}}.
\end{equation} 

\paragraph{Modes decomposition}
The algebraic relations \ref{algebra_glp} can also be directly written for the modes of the Drinfeld currents. In particular, introducing 
\begin{equation}
\psi_\o^\pm(z)=\psi_{\o,0}^\pm\exp\left(\pm\sum_{k\geq1}z^{\mp k}a_{\o,\pm k}\right),
\end{equation}
we find,
\begin{equation}
\psi_{\o,0}^+x_{\o'}^\pm(z)=q^{\pm \b_{\o\o'}}x_{\o'}^\pm(z)\psi_{\o,0}^+,\quad[a_{\o,k},a_{\o',l}]=(q^{kc}-q^{-kc})c_{\o\o'}^{(k)}\d_{k+l},\quad [a_{\o,k},x^\pm_{\o',l}]=\pm q^{\mp|k|c/2}c_{\o\o'}^{(k)}x_{\o',k+l}^\pm,
\end{equation} 
where the coefficients $c_{\o\o'}^{(k)}$ appear in the expansion of $\log g_{\o\o'}(z)$:\footnote{Alternatively,
\begin{equation}
c_{\o\o'}^{(k)}=\dfrac1k\left[(q_3^k-q_3^{-k})\d_{\o,\o'}+(q_2^k-q_1^{-k})\d_{\o,\o'-1}+(q_1^k-q_2^{-k})\d_{\o,\o'+1}\right]=k(\b_{\o\o'}^{[k]}-\b_{\o'\o}^{[-k]}),
\end{equation}
where $\b_{\o\o'}^{[k]}=(1+q_3^k)\d_{\o\o'}-q_1^{-k}\d_{\o,\o'-1}-q_2^{-k}\d_{\o,\o'+1}$ is the mass-deformed Cartan matrix of Kimura and Pestun \cite{Kimura2015} with the mass $\mu_e=q_1$ associated to each link $e:\o\to\o+1$ of the necklace quiver.}
\begin{equation}
[g_{\o\o'}(z)]_\pm=q^{\pm \b_{\o\o'}}\exp\left(\pm\sum_{k>0}z^{\mp k}c_{\o\o'}^{(\pm k)}\right),\quad c_{\o\o'}^{(k)}=c_{\o'\o}^{(-k)}=\dfrac1k\k^{-km_{\o\o'}}(q^{k\b_{\o\o'}}-q^{-k\b_{\o\o'}}),
\end{equation} 
where $[\cdots]_\pm$ denotes the expansion in powers of $z^{\mp1}$. In addition to the central charge $c$, it is possible to define a second central charge using the zero modes of the Cartan currents:
\begin{equation}
\prod_{\o=0}^{p-1}\psi_{\o,0}^\pm=q^{\mp\bc}.
\end{equation} 
Finally, the algebra can be supplemented with the following grading operators,
\begin{align}
\begin{split}
&q^d x_{\o}^\pm(z)q^{-d}=x^\pm_\o(q^{-1}z),\quad q^d\psi^\pm_{\o}(z)q^{-d}=\psi_{\o}^\pm(q^{-1}z),\\
&q^{\bd_\o} x_{\o'}^\pm(z)q^{-\bd_\o}=q^{\pm\d_{\o,\o'}}x_{\o'}^\pm(z),\quad q^{\bd_\o}\psi_{\o'}^\pm(z)q^{-\bd_\o}=\psi_{\o'}^\pm(z).
\end{split}
\end{align}

\paragraph{Coproduct} The Drinfeld coproduct for the quantum toroidal $\glp$ algebra takes the following form,
\begin{align}
\begin{split}\label{Drinfeld_glp}
&\D(x_\o^+(z))=x_\o^+(z)\otimes 1+\psi_\o^-(q^{c_{(1)}/2}z)\otimes x_\o^+(q^{c_{(1)}}z),\\
&\D(x_\o^-(z))=x_\o^-(q^{c_{(2)}}z)\otimes \psi_\o^+(q^{c_{(2)}/2}z)+1\otimes x_\o^-(z),\\
&\D(\psi_\o^\pm(z))=\psi_\o^\pm(q^{\pm c_{(2)}/2}z)\otimes\psi_\o^\pm(q^{\mp c_{(1)}/2}z),
\end{split}
\end{align}
with $d$, $\bd_\o$, and $c$ co-commutative. This coproduct, and the algebraic relations \ref{algebra_glp}, are the same as those presented in \cite{Awata2017}, upon the change of notations
\begin{equation}
x_\o^+(z)\to E_i(z),\quad x_\o^-(z)\to F_i(z),\quad \psi_\o^\pm(z)\to K_\o^\pm(\qf^{c/2}z),\quad q^{c_{(1)}}\to C_1,\quad q^{c_{(2)}}\to C_2.
\end{equation}

\subsection{Horizontal representation}
Representations of this type have central charge $c=1$, they have been constructed by Saito in \cite{Saito1996} under the name \textit{vertex representations}. We review here this construction.

For $c\neq0$, the Cartan modes $a_{\o,k}$ define $p$ coupled Heisenberg subalgebras. For later convenience we introduce the rescaled modes
\begin{equation}\label{com_alpha}
\a_{\o,k}=\dfrac{k}{q^k-q^{-k}}\rho^{(H)}(a_{\o,k})\implies [\a_{\o,k},\a_{\o',l}]=k\d_{k+l}\dfrac{q^{k\b_{\o\o'}}-q^{-k\b_{\o\o'}}}{q^k-q^{-k}}\k^{-km_{\o\o'}}.
\end{equation}
The representation of the currents $x_\o^\pm(z)$ and $\psi_\o^\pm(z)$ can be factorized into two commuting parts: a \textit{zero mode} part ($X_\o^\pm(z)$, $Y_\o^\pm(z)$), and a vertex operator part ($\eta_\o^\pm(z)$, $\vphi_\o^\pm(z)$) built over the Cartan modes $\a_{\o,k}$:
\begin{equation}
\rho^{(H)}(x_\o^\pm(z))=X_\o^\pm(z)\eta_\o^\pm(z),\quad \rho^{(H)}(\psi_\o^\pm(z))=Y_\o^\pm\vphi_\o^\pm(z).
\end{equation}

We focus first on the vertex operators part, it writes
\begin{equation}
\eta_\o^\pm(z)=:\exp\left(\mp\sum_{k\in\mathbb{Z}}\dfrac{z^{-k}}{k}q^{\mp|k|/2}\a_{\o,k}\right):,\quad \vphi_\o^\pm(z)=\exp\left(\pm\sum_{k>0}\dfrac{z^{\mp k}}{k}(q^k-q^{-k})\a_{\o,\pm k}\right),
\end{equation}
note that $\vphi_\o^\pm(z)=:\eta_\o^+(q^{\pm1/2}z)\eta_\o^-(q^{\mp1/2}z):$. The Fock vacuum $\ket{\vac}$ is annihilated by positive modes $\a_{\o,k>0}$, and we define accordingly the normal ordering $:\cdots:$ by writing positive modes on the right. It is a matter of simple algebra to derive the following normal-ordering relations:%\footnote{It would seem more natural here to define instead $S_{\o\o'}(z)\to S_{\o\o'}(qz)$.}
\begin{align}
\begin{split}
&\eta_\o^+(z)\eta_{\o'}^+(w)=S_{\o'\o}(w/z)^{-1}:\eta_\o^+(z)\eta_{\o'}^+(w):,\quad\eta_\o^-(z)\eta_{\o'}^-(w)=S_{\o'\o}(q^2w/z)^{-1}:\eta_\o^-(z)\eta_{\o'}^-(w):,\\
&\eta_\o^\pm(z)\eta_{\o'}^\mp(w)=S_{\o'\o}(qw/z):\eta_\o^\pm(z)\eta_{\o'}^\mp(w):,\\
&\vphi_\o^+(q^{\mp1/2}z)\eta_{\o'}^\pm(w)=\left(\dfrac{S_{\o'\o}(q^2w/z)}{S_{\o'\o}(w/z)}\right)^{\pm1}:\vphi_\o^+(q^{\mp1/2}z)\eta_{\o'}^\pm(w):,\\
&\eta_{\o}^\pm(z)\vphi_{\o'}^-(q^{\pm1/2}w)=\left(\dfrac{S_{\o'\o}(q^2w/z)}{S_{\o'\o}(w/z)}\right)^{\pm1}:\eta_{\o}^\pm(z)\vphi_{\o'}^-(q^{\pm1/2}w):,\\
&\vphi_\o^+(z)\vphi_{\o'}^-(w)=\left(\dfrac{S_{\o'\o}(qw/z)^2}{S_{\o'\o}(q^{-1}w/z)S_{\o'\o}(q^3w/z)}\right):\vphi_\o^+(z)\vphi_{\o'}^-(w):,
\end{split}
\end{align}
with the function
\begin{equation}\label{S_glp}
S_{\o\o'}(z)=\exp\left(\sum_{k>0}\dfrac1kz^kq^{-k}\k^{km_{\o\o'}}\dfrac{q^{k\b_{\o\o'}}-q^{-k\b_{\o\o'}}}{q^k-q^{-k}}\right).
\end{equation} 
In fact, it is possible to resum the infinite series and write the matrix elements $S_{\o\o'}(z)$ as simple rational functions:
\begin{equation}\label{S_glp_II}
S_{\o\o'}(z)=\prod_{r=0}^{|\b_{\o\o'}|-1}(1-\k^{m_{\o\o'}}q^{2r-|\b_{\o\o'}|}z)^{-\sign(\b_{\o\o'})}=\dfrac{(1-q_1z)^{\d_{\o,\o'-1}}(1-q_2z)^{\d_{\o,\o'+1}}}{(1-z)^{\d_{\o,\o'}}(1-q_1q_2z)^{\d_{\o,\o'}}}.
\end{equation} 
We then observe the crossing symmetry,
\begin{equation}\label{rel_S_glp}
S_{\o\o'}(q^2/z)=f_{\o\o'}(z)S_{\o'\o}(z),\quad f_{\o\o'}(z)=F_{\o\o'}z^{\b_{\o\o'}},\quad F_{\o\o'}=(-q)^{-\b_{\o\o'}}\k^{-m_{\o\o'}\b_{\o\o'}},
\end{equation} 
and $F_{\o\o'}F_{\o'\o}=q^{-2\b_{\o\o'}}$, $f_{\o\o'}(q^2/z)f_{\o'\o}(z)=1$. The structure function $g_{\o\o'}(z)$ can be written as a ratio of functions $S_{\o\o'}(z)$ with shifted arguments,
\begin{equation}\label{rel_g_S_glp}
g_{\o\o'}(z)=q^{-\b_{\o\o'}}\dfrac{S_{\o\o'}(z)}{S_{\o\o'}(q^2z)}=f_{\o'\o}(qz)\dfrac{S_{\o\o'}(z)}{S_{\o'\o}(z^{-1})}.
\end{equation} 

We now turn to the analysis of the zero-modes. In \cite{Saito1996}, Saito introduces the symbols $e^{\a_\o}$ associated to the roots $\a_\o$, and obeying the commutation relations $e^{\a_\o}e^{\a_{\o'}}=(-1)^{\b_{\o\o'}}e^{\a_{\o'}}e^{\a_\o}$ (in particular symbols attached to the same root commute). These symbols, together with the operators $a_{\o,0}$ and $\p_{\a_\o}$ act on states parameterized by a root $\a=\sum_{\o\in\Zp} r_\o\a_\o$ ($r_\o\in\mZ$) and a fundamental weight $\L_{\o_0}$,
\begin{align}
\begin{split}\label{zm_glp}
&e^{\a_\o}\ket{\a,\L_{\o_0}}=\prod_{\o'<\o}(-1)^{r_{\o'}\b_{\o\o'}}\ket{\a+\a_\o,\L_{\o_0}},\quad \p_{\a_\o}\ket{\a,\L_{\o_0}}=\la \a_\o^\vee,\b+\L_{\o_0}\ra\ \ket{\a,\L_{\o_0}},\\
&z^{a_{\o,0}}\ket{\a,\L_{\o_0}}=z^{\la \a_\o^\vee,\a+\L_{\o_0}\ra}\prod_{\o'}\k^{r_{\o'}\b_{\o\o'}m_{\o\o'}/2}\ket{\a,\L_{\o_0}}.
\end{split}
\end{align}
In this representation,
\begin{equation}
q^{\p_{\a_\o}}e^{\a_{\o'}}=q^{\b_{\o\o'}}e^{\a_{\o'}}q^{\p_{\a_\o}},\quad z^{a_{\o,0}}e^{\a_{\o'}}=z^{\b_{\o\o'}}\k^{m_{\o\o'}\b_{\o\o'}/2}e^{\a_{\o'}}z^{a_{\o,0}}.
\end{equation} 
Thus, introducing $X_\o^\pm(z)=e^{\pm\a_\o}z^{1\pm a_{\o,0}}$ and $Y_\o^\pm=q^{\pm\p_{\a_\o}}$, we find the algebraic relations%\footnote{The extra power $z$ in the definition of $X_\o^\pm(z)$ is necessary to kill the negative powers of $z$ in the commutator $[x_\o^+,x_{\o'}^-]$.}
\begin{align}
\begin{split}
&X_\o^\pm(z)X_{\o'}^\pm(w)=f_{\o'\o}(qz/w)X_{\o'}^\pm(w)X_\o^\pm(z),\quad X_\o^\pm(z)X_{\o'}^\mp(w)=f_{\o'\o}(qz/w)^{-1}X_{\o'}^\mp(w)X_\o^\pm(z),\\
&Y_\o^+X_{\o'}^\pm(w)=q^{\pm \b_{\o\o'}}X_{\o'}^\pm(w)Y_\o^+,\quad Y_\o^-X_{\o'}^\pm(w)=q^{\mp \b_{\o\o'}}X_{\o'}^\pm(w)Y_\o^-,\quad [Y_\o^+,Y_{\o'}^-]=0.
\end{split}
\end{align}
It is easy to verify that these are indeed the factors needed to reproduce the algebraic relations \ref{algebra_glp}. The only difficulty appears in the verification of the commutation relation $[x_\o^+,x_{\o'}^-]$ for which we need to use the property $z^{a_{\o,0}}w^{-a_{\o,0}}=z^{\p_{\a_\o}}w^{-\p_{\a_\o}}$ to treat the zero mode dependence. The value of the central charge $\bc$ can be recovered by noticing that
\begin{equation}
\sum_{\o\in\Zp}\la \a_\o^\vee,\b+\L_{\o_0}\ra=1\implies \prod_{\o\in\Zp} q^{\p_{\a_\o}}=q.
\end{equation} 
This representation has been extended to higher level $\bc$ in \cite{Awata2017}. Note however that in the definition of the $(\nu_1,\nu_2)$-deformed horizontal representation, a set of $4p$ Heisenberg algebras will be employed to define to the zero-modes $X_\o^\pm$ and $Y_\o^\pm$ instead of the symbols introduced in \ref{zm_glp}.

\subsection{Vertical representations}\label{AppA3}
The vertical representations have central charge $c=0$ and thus the Cartan currents $\psi_{\omega}^\pm(z)$ commute. They are diagonal in the basis of states $\dket{\bl}$ labeled by $m$-tuple Young diagram $\bl=(\l^{(1)},\cdots,\l^{(m)})$. The representation depends on an $m$-vector of weights $\bv=(v_1,\cdots,v_m)$ and a choice of coloring $c_\a$ for each component $v_\a$. We denote $m_\o=|C_\o(m)|$ the number of weights $v_\a$ of color $c_\a=\o$ (obviously, $m=\sum_{\o\in\Zp} m_\o$). The action of the Drinfeld currents on the states $\dket{\bl}$ reads
\begin{align}
\begin{split}\label{vertical}
&\rho^{(V)}(x_\omega^+(z))\dket{\bl}=(qz)^{-\gbY}\prod_{\mAbox\in\bl}(-\k)^{-m_{\o c(\sAbox)}/2}\sum_{\mAbox\in A_\omega(\boldsymbol\l)}\d(z/\chi_{\sAbox})\res_{z=\chi_{\sAbox}}\dfrac1{z\tilde{\CY}\oY(z)} \dket{\bl+\Abox},\\
&\rho^{(V)}(x_\omega^-(z))\dket{\bl}=qz^{\gbY+2}\prod_{\mAbox\in\bl}(-\k)^{m_{\o c(\sAbox)}/2} \sum_{\mAbox\in R_\omega(\boldsymbol\l)}\d(z/\chi_{\sAbox})\res_{z=\chi_{\sAbox}}z^{-1}\tilde{\CY}\oY(q_3^{-1}z)\dket{\bl-\Abox},\\
&\rho^{(V)}(\psi_\omega^\pm(z))\dket{\bl}=\left[\tilde{\Psi}\oY(z)\right]_\pm\dket{\bl}.
\end{split}
\end{align}
In the first two lines, summations are performed over the set of boxes of color $\o$ that can be added ($A_\o(\bl)$) to or removed from ($R_\o(\bl)$) the $m$-tuple Young diagram $\bl$. The summands are expressed in terms of residues involving the functions $\tilde{\CY}\oY(z)$, and the action of the Cartan is given as an expansion of the functions $\tilde{\Psi}\oY(z)$ in powers of $z^{\mp1}$. These two sets of functions are defined as follows:%\footnote{The following \textit{shell formula} is essential to verify the algebraic relations,
% \begin{equation}
% \CYY(z)=\dfrac{\prod_{\mAbox\in A_\o(\bl)}(1-\chi_{\sAbox}/z)}{\prod_{\mAbox\in R_\o(\bl)}(1-\chi_{\sAbox}/(q_3z))}.
% \end{equation}}
\begin{align}
\begin{split}\label{def_PsiY}
&\tilde{\Psi}\oY(z)=q^{-m_{\o}}\prod_{\a\in C_\o(m)}\dfrac{1-q^2v_\a/z}{1-v_\a/z}\prod_{\mAbox\in\bl}g_{\o c(\sAbox)}(z/\chi_{\sAbox})=q^{-\gbY}\dfrac{\tilde{\CY}\oY(q^{-2}z)}{\tilde{\CY}\oY(z)},\\
&\tilde{\CY}\oY(z)=\prod_{\a\in C_\o(m)}(1-v_\a/z)\prod_{\mAbox\in\bl}S_{c(\sAbox)\o}(\chi_{\sAbox}/z),\quad \gbY=m_\o-\sum_{\mAbox\in\bl}\b_{\o c(\sAbox)}.
\end{split}
\end{align}
The zero modes of the Cartan act as
\begin{equation}
\rho^{(V)}(\psi_{\o,0}^\pm)\dket{\bl}=q^{\mp\gbY}\dket{\bl}.
\end{equation} 
and, taking the product over the index $\o$, we deduce the level $\rho^{(V)}(\bc)=m$.

\subsection{Deformation of the algebra}\label{AppA4}
In order to define the algebraic engineering for the gauge theory on $(\nu_1,\nu_2)$-deformed $\Zp$-orbifolds, the physical quantity we need to reproduce is the scattering function $S_{\o\o'}(z)$ defined in \ref{def_S}. In this scope, it is easier to deform first the horizontal representations, and reproduce the commutation of the Heisenberg subalgebras \ref{def_sigma} using the $\glp$ formula \ref{com_alpha}. It leads to identify the Cartan matrix with the matrix $\b_{\o\o'}$ defined in \ref{def_beta}. Furthermore, the factor $\k^{m_{\o\o'}}$ has to be replaced with a more general matrix $\k_{\o\o'}$ that reads\footnote{We could also express the coefficients $\s_{\o\o'}^{[k]}=k\b_{\o\o'}\k_{\o\o'}^{-k}=kq_3^{k/2}\b_{\o\o'}^{[k]}$ in terms of the mass-deformed Cartan matrix $\b_{\o\o'}^{[k]}=\d_{\o,\o'}+q_3^k\d_{\o,\bo'}-q_1^{-k}\d_{\o,\o'+\nu_1}-q_2^{-k}\d_{\o,\o'+\nu_2}$.}
\begin{equation}
\k_{\o\o'}=q^{-1}q_3^{\d_{\o,\bo'}}q_1^{-\d_{\o,\o'+\nu_1}}q_2^{-\d_{\o,\o'+\nu_2}}.
\end{equation}
Using this identification, the crossing symmetry relation \ref{crossing} reduces to \ref{rel_S_glp} for $\nu_3=0$, and the formula \ref{rel_S_glp} for $F_{\o\o'}$ reproduces the definition \ref{def_beta}. The function $S_{\o\o'}(z)$ defined in \ref{def_S} is recovered by replacing the expression \ref{S_glp} with
\begin{equation}
S_{\o\o'}(z)=\exp\left(\sum_{k>0}\dfrac{z^k}{k}q^{-k}\k_{\o'\o}^{-k}\dfrac{q^{k\b_{\o'\o}}-q^{-k\b_{\o'\o}}}{q^k-q^{-k}}\right).
\end{equation} 

The definition of the deformed structure functions $g_{\o\o'}(z)$ is a little more difficult because of the freedom in defining the \textit{zero modes} factors. Comparing with the formula \ref{rel_g_S_glp} for the case of $\glp$, the most natural choice would be
\begin{equation}
g_{\o\o'}^{(1)}(z)=f_{\o\o'}(qz)\dfrac{S_{\o\o'}(z)}{S_{\o'\o}(z^{-1})}=q^{\b_{\o\o'}}F_{\o\o'}^2\prod_{i=1,2,3}\dfrac{(1-q_iz)^{\d_{\o,\o'-\nu_i}}}{(1-q_i^{-1}z)^{\d_{\o,\o'+\nu_i}}}.
\end{equation} 
Unfortunately, with this definition, the identity $g^{(1)}_{\o\o'}(z)g^{(1)}_{\o'\o}(z^{-1})=1$ is NOT satisfied, yet it is necessary for the consistency of the algebraic relations. It prompted us to propose instead the definition given in \ref{def_g} where the factor $f_{\o\o'}(qz)$ is missing. Unfortunately, this redefinition of the structure functions $g_{\o\o'}(z)$ breaks the natural symmetry between positive and negative currents, and makes the definition of the central charge $\bc$ more difficult. Note also that another possibility could have been to define
\begin{equation}
g_{\o\o'}^{(2)}(z)=\prod_{i=1,2,3}\dfrac{(1-q_iz)^{\d_{\o,\o'-\nu_i}}}{(1-q_iz^{-1})^{\d_{\o,\o'+\nu_i}}},
\end{equation} 
but it would require us to modify the definition of the functions $S_{\o\o'}(z)$:
\begin{equation}
g_{\o\o'}^{(2)}(z)=-\dfrac{S_{\o\o'}^{(2)}(z)}{S_{\o'\o}^{(2)}(z^{-1})},\quad S_{\o\o'}^{(2)}(z)=\dfrac{(1-q_1z)^{\d_{\o,\o'-\nu_1}}(1-q_2z)^{\d_{\o,\o'-\nu_2}}}{(1-z^{-1})^{\d_{\o,\o'}}(1-q_3z^{-1})^{\d_{\o,\o'-\nu_1-\nu_2}}}.
\end{equation}
In fact, the proper generalization of the quantum toroidal $\glp$ algebra's structure functions is found in appendix \ref{AppE} (formula \ref{def_tg}). It takes into account the twist of the Drinfeld currents, and involves half-integer powers of the spectral variable $z$.

\section{Shell formula}\label{AppB0}
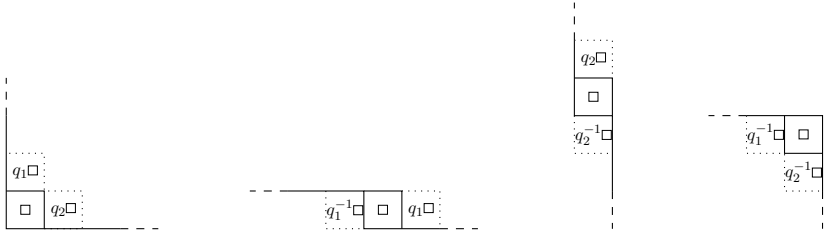
\begin{figure}
\begin{center}
\begin{tikzpicture}[scale=0.5]
\draw[dashed] (3,0) -- (4,0);
\draw[dashed] (0,3) -- (0,4);
\draw (3,0) -- (0,0) -- (0,3);
\draw (1,0) -- (1,1) -- (0,1);
\draw[dotted] (1,0) -- (2,0) -- (2,1) -- (1,1) -- (1,2) -- (0,2);
\node[scale=0.6] at (0.5,0.5) {$\Abox$};
\node[scale=0.6] at (1.5,0.5) {$q_2\Abox$};
\node[scale=0.6] at (0.5,1.5) {$q_1\Abox$};
\end{tikzpicture}
\hspace{1cm}
\begin{tikzpicture}[scale=0.5]
\draw[dashed] (-2,1) -- (-3,1);
\draw[dashed] (2,0) -- (3,0);
\draw (-2,1) -- (0,1) -- (0,0) -- (2,0) ;
\draw (0,1) -- (1,1) -- (1,0) ;
\draw[dotted] (1,1) -- (2,1) -- (2,0);
\draw[dotted] (-1,1) -- (-1,0) -- (0,0) ;
\node[scale=0.6] at (0.5,0.5) {$\Abox$};
\node[scale=0.6] at (1.5,0.5) {$q_1\Abox$};
\node[scale=0.6] at (-0.5,0.5) {$q_1^{-1}\Abox$};
\end{tikzpicture}
\hspace{1cm}
\begin{tikzpicture}[scale=0.5]
\draw[dashed] (1,-2) -- (1,-3);
\draw[dashed] (0,2) -- (0,3);
\draw (0,2) -- (0,0) -- (1,0) -- (1,-2) ;
\draw (0,1) -- (1,1) -- (1,0) ;
\draw[dotted] (0,2) -- (1,2) -- (1,1);
\draw[dotted] (0,0) -- (0,-1) -- (1,-1);
\node[scale=0.6] at (0.5,0.5) {$\Abox$};
\node[scale=0.6] at (0.5,1.5) {$q_2\Abox$};
\node[scale=0.6] at (0.5,-0.5) {$q_2^{-1}\Abox$};
\end{tikzpicture}
\hspace{1cm}
\begin{tikzpicture}[scale=0.5]
\draw[dashed] (-2,2) -- (-3,2);
\draw[dashed] (0,0) -- (0,-1);
\draw (-2,2) -- (0,2) -- (0,0) ;
\draw (-1,2) -- (-1,1) -- (0,1) ;
\draw[dotted] (-2,2) -- (-2,1) -- (-1,1) -- (-1,0) -- (0,0) ;
\node[scale=0.6] at (-0.5,1.5) {$\Abox$};
\node[scale=0.6] at (-1.5,1.5) {$q_1^{-1}\Abox$};
\node[scale=0.6] at (-0.5,0.5) {$q_2^{-1}\Abox$};
\end{tikzpicture}
\end{center}
\caption{Generic and degenerate configurations of a box added to the corner of a Young diagram}
\label{fig_Young}
\end{figure}

We provide here a short derivation of the shell formula \ref{shell_Y} for the functions $\CYY(z)$. Since $K_\o(\bl)$ is a direct sum of $K_\o(\l^{(\a)})$, the function $\CYY(z)$ factorizes into contributions of the individual Young diagrams $\CY^{[\l^{(\a)}]}_\o(z)$. Thus, it is possible to focus on the case of a single Young diagram $\l^{(\a)}$ corresponding to a weight $v_\a$ of color $c_\a$. The proof will be done by recursion on the number of boxes. We start with an empty Young diagram, for which $R_\o(\vac)=\vac$. The box $\Abox=(1,1)$ is of color $c(\Abox)=c_\a$, thus $A_\o(\vac)=\{(1,1)\}$ if $\o=c_\a$ and $A_\o(\vac)=\vac$ otherwise. Accordingly, we recover $\CY_\o^{[\vac]}(z)=(1-v_\a/z)^{\d_{\o,c_\a}}$.

Now, let's add a box $\Abox$ to $\l^{(\a)}$. From the definition \ref{def_Psi}, we have
\begin{equation}\label{var_CYY}
\dfrac{\CY_\o^{[\l^{(\a)}+\mAbox]}(z)}{\CY_\o^{[\l^{(\a)}]}(z)}=\dfrac{(1-q_1\chi_{\sAbox}/z)^{\d_{c(\sAbox),\o-\nu_1}}(1-q_2\chi_{\sAbox}/z)^{\d_{c(\sAbox),\o-\nu_2}}}{(1-\chi_{\sAbox}/z)^{\d_{c(\sAbox),\o}}(1-q_1q_2\chi_{\sAbox}/z)^{\d_{c(\sAbox),\o-\nu_1-\nu_2}}}.
\end{equation} 
There are four possible configurations for adding box in $\l^{(\a)}$, all represented in figure \ref{fig_Young}. We start with the generic case, for which
\begin{align}
\begin{split}
&A_\o(\l^{(\a)}+\Abox)=\left(A_\o(\l^{(\a)})\setminus\{\Abox\mid c(\Abox)=\o\}\right)\cup\{q_1\Abox\mid c(\Abox)=\o-\nu_1\}\cup\{q_2\Abox\mid c(\Abox)=\o-\nu_2\},\\
&R_{\o-\nu_1-\nu_2}(\l^{(\a)}+\Abox)=R_{\o-\nu_1-\nu_2}(\l^{(\a)})\cup\{\Abox\mid c(\Abox)=\o-\nu_1-\nu_2\}.
\end{split}
\end{align}
We employed here the shortcut notation $q_1^{\pm1}\Abox$ ($q_2^{\pm1}\Abox$) to designate the box of coordinate $(i\pm1,j)$ (resp. $(i,j\pm1)$) next to $\Abox=(i,j)$. In this generic case, the factors induced by the variation of the content of the sets $A_\o$ and $R_{\o-\nu_1-\nu_2}$ reproduce the extra factor $S_{c(\sAbox)\o}(\chi_{\sAbox}/z)$ in the RHS of \ref{var_CYY}.

We now turn to the first case of the degenerate configurations represented on figure \refOld{fig_Young}. In this case, only one more box can be added to $A_\o(\l^{(\a)}+\Abox)$. On the other hand, the addition of the box $\Abox$ prevents the removal of the box $q_1^{-1}\Abox$. As a result,
\begin{align}
\begin{split}
&A_\o(\l^{(\a)}+\Abox)=\left(A_\o(\l^{(\a)})\setminus\{\Abox\mid c(\Abox)=\o\}\right)\cup\{q_1\Abox\mid c(\Abox)=\o-\nu_1\},\\
&R_{\o-\nu_1-\nu_2}(\l^{(\a)}+\Abox)=\left(R_{\o-\nu_1-\nu_2}(\l^{(\a)})\cup\{\Abox\mid c(\Abox)=\o-\nu_1-\nu_2\}\right)\setminus\{q_1^{-1}\Abox\mid c(\Abox)=\o-\nu_2\}.
\end{split}
\end{align}
Once again, we observe the agreement between the variation of the RHS \ref{shell_Y} and the recursion relation \ref{var_CYY}. The other two cases are treated in the same way.

\section{Representations of the extended algebra}\label{AppC}
\subsection{Vertical representation}\label{AppB1}
The vertical representation is of the highest weight type. The highest state $\dket{\vac}$, also called \textit{vacuum state}, is annihilated by the currents $x_\o^-(z)$, while $x_\o^+(z)$ create excitations. The excited states $\dket{\bl}$ are parameterized by an $m$-tuple Young diagram $\bl$. The weights $\bv=(v_1,\cdots,v_m)$ parameterize the action of the Cartan $\psi_\o^\pm(z)$ on the vacuum state. The two Cartan currents commute, they are diagonal in the basis $\dket{\bl}$, with the eigenvalue $[\PsiY(z)]_\pm$ where $\pm$ denotes an expansion in powers of $z^{\mp1}$. The action of $x_\o^\pm(z)$ add/remove a box of color $\o$. In order to produce the Dirac $\d$-function in the commutator $[x^+,x^-]$, it is natural to assume that modes $x_{\o,k}^\pm$ depends on the index $k$ only through a factor of $\chi_{\sAbox}^k$ where $\Abox$ is the box that is added/removed. Taking all these assumptions in consideration, we arrive at the following ansatz:
\begin{align}
\begin{split}\label{ansatz_vert}
&x_\omega^+(z)\dket{\bl}=\sum_{\mAbox\in A_\omega(\boldsymbol\l)}\d(z/\chi_{\sAbox}) A^{+[\bl]}_\o(x)  \dket{\bl+\Abox},\\
&x_\omega^-(z)\dket{\bl}=  \sum_{\mAbox\in R_\omega(\boldsymbol\l)}\d(z/\chi_{\sAbox})A^{-[\bl]}_\o(x) \dket{\bl-\Abox},\\
&\psi_\omega^\pm(z)\dket{\bl}=\left[\PsiY(z)\right]_\pm\dket{\bl},
\end{split}
\end{align}
where $A^{\pm[\bl]}_\o(x)$ are the coefficients to be determined.

When the central charge $c$ is vanishing, the algebra \ref{algebra} simplifies drastically,
\begin{align}\label{algebra_vert}
\begin{split}
&x_\o^\pm(z)x_{\o'}^\pm(w)=g_{\o\o'}(z/w)^{\pm1}x_{\o'}^\pm(w)x_\o^\pm(z),\quad [\psi_\o^+(z),\psi_{\o'}^-(w)]=[\psi_\o^\pm(z),\psi_{\o'}^\pm(w)]=0,\\
&\psi^+_\o(z)x^\pm_{\o'}(w)=g_{\o\o'}(z/w)^{\pm1}x^\pm_{\o'}(w)\psi^+_\o(z),\quad \psi^-_\o(z)x^\pm_{\o'}(w)=g_{\o\o'}(z/w)^{\pm1} x^\pm_{\o'}(w)\psi^-_\o(z),\\
&[x_\o^+(z),x_{\o'}^-(w)]=\kO\d_{\o,\o'}\d(z/w)\left[\psi_\o^+(z)-\psi^-_{\o}(z)\right].
\end{split}
\end{align}
Plugging in the ansatz \ref{ansatz_vert}, and the expression \ref{def_Psi} for $\PsiY(z)$, we find that these relations are satisfied provided that
\begin{align}\label{3constraints}
\begin{split}
&A^{-[\bl]}_\o(x)A^{+[\bl-\Abox]}_\o(x)=\kO\res_{z=\chi_{\sAbox}}z^{-1}\PsiY(z),\quad \mAbox\in R_\o(\l),\\
&A^{+[\bl]}_\o(x)A^{-[\bl+\Abox]}_\o(x)=-\kO\res_{z=\chi_{\sAbox}}z^{-1}\PsiY(z),\quad \mAbox\in A_\o(\l),\\
&\dfrac{A_\o^{\pm[\bl\pm x]}(y)}{A_\o^{\pm[\bl]}(y)}=g_{\o\o'}(\chi_y/\chi_{\sAbox})^{\pm1}\dfrac{A_{\o'}^{\pm[\bl\pm y]}(x)}{A_{\o'}^{\pm[\bl]}(x)},\quad c(\sAbox)=\o',\quad c(y)=\o.
\end{split}
\end{align}
The first two relations come from the projection of the commutator $[x^+,x^-]$ on the basis $\dket{\bl}$, decomposing the RHS as
\begin{equation}\label{poles_Psi}
\left[\PsiY(z)\right]_+-\left[\PsiY(z)\right]_-=\sum_{\mAbox\in A_\o(\bl)}\d(z/\chi_{\sAbox})\res_{z=\chi_{\sAbox}}z^{-1}\PsiY(z)+\sum_{\mAbox\in R_\o(\bl)}\d(z/\chi_{\sAbox})\res_{z=\chi_{\sAbox}}z^{-1}\PsiY(z).
\end{equation} 
The last equation in \ref{3constraints} arises from the exchange relations $x^\pm x^\pm$. Then, it is simply a matter of calculation to check that the following coefficients do indeed satisfy the relations \ref{3constraints},
\begin{align}
\begin{split}
&A^{+[\bl]}_\o(\Abox)=F^{1/2}\res_{z=\chi_{\sAbox}}z^{-1}\CYY(z)^{-1}=\kO\CY_\o^{[\bl+\mAbox]}(\chi_{\sAbox})^{-1},\\
&A^{-[\bl]}_\o(\Abox)=\mr{f}_{\bo}^{[\bl]}(q_3^{-1}z)\res_{z=\chi_{\sAbox}}z^{-1}\CY_{\bo}^{[\bl]}(q_3^{-1}z)=-\kO F^{1/2}\mr{f}_{\bo}^{[\bl]}(q_3^{-1}z)\CY_{\bo}^{[\bl-\mAbox]}(q_3^{-1}\chi_{\sAbox}).
\end{split}
\end{align}

\subsection{Horizontal representation}\label{AppB2}
Here the strategy is to start by computing the algebraic relations satisfied by the vertex operators $\eta_\o^\pm$ and $\vphi_\o^\pm$, compare them with \ref{algebra}, and introduce the zero-modes factors to compensate unwanted factors. Using the definition \ref{def_eta_vphi}, we can compute the normal-ordering relations
\begin{align}
\begin{split}
&\eta_\o^+(z)\eta_{\o'}^+(w)=S_{\o'\o}(w/z)^{-1}:\eta_\o^+(z)\eta_{\o'}^+(w):,\quad \eta_\o^+(z)\eta_{\o'}^-(w)=S_{\o'\o}(w/z):\eta_\o^+(z)\eta_{\o'}^-(w):,\\
&\eta_\o^-(z)\eta_{\o'}^-(w)=S_{\o'\bo}(q_3w/z)^{-1}:\eta_\o^-(z)\eta_{\o'}^-(w):=f_{\o'\o}(z/w)^{-1}S_{\o\o'}(z/w)^{-1}:\eta_\o^-(z)\eta_{\o'}^-(w):,\\
&\eta_\o^-(z)\eta_{\o'}^+(w)=S_{\o'\ \bo}(q_3w/z):\eta_\o^-(z)\eta_{\o'}^+(w):=f_{\o'\o}(z/w)S_{\o\o'}(z/w):\eta_\o^-(z)\eta_{\o'}^+(w):,
\end{split}
\end{align}
and, since $\vphi_\o^+(z)=:\eta_\o^+(z)\eta_\o^-(z):$ and $\vphi_\o^-(z)=:\eta_{\bo}^+(q_3^{-1}z)\eta_\o^-(z):$,
\begin{align}
\begin{split}
&\vphi_\o^+(z)\eta_{\o'}^\pm(w)=f_{\o'\o}(z/w)^{\pm1}g_{\o\o'}(z/w)^{\pm1}:\vphi_\o^+(z)\eta_{\o'}^\pm(w):,\\
&\eta_{\o}^+(z)\vphi_{\o'}^-(w)=f_{\bar\o'\o}(q_3z/w)g_{\o\bar\o'}(q_3z/w):\eta_{\o}^+(z)\vphi_{\o'}^-(w):\\
&\eta_{\o}^-(z)\vphi_{\o'}^-(w)=f_{\o'\o}(z/w)^{-1}g_{\o\o'}(z/w)^{-1}:\eta_{\o}^-(z)\vphi_{\o'}^-(w):,\\
&\vphi_\o^+(z)\vphi_{\o'}^-(w)=\dfrac{f_{\bo'\o}(q_3z/w)}{f_{\o'\o}(z/w)}\dfrac{g_{\o\bar\o'}(q_3z/w)}{g_{\o\o'}(z/w)}:\vphi_\o^+(z)\vphi_{\o'}^-(w):.
\end{split}
\end{align}
We deduce the algebraic relations between vertex operators. Comparing them with the currents algebra \ref{algebra} at $c=1$, we observe that the latter are satisfied provided that we set
\begin{equation}
\rho^{(H)}_{u}(x_\o^\pm(z))=X_\o^\pm(z)\eta_\o^\pm(z),\quad \rho^{(H)}_{u}(\psi_\o^\pm(z))=Y_\o^\pm(z)\vphi_\o^\pm(z),
\end{equation}
with\footnote{Note that these relations imply
\begin{equation}
Y_\o^+(z)Y_{\o'}^-(w)=\dfrac{f_{\o'\o}(z/w)}{f_{\bo'\o}(q_3z/w)}Y_{\o'}^-(w)Y_\o^+(z).
\end{equation}}
\begin{align}\label{rel_X_Y}
\begin{split}
&X_\o^+(z)X^+_{\o'}(w)=X^+_{\o'}(w)X_\o^+(z),\quad X_\o^-(z)X^-_{\o'}(w)=\dfrac{f_{\o'\o}(z/w)}{f_{\o\o'}(w/z)}X^-_{\o'}(w)X_\o^-(z),\\
&X_\o^+(z)X^-_{\o'}(w)=:X_\o^+(z)X^-_{\o'}(w):=f_{\o\o'}(w/z)X^-_{\o'}(w)X_\o^+(z),\\
&Y_\o^+(z)X_{\o'}^\pm(w)=f_{\o'\o}(z/w)^{\mp1}X_{\o}^\pm(z)Y_{\o'}^+(w),\\
&X_{\o}^+(z)Y_{\o'}^-(w)=f_{\bo'\o}(q_3z/w)^{-1}Y_{\o'}^-(w)X_{\o}^+(z),\quad X_{\o}^-(z)Y_{\o'}^-(w)=f_{\o'\o}(z/w)Y_{\o'}^-(w)X_{\o}^-(z),\\
&Y_\o^+(z)=F^{-1/2}:X_\o^+(z)X_\o^-(z):,\quad Y_\o^-(z)=F^{1/2}:X_{\bar\o}^+(q_3^{-1}z)X_\o^-(z):.
\end{split}
\end{align}
The last two relations come from the commutator $[x^+,x^-]$, they have been obtained using the pole decomposition of the function $S_{\o'\o}(w/z)$ which brings
\begin{equation}\label{decomp_S_g}
\left[S_{\o'\o}(w/z)\right]_+-\left[S_{\o'\o}(w/z)\right]_-=\kO\left[\d_{\o,\o'}\d(z/w)F^{-1/2}-\d_{\o,\bar\o'}\d(q_3 z/w)F^{1/2}\right].
\end{equation}

The relations \ref{rel_X_Y} are satisfied if we set
\begin{equation}
X_\o^+(z)=Q_\o(z),\quad X^-_\o(z)=Q_\o(z)^{-1}P_{\bo}(q_3^{-1}z),\quad Y^+_\o(z)=F^{-1/2}P_{\bo}(q_3^{-1}z),\quad Y^-_\o(z)=F^{1/2}\dfrac{Q_{\bo}(q_3^{-1}z)}{Q_{\o}(z)}P_{\bo}(q_3^{-1}z),
\end{equation} 
where $Q_\o(z)$ and $P_{\o'}(w)$ obey \ref{rel_PQ}. These operators can be constructed in terms of $2p$ Heisenberg algebras $[p_\o,q_{\o'}]=\d_{\o,\o'}$ and $[\tp_\o,\tq_{\o'}]=\d_{\o,\o'}$ by setting
\begin{equation}\label{def_PQ}
Q_\o(z)=e^{q_\o+\tq_\o\log z},\quad P_\o(z)=z^{-\sum_{\o'}\b_{\o\o'}p_{\o'}}e^{\sum_{\o'}\b_{\o\o'}\tp_{\o'}}(-1)^{p_\o}(-q_3)^{-p_{\o+\nu_3}}(-q_1)^{-p_{\o-\nu_1}}(-q_2)^{-p_{\o-\nu_2}}.
\end{equation} 
Combining the operators $X_\o^\pm$, $Y_\o^\pm$ and the vertex operators $\eta_\o^\pm$, $\psi_\o^\pm$, we find the representation \ref{rep_H}. The dependence in the weights $u_\o$ and levels $n_\o$ is recovered using the freedom to shift the operators $q_\o$, $\tq_\o$ as $q_\o\to q_\o+\log(u_\o)$, $\tq_\o\to\tq_\o-n_\o$.

It remains to compute the central charge $\bc$. The zero modes of the Cartan currents write
\begin{equation}
\rho^{(H)}(\psi^+_{\o,0})=F^{-1/2}P_{\bo}(q_3^{-1})\quad \rho^{(H)}(\psi_{\o,0}^-)=F^{1/2}\dfrac{u_{\bo}}{u_\o}q_3^{n_{\bo}}\dfrac{Q_{\bo}(q_3^{-1})}{Q_{\o}(1)}P_{\bo}(q_3^{-1}).
\end{equation} 
We deduce that
\begin{equation}
\rho^{(H)}\left(\prod_{\o\in\Zp}\psi_{\o,0}^+(\psi_{\o,0}^-)^{-1}\right)=q_3^{-n-\tq},\quad \tq=\sum_{\o\in\Zp}\tq_\o.
\end{equation} 
Since $[\tq_\o,P_{\o'}(w)]=\b_{\o'\o}P_{\o'}(w)$, the operator $\tq$ commute with $P_\o(z)$, thus it is central in this representation. Moreover, since $Q_\o(z)$ acts trivially on the dual state $\bra{\vac}$, we have $\tq=0$. Finally, we also have to take into account the non-commutation of the zero modes which brings the extra factor
\begin{equation}
\prod_{\superp{\o,\o'=0}{\o\leq\o'}}^{p-1}\dfrac{F_{\o'\o}}{F_{\bo'\o}}=\prod_{\superp{\o,\o'=0}{\o\leq\o'}}^{p-1}q_3^{\b_{\o\o'}}F_{\o\o'}F_{\o'\o}=\prod_{\superp{\o,\o'=0}{\o\leq\o'}}^{p-1}q_3^{\b_{\o\o'}}\times F^{-p}\prod_{\o,\o'=0}^{p-1}F_{\o\o'}=\prod_{\superp{\o,\o'=0}{\o\leq\o'}}^{p-1}q_3^{\b_{\o\o'}}.
\end{equation} 
Since $\b_{\o\o'}$ is circulant, it is easy to compute
\begin{equation}
\sum_{\superp{\o,\o'=0}{\o\leq\o'}}^{p-1}\b_{\o\o'}=
\left\{
\begin{array}{cc}
p & (\nu_1+\nu_2<p),\\
0 & \text{(else)},
\end{array}
\right.
\end{equation} 
assuming $0\leq\nu_1,\nu_2\leq p-1$. This gives us the value of the central charge $\bc$.

\section{Automorphisms, gradings and modes expansion}\label{AppD}
\subsection{Automorphisms and gradings}\label{sec_gradings}
The algebraic relations \ref{algebra} can be supplemented with the grading operators $d$ and $\bd_\o$ ($\o\in\mZ_p$) acting on the currents as
\begin{align}
\begin{split}
&e^{\a d} x_{\o}^\pm(z)e^{-\a d}=x^\pm(e^\a z),\quad e^{\a d}\psi^\pm_{\o}(z)e^{-\a d}=\psi_{\o}^\pm(e^{\a}z),\\
&e^{\a\bd_\o} x_{\o'}^\pm(z)e^{-\a\bd_\o}=e^{\pm\a\d_{\o,\o'}}x_{\o'}^\pm(z),\quad e^{\a\bd_\o}\psi_{\o'}^+(z)e^{-\a\bd_\o}=\psi_{\o'}^+(z),\quad e^{\a\bd_\o}\psi_{\o'}^-(z)e^{-\a\bd_\o}=e^{\a(\d_{\o,\o'-\nu_3c}-\d_{\o\o'})}\psi_{\o'}^-(z),
\end{split}
\end{align}
for any parameter $\a\in\mC$. The grading operator $d$ reflects the invariance of the algebra under rescaling of the variable $z\to e^{\a}z$, it defines the automorphisms $\t_\a$ acting on an element $x$ of the algebra as $\t_\a(x)=e^{\a d}xe^{-\a d}$. Similarly, the grading operators $\bd_\o$ defines the automorphisms $\bar{\t}_{\o,\a}(x)=e^{\a\bd_\o}xe^{-\a\bd_\o}$ associated to the invariance under the following rescaling of the currents for a fixed $\o$:
\begin{equation}
x_\o^\pm(z)\to e^{\pm\a}x_\o^\pm(z),\quad \psi_\o^-(z)\to e^{-\a}\psi^-_\o(z),\quad \psi_{\o+\nu_3c}^-(z)\to e^{\a}\psi_{\o+\nu_3c}^-(z),
\end{equation} 
while the currents $x_{\o'\neq\o}^\pm(z)$, $\psi_{\o'}^+(z)$ and $\psi_{\o'\neq\o,\o+\nu_3c}^-(z)$ remain invariant.%\footnote{Note that when $c=0$ or $\nu_3=0$, all the currents $\psi_{\o'}^-(z)$ become invariant.}

In addition to the automorphisms $\t_\a$ and $\bar\t_{\o,\a}$, the algebraic relations are invariant under a third class of automorphisms $\tilde{\t}_{\o,\a}(x)=e^{\a\td_\o}xe^{-\a\td_\o}$ defined as
\begin{align}
\begin{split}
&e^{\a\td_\o} x_{\o'}^\pm(z)e^{-\a\td_\o}=z^{\pm\a\d_{\o,\o'}}x_{\o'}^\pm(z),\quad e^{\a\td_\o}\psi_{\o'}^+(z)e^{-\a\td_\o}=\psi_{\o'}^+(z),\\
&e^{\a\td_\o}\psi_{\o'}^-(z)e^{-\a\td_\o}=(q_3^{-c}z)^{\a\d_{\o,\o'-\nu_3c}}z^{-\a\d_{\o\o'}}\psi_{\o'}^-(z).
\end{split}
\end{align}
This transformation is the generalization of the element $\CT$ of the SL$(2,\mathbb{Z})$ group of automorphisms for the quantum toroidal algebra of $\mathfrak{gl}_1$ (or Ding-Iohara-Miki algebra) \cite{Bourgine2017b}. With a slight abuse of terminology, we will also call $\td_\o$ a grading operator.

\subsection{Modes expansion}
In order to define properly the modes expansion of the currents $x_\o^\pm(z)$ and $\psi_\o^\pm(z)$, we need to remove some part of the zero modes factors. For this purpose, we use a twist by a combination of automorphisms to define the new currents $\tx_\o^\pm(z)$ and $\tpsi_\o^\pm(z)$ with proper modes expansion. First, we introduce the following combinations of grading operators,
\begin{equation}\label{expr_xi}
F_\o=(-1)^{\bd_\o}(-q_3)^{-\bd_{\o+\nu_3}}(-q_1)^{-\bd_{\o-\nu_1}}(-q_2)^{-\bd_{\o-\nu_2}},\quad \b_\o=-\sum_{\o'}\b_{\o\o'}\bd_{\o'},\quad D_\o=e^{\sum_{\o'}\b_{\o\o'}\td_{\o'}},
\end{equation} 
such that
\begin{align}
\begin{split}
&F_\o x_{\o'}^\pm(w)F_\o^{-1}=F_{\o\o'}^{\pm1}\ x_{\o'}^\pm(w),\quad z^{\b_\o}x_{\o'}^\pm(w)z^{-\b_\o}=z^{\mp\b_{\o\o'}}x_{\o'}^\pm(w),\quad D_\o x_{\o'}^\pm(w)D_\o^{-1}=w^{\pm\b_{\o\o'}}\ x_{\o'}^\pm(w),\\
&F_\o \psi_{\o'}^-(w)F_\o^{-1}=\dfrac{F_{\o\o'-\nu_3c}}{F_{\o\o'}}\ \psi_{\o'}^-(w),\quad z^{\b_\o}\psi_{\o'}^-(w)z^{-\b_\o}=z^{\b_{\o\o'}-\b_{\o\o'-\nu_3c}}\psi_{\o'}^-(w),\\
&D_\o \psi_{\o'}^-(w)D_\o^{-1}=q_3^{-c\b_{\o,\o'-\nu_3c}}w^{-\b_{\o\o'}+\b_{\o\o'-\nu_3c}}\ \psi_{\o'}^-(w),
\end{split}
\end{align}
and $\psi_{\o'}^+(w)$ remains invariant. Defining $\xi_\o(z)=z^{\b_\o}D_\o F_\o$, we find
\begin{align}
\begin{split}\label{rel_xi}
&\xi_\o(z)x^\pm_{\o'}(w)=f_{\o\o'}(w/z)^{\pm1}x^\pm_{\o'}(w)\xi_\o(z),\quad [\xi_\o(z),\xi_{\o'}(w)]=0,\\
&\xi_\o(z)\psi_{\o'}^+(w)=\psi_{\o'}^+(w)\xi_\o(z),\quad \xi_\o(z)\psi_{\o'}^-(w)=\dfrac{f_{\o\ \o'-\nu_3c}(q_3^{-c}w/z)}{f_{\o\o'}(w/z)}\psi_{\o'}^-(w)\xi_\o(z).
\end{split}
\end{align}

The operator $\xi_\o(z)$ is used to define the twisted currents
\begin{equation}
x_\o^+(z)=\tx_\o^+(z),\quad x_\o^-(z)=\tx_\o^-(z)\xi_{\bo}(q_3^{-1}z),\quad \psi_\o^+(z)=\tpsi_\o^+(z)\xi_{\bo}(q_3^{-1}z),\quad\psi_\o^-(z)=\tpsi_\o^-(z)\xi_{\o-\nu_3c}(q_3^{-c}z),
\end{equation}
that satisfy the following algebraic relations,
\begin{align}
\begin{split}\label{algebra_tilde}
&\tx_\o^+(z)\tx_{\o'}^+(w)=g_{\o\o'}(z/w)\tx_{\o'}^+(w)\tx_\o^+(z),\quad \tx_\o^-(z)\tx_{\o'}^-(w)=\dfrac{f_{\o\o'}(w/z)}{f_{\o'\o}(z/w)}g_{\o\o'}(z/w)^{-1}\tx_{\o'}^-(w)\tx_\o^-(z),\\
&\tpsi^+_\o(z)\tx^\pm_{\o'}(w)=f_{\o'\o}(z/w)^{\pm1}g_{\o\o'}(z/w)^{\pm1}\tx^\pm_{\o'}(w)\tpsi^+_\o(z),\\
&\tpsi^-_\o(z)\tx^+_{\o'}(w)=f_{\o-\nu_3c\ \o'}(q_3^cw/z)^{-1}g_{\o-\nu_3c\ \o'}(q_3^{-c}z/w)\tx^+_{\o'}(w)\tpsi^-_\o(z),\\
&\tpsi^-_\o(z)\tx^-_{\o'}(w)=f_{\o\o'}(w/z)g_{\o\o'}(z/w)^{-1}\tx^-_{\o'}(w)\tpsi^-_\o(z),\\
&\tpsi_\o^+(z)\tpsi_{\o'}^-(w)=\dfrac{f_{\o'-\nu_3c\ \o}(q_3^{c} z/w)g_{\o\o'-\nu_3c}(q_3^{c} z/w)}{f_{\o'\o}(z/w)g_{\o\o'}(z/ w)}\tpsi_{\o'}^-(w)\tpsi_\o^+(z),\\
&\tx_\o^+(z)\tx_{\o'}^-(w)-f_{\o\o'}(w/z)^{-1}\tx_{\o'}^-(w)\tx_\o^+(z)=\kO\d_{\o,\o'}\d\left(\dfrac{z}{ w}\right)\tpsi_\o^+(z)-\kO\d_{\o,\o'-\nu_3c}\d\left(\dfrac{q_3^{c} z}{w}\right)\tpsi^-_{\o'}(q_3^cz)\dfrac{\xi_{\o}(z)}{\xi_{\bo'}(q_3^{c-1}z)}.
\end{split}
\end{align}
The operators $\xi_\o(z)$ do not fully decouple from the twisted algebra as it appears in the commutation relation $[\tx^+,\tx^-]$. The exchange relations $\tpsi^\pm-\tx$ now have the correct behavior as $z^{\pm1}\to\infty$ to define the expansions
\begin{equation}\label{exp_tx_tpsi}
\tx_\o^\pm(z)=\sum_{k\in\mathbb{Z}}z^{-k}\tx_{\o,k}^\pm,\quad \tpsi_\o^+(z)=\tpsi_{\o,0}^+\exp\left(\sum_{k>0}z^{-k}a_{\o,k}\right),\quad \tpsi_\o^-(z)=\tpsi_{\o,0}^-z^{\ta_{\o,0}}\exp\left(\sum_{k>0}z^{k}a_{\o,-k}\right).
\end{equation}
The currents $\tpsi_\o^-(z)$ still conserve a zero mode dependence $\ta_{\o,0}$ that is required to reproduce the exchange relation \ref{rel_xi} with the grading operator $\xi_\o(z)$. From the asymptotic behavior of the algebraic relations, we deduce that this zero mode operator $\ta_{\o,0}$ commutes with all the twisted currents $\tx_\o^\pm$, $\tpsi_\o^\pm$ but not with the gradings $\xi_\o(z)$:
\begin{equation}\label{rel_a_xi}
\xi_\o(z)w^{\ta_{\o',0}}=w^{\b_{\o\o'-\nu_3c}-\b_{\o\o'}}w^{\ta_{\o',0}}\xi_\o(z).
\end{equation} 
Note that this operator becomes central if $\nu_3=0$ or $c=0$. Expanding in powers of the spectral parameters, the exchange relations $\tpsi-\tx$ and $\tpsi-\tpsi$ given in \ref{algebra_tilde} provide the commutation relations between the modes,
\begin{align}
\begin{split}
&[a_{\o,k>0},a_{\o',l}]=\d_{k+l}(q_3^{-kc}c_{\o\o'-\nu_3c}^{(k)}-c_{\o\o'}^{(k)}),\quad [a_{\o,k>0},\tx_{\o',l}^\pm]=\pm c_{\o\o'}^{(k)}\tx_{\o',l+k}^\pm,\\
&[a_{\o,-k<0},\tx_{\o',l}^+]=q_3^{-kc}c_{\o\o'+\nu_3c}^{(-k)}\tx_{\o',l-k}^+,\quad [a_{\o,-k<0},\tx_{\o',l}^-]=-c_{\o\o'}^{(-k)}\tx_{\o',l-k}^+,\\
\end{split}
\end{align}
where\footnote{These coefficients appear in the expansions
\begin{equation}\label{exp_g}
[g_{\o\o'}(z)]_+=f_{\o'\o}(z)^{-1}\exp\left(\sum_{k>0}z^{-k}c_{\o\o'}^{(k)}\right),\quad [g_{\o\o'}(z)]_-=f_{\o\o'}(z^{-1})\exp\left(-\sum_{k>0}z^{k}c_{\o\o'}^{(-k)}\right).
\end{equation}}
\begin{equation}
c_{\o\o'}^{(k)}=c_{\o'\o}^{(-k)}=\dfrac1k\sum_{i=1,2,3}(q_i^k\d_{\o,\o'+\nu_i}-q_i^{-k}\d_{\o,\o'-\nu_i}).
\end{equation} 
In particular, when $c\neq0$, the modes $a_{\o,k}$ of the Cartan currents define $p$ Heisenberg subalgebras. This property is used to build the horizontal representation in appendix \ref{AppB2}. The exchange relations $\tx-\tx$ can also be written in terms of modes by projecting the following relations:
\begin{align}
\begin{split}
&z^{\b_{\o\o'}}\prod_{i=1,2,3}(w-q_i^{-1}z)^{\d_{\o,\o'+\nu_i}}\tx_\o^+(z)\tx_{\o'}^+(w)=F_{\o\o'}w^{\b_{\o'\o}}\prod_{i=1,2,3}(w-q_iz)^{\d_{\o\o'-\nu_i}}\tx_{\o'}^+(w)\tx_\o^+(z),\\
&z^{\b_{\o'\o}}\prod_{i=1,2,3}(w-q_iz)^{\d_{\o,\o'-\nu_i}}\tx_\o^-(z)\tx_{\o'}^-(w)=F_{\o'\o}^{-1}w^{\b_{\o\o'}}\prod_{i=1,2,3}(w-q_i^{-1}z)^{\d_{\o\o'+\nu_i}}\tx_{\o'}^-(w)\tx_\o^-(z).
\end{split}
\end{align}
A priori, the commutator $[\tx^+,\tx^-]$ could also be written in terms of modes, but the expression is rather cumbersome. On the other hand, the grading operators have simple actions on the modes:
\begin{align}\label{def_gradings}
\begin{split}
&[d,\tx_{\o',k}^\pm]=-k\tx_{\o',k}^\pm,\quad [d,a_{\o,k}]=-ka_{\o,k},\quad [d,\tpsi_{\o',0}^\pm]=0,\quad [d,\ta_{\o',0}]=0,\\
&[\bd_\o,\tx_{\o',k}^\pm]=\pm \d_{\o,\o'}\tx_{\o',k}^\pm,\quad[\bd_\o,a_{\o',k}]=0,\quad [\bd_\o,\psi_{\o',0}^+]=[\bd_\o,\ta_{\o',0}]=0,\quad[\bd_{\o},\tpsi_{\o',0}^-]=(\d_{\o,\o'-\nu_3c}-\d_{\o,\o'})\tpsi_{\o',0}^-,\\
&e^{\td_\o}\tx_{\o',k}^\pm e^{-\td_\o}=\tx_{\o',k\pm\d_{\o,\o'}}^\pm,\quad e^{\td_\o}\tpsi_{\o',0}^+e^{-\td_\o}=\tpsi_{\o',0}^+,\quad e^{\td_\o}\tpsi_{\o',0}^-e^{-\td_\o}=q_3^{-c\d_{\o,\o'-\nu_3c}}\tpsi_{\o',0}^-,\\
&[\td_\o,a_{\o',k}]=0,\quad e^{\td_\o}z^{\ta_{\o',0}}e^{-\td_\o}=z^{\d_{\o,\o'-\nu_3c}-\d_{\o,\o'}}z^{\ta_{\o',0}}.
\end{split}
\end{align}

\subsection{Coproduct}
The Hopf algebra structure can be extended to include the grading operators, provided we define the coproduct, counit and antipode as
\begin{align}
\begin{split}
&\D(d)=d\otimes1+1\otimes d,\quad \D(\bd_\o)=\bd_\o\otimes1+1\otimes\bd_{\o-\nu_3c_{(1)}},\\
&\D(\td_\o)=\td_\o\otimes1+1\otimes\td_{\o-\nu_3c_{(1)}}+(\log q_3)\ c\otimes\bd_{\o-\nu_3c_{(1)}},\\
&\e(d)=\e(\bd_\o)=\e(\td_\o)=0,\quad S(d)=-d,\quad S(\bd_\o)=-d_{\o+\nu_3c},\\
&S(\td_\o)=-\td_{\o+\nu_3c}+(\log q_3)\ c\bd_{\o+\nu_3c}.
\end{split}
\end{align}
We deduce, for the composite operators,
\begin{align}
\begin{split}
&\D(\b_\o)=\b_\o\otimes1+1\otimes\b_{\o-\nu_3c_{(1)}},\quad \e(\b_\o)=0,\quad S(\b_\o)=-\b_{\o+\nu_3c},\\
&\D(F_\o)=F_\o\otimes F_{\o-\nu_3c_{(1)}},\quad \e(F_\o)=1,\quad S(F_\o)=F_{\o+\nu_3c}^{-1},\\
&\D(D_\o)=D_\o\otimes q_3^{-c_{(1)}\b_{\o-\nu_3c_{(1)}}}D_{\o-\nu_3c_{(1)}},\quad \e(D_\o)=1,\quad S(F_\o)=q_3^{c\b_{\o+\nu_3c}}D_{\o+\nu_3c}^{-1},
\end{split}
\end{align}
and, finally,
\begin{equation}
\D(\xi_\o(z))=\xi_\o(z)\otimes\xi_{\o-\nu_3c_{(1)}}(q_3^{-c_{(1)}}z),\quad \e(\xi_\o(z))=1,\quad S(\xi_\o(z))=\xi_{\o+\nu_3c}(q_3^{c}z)^{-1}.
\end{equation} 

We can also compute the coproduct for the twisted currents,
\begin{align}
\begin{split}
&\D(\tx_\o^+(z))=\tx_\o^+(z)\otimes1+\tpsi^-_{\o+\nu_3c_{(1)}}(q_3^{c_{(1)}}z)\xi_\o(z)\otimes\tx_\o^+(z)\\
&\D(\tx_\o^-(z))=\xi_{\bo}(q_3^{-1}z)^{-1}\otimes\tx^-_{\o-\nu_3c_{(1)}}(q_3^{-c_{(1)}}z)+\tx_\o^-(z)\otimes\tpsi^+_{\o-\nu_3c_{(1)}}(q_3^{-c_{(1)}}z),\\
&\D(\tpsi_\o^+(z))=\tpsi_\o^+(z)\otimes\tpsi_{\o-\nu_3c_{(1)}}^+(q_3^{-c_{(1)}}z),\\
&\D(\tpsi_\o^-(z))=\tpsi_{\o-\nu_3c_{(2)}}^-(q_3^{-c_{(2)}}z)\otimes\tpsi_{\o-\nu_3c_{(1)}}^-(q_3^{-c_{(1)}}z)\xi_{\o-\nu_3c_{(1)}-\nu_3c_{(2)}}(q_3^{-c_{(1)}-c_{(2)}}z)\xi_{\o-2\nu_3c_{(1)}-\nu_3c_{(2)}}(q_3^{-2c_{(1)}-c_{(2)}}z)^{-1},
\end{split}
\end{align}
and deduce
\begin{align}
\begin{split}
&\D(a_{\o,k>0})=a_{\o,k}\otimes1+q_3^{ck}\otimes a_{\o-\nu_3c_{(1)},k},\quad \D(a_{\o,-k<0})=a_{\o-\nu_3c_{(2)},-k}\otimes q_3^{-kc}+q_3^{-kc}\otimes a_{\o-\nu_3c_{(1)},-k},\\
&\D(a_{\o,0})=a_{\o-\nu_3c_{(2)},0}\otimes1+1\otimes a_{\o-\nu_3c_{(1)},0}+1\otimes\left(\b_{\o-\n_3c_{(1)}-\nu_3c_{(2)}}-\b_{\o-2\n_3c_{(1)}-\nu_3c_{(2)}}\right),\\
\end{split}
\end{align}
together with the coproduct of the zero modes $\psi_{\o,0}^\pm$.

\subsection{Vertical representation}
In the vertical representation, the grading operators $\bd_\o$ and $\td_\o$ commute with the currents $\psi_\o^\pm(z)$, therefore they are diagonal in the basis $\dket{\bl}$. Their eigenvalues can be determined recursively using the relations with the currents $x^\pm(z)$, 
\begin{equation}
\rho_{\bv}^{(0,m)}(\bd_\o)\dket{\bl}=|K_\o(\bl)|\dket{\bl},\quad \rho_{\bv}^{(0,m)}(\td_\o)\dket{\bl}=\left(\sum_{\mAbox\in\bl}\d_{\o,c(\sAbox)}\log\chi_{\sAbox}\right)\dket{\bl},
\end{equation} 
where the eigenvalues on the vacuum have been chosen to be zero. Then, the representation of $\xi_\o(z)$ takes the simple form
\begin{equation}\label{vert_xi}
\rho^{(V)}(\xi_\o(z))\dket{\bl}=\mr{f}\oY(z)\dket{\bl},
\end{equation}
with the function $\mr{f}\oY(z)$ defined in \ref{def_Psi}. We find the representation of the twisted currents to be
\begin{align}\label{vertical_g}
\begin{split}
&\rho^{(V)}(\tx_\omega^+(z))\dket{\bl}=F^{1/2}\sum_{\mAbox\in A_\omega(\boldsymbol\l)}\d(z/\chi_{\sAbox}) \res_{z=\chi_{\sAbox}}z^{-1}\CYY(z)^{-1}  \dket{\bl+\Abox},\\
&\rho^{(V)}(\tx_\omega^-(z))\dket{\bl}=\sum_{\mAbox\in R_\omega(\boldsymbol\l)}\d(z/\chi_{\sAbox})\res_{z=\chi_{\sAbox}}z^{-1}\CY_{\bo}^{[\bl]}(q_3^{-1}z) \dket{\bl-\Abox},\\
&\rho^{(V)}(\tpsi_\omega^+(z))\dket{\bl}=\mr{f}_{\bo}^{[\bl]}(q_3^{-1}z)^{-1}\left[\PsiY(z)\right]_+\dket{\bl},\quad \rho^{(V)}(\tpsi_{\omega,0}^+)=1,\\
&\rho^{(V)}(\tpsi_\omega^-(z))\dket{\bl}=\mr{f}\oY(z)^{-1}\left[\PsiY(z)\right]_-\dket{\bl},\quad \rho^{(V)}(\tpsi_{\omega,0}^+)=\dfrac{\prod_{\a\in C_\bo(m)}(-q_3v_\a)}{\prod_{\a\in C_\o(m)}(-v_\a)},\\
&\rho^{(V)}(a_{\o,k})\dket{\bl}=\left(\sum_{\mAbox\in\bl}c_{\o c(\sAbox)}^{(k)}\chi_{\sAbox}^k\right)\dket{\bl},\quad \rho^{(V)}(\ta_{\o,0})=m_\o-m_\bo.
\end{split}
\end{align}

\subsection{Horizontal representation}
Computing the exchange relations of the operators $e^{\a p_\o}$ and $e^{\a\tp_\o}$ with the Drinfeld currents leads to the identification of the representation for grading operators
\begin{equation}
\rho^{(H)}(\bd_\o)=p_\o,\quad\rho^{(H)}(\td_\o)=\tp_\o\implies \rho^{(H)}(\xi_\o(z))=P_\o(z).
\end{equation} 
Thus, we find the representation for the twisted currents,
\begin{align}
\begin{split}
&\rho^{(H)}(\tx^+_\o(z))=u_\o z^{-n_\o}Q_\o(z)\eta_\o^+(z),\quad \rho_{u}^{(1,n)}(\tx^-_\o(z))=u_\o^{-1}z^{n_\o}Q_\o(z)^{-1}\eta_\o^-(z),\\
&\rho^{(H)}(\tpsi^+_\o(z))=F^{-1/2}\vphi_\o^+(z),\\
&\rho^{(H)}(\tpsi_\o^-(z))=F^{1/2}\dfrac{u_{\bo}}{u_\o}q_3^{n_{\bo}}z^{n_\o-n_{\bo}}\dfrac{Q_{\bo}(q_3^{-1}z)}{Q_{\o}(z)}\vphi_\o^-(z),
\end{split}
\end{align}
and the modes
\begin{align}
\begin{split}
&\rho^{(H)}(a_{\o,k>0})=-\dfrac1k(q_3^{-k/2}\a_{\o,k}-q_3^{k/2}\a_{\bo,k}),\quad \rho^{(H)}(a_{\o,-k<0})=-\dfrac1k(q_3^{-k}\a_{\bo,-k}-\a_{\o,-k}),\\
&\rho^{(H)}(\tpsi_{\o,0}^+)=F^{-1/2},\quad \rho^{(H)}(\tpsi_{\o,0}^-)=F^{1/2}\dfrac{u_{\bo}}{u_\o}q_3^{n_{\bo}-\tq_\bo}e^{q_\bo-q_\o},\quad \rho^{(H)}(\ta_{\o,0})=n_\o-n_{\bo}+\tq_{\bo}-\tq_\o.
\end{split}
\end{align}
We can verify that $\ta_{\o,0}$ commutes with the twisted currents, and satisfies the relation \ref{rel_a_xi} with the grading operator $\xi_\o(z)$.

\section{Relation with the quantum toroidal $\glp$ algebra}\label{AppE}
In order to reproduce the Drinfeld currents of the quantum toroidal $\glp$ algebra when $\nu_1=-\nu_2\to1$, we need to introduce another twist by grading elements, namely
\begin{align}\label{twist_currents_II}
\begin{split}
&\tx_\o^+(z)=x_\o^+(z)\xi_\o(z)^{-1/2},\quad \tx_\o^-(z)=x_\o^-(z)\xi_{\bo}(q_3^{-1}z)^{-1/2},\\
&\tpsi_\o^+(z)=\psi_\o^+(z)\xi_{\bo}(q_3^{-1/2}z)^{-1},\quad \tpsi_\o^-(z)=\psi_\o^-(z)\xi_{\o-\nu_3c}(q_3^{-(c+1)/2}z)^{-1}.
\end{split}
\end{align}
Note that we keep here the notation $\tx^\pm_\o(z)$, $\tpsi_\o^\pm(z)$ for the twisted currents, but the twist is different from the one studied in appendix \refOld{AppD}. All the grading operators appearing in the definition of $\xi_\o(z)$ commute, and the powers $\xi_\o(z)^\a$ can be defined without ambiguity. The twisted currents \ref{twist_currents_II} obey the same algebraic relations as in \ref{algebra} with the structure functions $g_{\o\o'}(z)$ replaced by new functions $\tg_{\o\o'}(z)$,
\begin{align}\label{algebra_II}
\begin{split}
&\tx_\o^\pm(z)\tx_{\o'}^\pm(w)=\tg_{\o\o'}(z/w)^{\pm1}\tx_{\o'}^\pm(w)\tx_\o^\pm(z),\quad \tpsi^+_\o(z)\tx^\pm_{\o'}(w)=\tg_{\o\o'}(z/w)^{\pm1}\tx^\pm_{\o'}(w)\tpsi^+_\o(z),\\
&\tpsi^-_\o(z)\tx^+_{\o'}(w)=\tg_{\o-\nu_3c\ \o'}(q_3^{-c}z/w)\tx^+_{\o'}(w)\tpsi^-_\o(z),\quad \tpsi^-_\o(z)\tx^-_{\o'}(w)=\tg_{\o\o'}(z/w)^{-1}\tx^-_{\o'}(w)\tpsi^-_\o(z),\\
&\tpsi_\o^+(z)\tpsi_{\o'}^-(w)=\dfrac{\tg_{\o\o'-\nu_3c}(q_3^{c} z/w)}{\tg_{\o\o'}(z/w)}\tpsi_{\o'}^-(w)\tpsi_\o^+(z),\quad [\tpsi_\o^\pm(z),\tpsi_{\o'}^\pm(w)]=0,\\
&[\tx_\o^+(z),\tx_{\o'}^-(w)]=\kO\left[F^{1/2}\d_{\o,\o'}\d(z/w)\tpsi_\o^+(z)-f_{\o\o'}(q_3^c)^{1/2}\d_{\o,\o'-\nu_3c}\d(q_3^{c} z/w)\tpsi^-_{\o+\nu_3c}(q_3^{c}z)\right].
\end{split}
\end{align}
The new structure functions write
\begin{equation}\label{def_tg}
\tg_{\o\o'}(z)=\left(\dfrac{f_{\o'\o}(z)}{f_{\o\o'}(z^{-1})}\right)^{1/2}g_{\o\o'}(z)=(-1)^{\d_{\o,\o'}}\prod_{i=1,2,3}\dfrac{\left((-q_iz)^{1/2}+(-q_iz)^{-1/2}\right)^{\d_{\o,\o'-\nu_i}}}{\left((-q_i^{-1}z)^{1/2}+(-q_i^{-1}z)^{-1/2}\right)^{\d_{\o,\o'+\nu_i}}}.
\end{equation} 
They possess the necessary property $\tg_{\o\o'}(z)\tg_{\o'\o}(z^{-1})=1$, and exhibit the following asymptotics,
\begin{equation}
\tg_{\o\o'}(z)\superpsim{0}(F_{\o\o'}F_{\o'\o})^{1/2}z^{-(\b_{\o\o'}-\b_{\o'\o})/2},\quad \tg_{\o\o'}(z)\superpsim{\infty}(F_{\o\o'}F_{\o'\o})^{-1/2}z^{(\b_{\o\o'}-\b_{\o'\o})/2}.
\end{equation} 
When $\nu_3=0$, the matrix $\b_{\o\o'}$ is symmetric, and the $z$-dependence disappears from the asymptotics. In this case, zero modes are no longer needed for the Drinfeld currents $\tpsi_\o^\pm(z)$. In fact, when $\nu_1=-\nu_2=1$, the function $\tg_{\o\o'}(z)$ reduces to the structure function of quantum toroidal $\glp$ given in \ref{def_g_glp}. On the other hand, when $\nu_3\neq 0$, the algebraic relations involve half-integer powers of $z$, and a choice of branch cut for the square root is necessary. We will ignore these issues in this appendix as we have only in mind the application to the case $\nu_3=0$.

The relations \ref{algebra_II} do not reduce to the algebraic relation \ref{algebra_glp} of the quantum toroidal $\glp$ algebra yet. An extra shift involving the central charge $c$ is necessary,
\begin{equation}\label{shift_currents}
\tx_\o^\pm(z)\to \tx_\o^\pm(q_3^{\mp c/4}),\quad \tpsi_\o^-(z)\to\tpsi_\o^-(q_3^{c/2}z),
\end{equation} 
the current $\tpsi_\o^+(z)$ remaining invariant. Then, the shifted currents satisfy the relations \ref{algebra_glp} when $\nu_3=0$, with the exception of the Serre relations that still have to be imposed by hand.

The coproduct of the twisted currents, without taking into account the previous shift, takes the form
\begin{align}
\begin{split}
\D(\tx_\o^+(z))&=\tx_\o^+(z)\otimes\xi_{\o-\nu_3c_{(1)}}(q_3^{-c_{(1)}}z)^{-1/2}\\
&+\tpsi^-_{\o+\nu_3c_{(1)}}(q_3^{c_{(1)}}z)\xi_{\bo+\nu_3c_{(1)}}(q_3^{c_{(1)}-1}z)^{1/2}\otimes\tx_\o^+(z)\xi_\o(z)^{1/2}\xi_{\o-\nu_3c_{(1)}}(q_3^{-c_{(1)}}z)^{-1/2},\\
\D(\tx_\o^-(z))&=\tx_\o^-(z)\otimes\tpsi^+_{\o-\nu_3c_{(1)}}(q_3^{-c_{(1)}}z)\xi_{\o-\nu_3c_{(1)}}(q_3^{-c_{(1)}}z)^{1/2}\\
&+\xi_{\bo}(q_3^{-1}z)^{-1/2}\otimes\tx_{\o-\nu_3c_{(1)}}^-(q_3^{-c_{(1)}}z)\\
\D(\tpsi_\o^+(z))&=\tpsi_\o^+(z)\otimes\tpsi^+_{\o-\nu_3c_{(1)}}(q_3^{-c_{(1)}}z)\\
\D(\tpsi_\o^-(z))&=\tpsi_{\o-\nu_3c_{(2)}}^-(q_3^{-c_{(2)}}z)\xi_{\bo-\nu_3c_{(2)}}(q_3^{-c_{(2)}-1}z)^{1/2}\xi_{\bo}(q_3^{-1}z)^{-1/2}\\
&\otimes\tpsi_{\o-\nu_3c_{(1)}}^-(q_3^{-c_{(1)}}z)\xi_{\o-\nu_3c_{(1)}-\n_3c_{(2)}}(q_3^{-c_{(1)}-c_{(2)}}z)^{1/2}\xi_{\o-2\nu_3c_{(1)}-\nu_3c_{(2)}}(q_3^{-2c_{(1)}-c_{(2)}}z)^{-1/2}\\
\end{split}
\end{align}
In order to recover the Drinfeld coproduct of the quantum toroidal $\glp$ algebra, we need to remove the dependence in the grading elements. This is done using a twist by the two-tensor $\Xi$, i.e. we define $\tD=\Xi^{-1}\D\Xi$ with
\begin{align}
\begin{split}
\Xi=&\exp\left(\hf\sum_{\o,\o'}\b_{\o-\nu_3c_{(1)},\o'}\left(\td_\o\otimes\bd_{\o'}-\bd_\o\otimes\td_{\o'}-c_{(1)}(\log q_3)\bd_\o\otimes\bd_{\o'}\right)\right)\\
&\times (-1)^{-\frac12\bd^\otimes_{\nu_3c_{(1)}}}(-q_3)^{\frac12\bd^\otimes_{\nu_3c_{(1)}-\nu_3}}(-q_1)^{\frac12\bd^\otimes_{\nu_3c_{(1)}+\nu_1}}(-q_2)^{\frac12\bd^\otimes_{\nu_3c_{(1)}+\nu_2}},
\end{split}
\end{align}
and the shortcut notation $\bd^\otimes_\a=\sum_{\o} \bd_{\o+\a}\otimes\bd_\o$. Then, the twisted coproduct $\tD$ of the twisted currents $\tx^\pm_\o(z)$, $\tpsi_\o^\pm(z)$ reproduces precisely the formulas \ref{Drinfeld_coproduct} for the coproduct $\D$ of the untwisted currents. However, this is not quite the usual coproduct of quantum toroidal $\glp$ algebra, and, together with the shifts \ref{shift_currents}, an extra twist by the two-tensor
\begin{equation}
\G=q_3^{\frac14(d\otimes c+3c\otimes d)}
\end{equation} 
is needed to recover the coproduct \ref{Drinfeld_glp} for the shifted currents.

The aim of this appendix is to recover the intertwining operators of the quantum toroidal $\glp$ algebra. Since the intertwining relations \ref{intw} are also satisfied by the grading operators, the twist \ref{twist_currents_II} of the currents has no effect: it is equivalent to solve these relations for the currents $x_\o^\pm(z)$, $\psi_\o^\pm(z)$, or for their twisted version. On the other hand, the twist of the coproduct by the two-tensor $\Xi$ has to be taken into account. It is easily seen that if the intertwiner $\Phi$ and $\Phi^\ast$ satisfy the relations \ref{intw} for the coproduct $\D$, then the following intertwiners satisfy the same relations for the twisted coproduct $\tD$,
\begin{equation}
\tPhi=\Phi\left(\rho^{(V)}\otimes\rho^{(H)}\ \Xi\right),\quad \tPhi^\ast=\left(\rho^{(V)}\otimes\rho^{(H)}\ \Xi'^{-1}\right)\Phi^\ast,
\end{equation} 
where $\Xi'$ denotes the permutation of the two-tensor $\Xi$. The two-tensor $\Xi$ is diagonal in the vertical basis $\dket{\bl}$, and these relations can be projected on this basis. As result, we find the relations $\tPhi_\bl=\Phi_\bl\Xi_\bl$, and $\tPhi_\bl^\ast=\Xi_\bl^\ast\Phi_\bl^\ast$ for the components $\bl$, with
\begin{align}
\begin{split}
&\Xi_\bl=\left(\rho^{(V)}\otimes\rho^{(H)}\ \Xi\right)\left(\dket{\bl}\otimes 1\right)=\prod_{\sAbox\in\bl}P_{c(\sAbox)}(\chi_{\sAbox})^{-1/2},\\
&\Xi_\bl^\ast=\left(\dbra{\bl}\otimes 1\right)\left(\rho^{(V)}\otimes\rho^{(H)}\ \Xi'^{-1}\right)=\prod_{\sAbox\in\bl}P_{c(\sAbox)}(q_3^{-1}\chi_{\sAbox})^{-1/2}.
\end{split}
\end{align}
Finally, we also need to take into account the twist by the two tensor $\G$, and multiply by
\begin{equation}
\left(\rho^{(V)}\otimes\rho^{(H)}\ \G\right)=q_3^{\frac14\rho^{(V)}(d)},\quad \left(\rho^{(V)}\otimes\rho^{(H)}\ \G'^{-1}\right)=q^{-\frac34\rho^{(V)}(d)}.
\end{equation} 
The vertical representation of the grading operator $e^{\a d}$ is a shift operator acting on the weights $\bv$, sending $\bv\to e^{\a}\bv$.

\section{Derivation of the vertex operators}\label{AppF}
\subsection{Definition of the vacuum components}
Before sketching the derivation of the solution for the intertwining relations, we would like to provide a bold argument for the definition of the vacuum components $\Phi_\vac$ and $\Phi_\vac^\ast$ entering in the definition \ref{def_Phi} of the intertwiners. In fact, the full partition function of the gauge theory, including classical, one-loop and instantons contributions, has a nice description in terms of the melting crystal picture \cite{Nekrasov2003,Bourgine2017}. Indeed, the one-loop contribution can be written as a double product over the boxes of completely filled (infinite) Young diagrams $\blinf=\{(\a,i,j)\diagup\a=1\cdots m, i=1\cdots\infty, j=1\cdots\infty\}$, assuming a $\z_2$-regularization for the infinite product. Then, the instanton correction of order $O(\qf^k)$ is obtained by removing $k$ boxes to $\blinf$, taking the double product over $\blc=\blinf\setminus\bl=\{(\a,i,j)\diagup\a=1\cdots m,i=\l_j^{(\a)}+1\cdots\infty,j=1\cdots\infty\}$ and summing over the configurations $\bl$ of $k=|\bl|$ boxes. The vacuum component $\Phi_\vac$ of the intertwiner $\Phi$ is associated to this infinite product over boxes in $\blinf$, so that formally
\begin{equation}
\Phi_\vac\simeq:\prod_{\mAbox\in\blinf}\eta_{c(\sAbox)}^+(\chi_{\sAbox})^{-1}:,\quad \Phi_\bl\simeq t_\bl:\prod_{\mAbox\in\blc}\eta_{c(\sAbox)}^+(\chi_{\sAbox})^{-1}:,
\end{equation} 
and similarly for $\Phi_\bl^\ast$, replacing $\eta_{c(\sAbox)}^+(\chi_{\sAbox})$ with $\eta_{c(\sAbox)+\nu_3}^-(q_3\chi_{\sAbox})$.

In order to develop this idea, we may introduce a very crude cut-off $N$ such that $\blinf$ is obtained as the limit $N\to\infty$ of $m$ Young diagrams consisting of squares of size $(pN)\times(pN)$, i.e. $\bl_N=\{(\a,i,j)\diagup\a=1\cdots m, i=1\cdots pN, j=1\cdots pN\}$. Then, we may consider the product over boxes $(\a,i,j)\in\bl_N$ and decompose the indices $(i,j)$ as $i=\bi+1+k_ip$, $j=\bj+1+k_jp$ with $\bi,\bj=0\cdots p-1$ and $k_i,k_j=0\cdots N-1$. We end up with
\begin{align}
\begin{split}
:\prod_{\mAbox\in\bl_N}\eta_{c(\sAbox)}^+(\chi_{\sAbox})^{-1}:=:\prod_{\a=1}^{m}\prod_{\bi,\bj=0}^{p-1}&\exp\left(-\sum_{k>0}\dfrac{(v_\a q_1^{\bi}q_2^{\bj})^k}{k}\sum_{k_i,k_j=0}^{N-1}q_1^{pk_ik}q_2^{pk_jk}\a_{c_\a+\bi\nu_1+\bj\nu_2,-k}\right)\\
&\times\exp\left(\sum_{k>0}\dfrac{(v_\a q_1^{\bi}q_2^{\bj}\g)^{-k}}{k}\sum_{k_i,k_j=0}^{N-1}q_1^{-pk_ik}q_2^{-pk_jk}\a_{c_\a+\bi\nu_1+\bj\nu_2,k}\right):,
\end{split}
\end{align}
Performing the sum over $k_i$ and $k_j$, we find
\begin{align}
\begin{split}
:\prod_{\mAbox\in\bl_N}\eta_{c(\sAbox)}^+(\chi_{\sAbox})^{-1}:=:\prod_{\a=1}^{m}\prod_{\bi,\bj=0}^{p-1}&\exp\left(-\sum_{k>0}\dfrac{(v_\a q_1^{\bi}q_2^{\bj})^k}{k}\dfrac{1-q_1^{pkN}}{1-q_1^{pk}}\dfrac{1-q_2^{pkN}}{1-q_2^{pk}}\a_{c_\a+\bi\nu_1+\bj\nu_2,-k}\right)\\
&\times\exp\left(\sum_{k>0}\dfrac{(v_\a q_1^{\bi}q_2^{\bj}\g)^{-k}}{k}\dfrac{1-q_1^{-pkN}}{1-q_1^{-pk}}\dfrac{1-q_2^{-pkN}}{1-q_2^{-pk}}\a_{c_\a+\bi\nu_1+\bj\nu_2,k}\right):,
\end{split}
\end{align}
At this stage, the limit $N\to\infty$ is ill-defined because the first exponential converges when $|q_1|,|q_2|<1$ while the second exponential for $|q_1|,|q_2|>1$. However, we notice that each color can be treated independently, and their contribution written in terms of the vacuum component for the intertwiner describing instantons on a omega-background with no orbifold \cite{Awata2011,Bourgine2017b}, with the replacement $\e_1,\e_2\to p\e_1,p\e_2$. Thus, we can borrow the corresponding operator and simply define 
\begin{align}\label{def_Phi_vac}
\begin{split}
\Phi_\vac=:\prod_{\a=1}^{m}\prod_{\bi,\bj=1}^{p-1}&\exp\left(-\sum_{k>0}\dfrac{(v_\a q_1^{\bi}q_2^{\bj})^k}{k(1-q_1^{pk})(1-q_2^{pk})}\a_{c_\a+\bi\nu_1+\bj\nu_2,-k}\right)\exp\left(\sum_{k>0}\dfrac{(v_\a q_1^{\bi}q_2^{\bj}\g)^{-k}}{k(1-q_1^{-pk})(1-q_2^{-pk})}\a_{c_\a+\bi\nu_1+\bj\nu_2,k}\right):.
\end{split}
\end{align}
The appearance of quantities defined on the background $\mC_{p\e_1}\times\mC_{p\e_2}\times S_R^1$ is reminiscent of the surface defect interpretation of the orbifold developed in \cite{Nekrasov_BPS4,Jeong2018}. It may also be related to the \textit{abelianization} procedure described in the case of $\glp$ (unrefined, i.e. $q_3=1$) in \cite{Awata2017}.

Using the definition \ref{def_Phi_vac}, we obtain the following normal-ordering relations\footnote{We have used the following property to perform the sum over indices $\bi,\bj$:
\begin{equation}
\dfrac1k\sum_{\bi,\bj=0}^{p-1}(q_1^{\bi}q_2^{\bj})^kq_3^{-k/2}\s_{\o,c_\a+\bi\nu_1+\bj\nu_2}^{(k)}=(1-q_1^{pk})(1-q_2^{pk})\d_{\o,c_\a}.
\end{equation}}
\begin{align}
\begin{split}
&\eta_\o^+(z)\Phi_{\vac}=\prod_{\a\in C_\o(m)}(1-v_\a/z)^{-1}:\eta_\o^+(z)\Phi_{\vac}:,\quad\Phi_{\vac}\eta_\o^+(z)=\prod_{\a\in C_\bo(m)}(1-z/(q_3v_\a))^{-1}:\eta_\o^+(z)\Phi_{\vac}:,\\
&\eta_\o^-(z)\Phi_{\vac}=\prod_{\a\in C_\bo(m)}(1-q_3v_\a/z):\eta_\o^-(z)\Phi_{\vac}:,\quad\Phi_{\vac}\eta_\o^-(z)=\prod_{\a\in C_\bo(m)}(1-z/(q_3v_\a)):\eta_\o^-(z)\Phi_{\vac}:.
\end{split}
\end{align}
Since $\vphi_\o^\pm(z)$ can be expressed in terms of $\eta_\o^\pm(z)$, we easily deduce the normal-ordering relations for these vertex operators as well. This argument can also be applied to $\Phi_\vac^\ast$, it leads to define
\begin{align}
\begin{split}
\Phi_\vac^\ast=:\prod_{\a=1}^{m}\prod_{\bi,\bj=0}^{p-1}&\exp\left(\sum_{k>0}\dfrac{(v_\a q_1^{\bi}q_2^{\bj}q_3)^k}{k(1-q_1^{pk})(1-q_2^{pk})}\a_{c_\a+(\bi-1)\nu_1+(\bj-1)\nu_2,-k}\right)\\
\times&\exp\left(-\sum_{k>0}\dfrac{(v_\a q_1^{\bi}q_2^{\bj}\g)^{-k}}{k(1-q_1^{-pk})(1-q_2^{-pk})}\a_{c_\a+\bi\nu_1+\bj\nu_2,k}\right):,
\end{split}
\end{align}
and we obtain
\begin{align}
\begin{split}
&\eta_\o^+(z)\Phi_{\vac}^\ast=\prod_{\a\in C_\bo(m)}(1-q_3v_\a/z):\eta_\o^+(z)\Phi_{\vac}^\ast:,\quad\Phi_{\vac}^\ast\eta_\o^+(z)=\prod_{\a\in C_\bo(m)}(1-z/(q_3v_\a)):\eta_\o^+(z)\Phi_{\vac}^\ast:,\\
&\eta_\o^-(z)\Phi_{\vac}^\ast=\prod_{\a\in C_{\o+2\nu_1+2\nu_2}(m)}(1-q_3^2v_\a/z)^{-1}:\eta_\o^-(z)\Phi_{\vac}^\ast:,\quad\Phi_{\vac}^\ast\eta_\o^-(z)=\prod_{\a\in C_\bo(m)}(1-z/(q_3v_\a))^{-1}:\eta_\o^-(z)\Phi_{\vac}^\ast:.
\end{split}
\end{align}
We can also compute
\begin{align}
\begin{split}
&\Phi_\vac\Phi_{\vac}'=\CG(\bv'|\bv)^{-1}:\Phi_\vac\Phi_{\vac}':,\quad \Phi_\vac^\ast\Phi_{\vac}^{\ast\prime}=\CG(\bv'|q_3^{-1}\bv)^{-1}:\Phi_\vac^\ast\Phi_{\vac}^{\ast\prime}:,\\
&\Phi_\vac\Phi_{\vac}^{\ast\prime}=\CG(\bv'|q_3^{-1}\bv):\Phi_\vac\Phi_{\vac}^{\ast\prime}:,\quad \Phi_\vac^\ast\Phi_{\vac}'=\CG(\bv'|\bv):\Phi_\vac^\ast\Phi_{\vac}':,
\end{split}
\end{align}
where we $\CG(\bv|\bv')$ denotes the bifundamental contribution at one-loop expressed in terms of the function $\CG_{q_1,q_2}(z)$,\footnote{Note that when the weights are shifted as $\bv\to q_3\bv$, we have to shift the colors $c_\a\to c_\a+\nu_3$ accordingly. For instance,
\begin{equation}
\CG(\bv|q_3^{-1}\bv')=\prod_{\a=1}^m\prod_{\a'=1}^{m'}\prod_{\bi,\bj=0}^{p-1}\CG_{q_1^p,q_2^p}(v'_{\a'} q_1^{\bi+1}q_2^{\bj+1}/v_\a)^{\d_{c_\a,c'_{\a'}+(\bi+1)\nu_1+(\bj+1)\nu_2}}.
\end{equation}}
\begin{equation}\label{def_CG}
\CG(\bv|\bv')=\prod_{\a=1}^m\prod_{\a'=1}^{m'}\prod_{\bi,\bj=0}^{p-1}\CG_{q_1^p,q_2^p}(v_\a q_1^{\bi+1}q_2^{\bj+1}/v'_{\a'})^{-\d_{c'_{\a'},c_\a+(\bi+1)\nu_1+(\bj+1)\nu_2}},\quad \CG_{q_1,q_2}(z)=\exp\left(-\sum_{k=1}^\infty\dfrac1k\dfrac{z^k}{(1-q_1^{k})(1-q_2^{k})}\right).
\end{equation}

\subsection{Solution of intertwining relations}
Once projected on the vertical states using the decomposition  \ref{Phi_vert}, the intertwining relations \ref{intw} write
\begin{align}\label{intw_I}
\begin{split}
&x_\o^+(z)\Phi_\bl=\PsiY(z)\Phi_\bl x_\o^+(z)+\rho^{(V)}(x_\o^+(z))\cdot\Phi_\bl,\\
&x_\o^-(z)\Phi_\bl=\Phi_\bl x_\o^-(z)+\left[\rho^{(V)}(x_\o^-(z))\cdot\Phi_\bl\right]\psi_\o^+(z),\\
&\psi_\o^+(z)\Phi_\bl=\PsiY(z)\Phi_\bl\psi_\o^+(z),\quad \psi_\o^-(z)\Phi_\bl(z)=\Psi\boY(q_3^{-1}z)\Phi_\bl\psi_\o^-(z),
\end{split}
\end{align}
and
\begin{align}\label{intw_II}
\begin{split}
&x_\o^+(z)\Phi_\bl^\ast=\Phi_\bl^\ast x_\o^+(z)-\psi_{\o-\nu_1-\nu_2}^-(q_3z)\left[\rho^{(V)\ast}(x_\o^+(z))\cdot\Phi_\bl^\ast\right],\\
&\Psi\boY(q_3^{-1}z)x_\o^-(z)\Phi_\bl^\ast=\Phi_\bl^\ast x_\o^-(z)-\rho^{(V)\ast}(x_\bo^-(q_3^{-1}z))\cdot\Phi_\bl^\ast,\\
&\psi_\o^+(z)\Phi_\bl^\ast=\Psi\boY(q_3^{-1}z)^{-1}\Phi_\bl^\ast\psi_\o^+(z),\quad \psi_\o^-(z)\Phi_\bl^\ast(z)=\Psi\boY(q_3^{-1}z)^{-1}\Phi_\bl^\ast\psi_\o^-(z).
\end{split}
\end{align}
To lighten the notations, we have omitted the horizontal representations $\rho^{(H)}$ and $\rho^{(H')}$ and indicated the vertical action with a central dot. In order to show that the operators $\Phi_\bl$ and $\Phi_\bl^\ast$ defined in \ref{def_Phi} satisfy these relations, we need to compute the factors coming from the normal ordering of products with the Drinfeld currents in the horizontal representation. It is easier to treat separately the vertex operators part,
\begin{align}
\begin{split}
&\eta_\o^+(z)\Phi_{\bl}=\CYY(z)^{-1}:\eta_\o^+(z)\Phi_{\bl}:,\quad\Phi_{\bl}\eta_\o^+(z)=f_{\bo}^{[\bl]}(q_3^{-1}z)^{-1}\CY_{\bo}^{[\bl]}(q_3^{-1}z)^{-1}:\eta_\o^+(z)\Phi_{\bl}:,\\
&\eta_\o^-(z)\Phi_{\bl}=\CY_{\bo}(q_3^{-1}z):\eta_\o^-(z)\Phi_{\bl}:,\quad\Phi_{\bl}\eta_\o^-(z)=f_{\bo}^{[\bl]}(q_3^{-1}z)\CY_{\bo}^{[\bl]}(q_3^{-1}z):\eta_\o^-(z)\Phi_{\bl}:,\\
&\vphi_\o^+(z)\Phi_{\bl}=\mr{f}_{\bo}^{[\bl]}(q_3^{-1}z)^{-1}\PsiY(z):\vphi_\o^+(z)\Phi_{\bl}:,\quad\Phi_{\bl}\vphi_\o^+(z)=:\vphi_\o^+(z)\Phi_{\bl}:,\\
&\vphi_\o^-(z)\Phi_{\bl}=:\vphi_\o^-(z)\Phi_{\bl}:,\quad\Phi_{\bl}\vphi_\o^-(z)=f_{\bo}^{[\bl]}(q_3^{-1}z)\dfrac{\mr{f}_{\o+2\nu_1+2\nu_2}(q_3^{-2}z)}{f_{\o+2\nu_1+2\nu_2}(q_3^{-2}z)}\Psi_{\bo}^{[\bl]}(q_3^{-1}z)^{-1}:\vphi_\o^-(z)\Phi_{\bl}:,
\end{split}
\end{align}
%and
\begin{align}
\begin{split}
&\eta_\o^+(z)\Phi_{\bl}^\ast=\CY_{\bo}^{[\bl]}(q_3^{-1}z):\eta_\o^+(z)\Phi_{\bl}^\ast:,\quad\Phi_{\bl}^\ast\eta_\o^+(z)=f_{\bo}^{[\bl]}(q_3^{-1}z)\CY_{\bo}(q_3^{-1}z):\eta_\o^+(z)\Phi_{\bl}^\ast:,\\
&\eta_\o^-(z)\Phi_{\bl}^\ast=\CY_{\o+2\nu_1+2\nu_2}^{[\bl]}(q_3^{-2}z)^{-1}:\eta_\o^-(z)\Phi_{\bl}^\ast:,\quad\Phi_{\bl}^\ast\eta_\o^-(z)=f_{\bo}^{[\bl]}(q_3^{-1}z)^{-1}\CY_{\bo}(q_3^{-1}z)^{-1}:\eta_\o^-(z)\Phi_{\bl}^\ast:,\\
&\vphi_\o^+(z)\Phi_{\bl}^\ast=\mr{f}_{\o+2\nu_1+2\nu_2}^{[\bl]}(q_3^{-2}z)\Psi_{\bo}^{[\bl]}(q_3^{-1}z)^{-1}:\vphi_\o^+(z)\Phi_{\bl}^\ast:,\quad\Phi_{\bl}^\ast\vphi_\o^+(z)=:\vphi_\o^+(z)\Phi_{\bl}^\ast:,\\
&\vphi_\o^-(z)\Phi_{\bl}^\ast=:\vphi_\o^-(z)\Phi_{\bl}^\ast:,\quad\Phi_{\bl}^\ast\vphi_\o^-(z)=f_{\bo}^{[\bl]}(q_3^{-1}z)^{-1}\dfrac{f_{\o+2\nu_1+2\nu_2}^{[\bl]}(q_3^{-2}z)}{\mr{f}_{\o+2\nu_1+2\nu_2}^{[\bl]}(q_3^{-2}z)}\Psi_{\bo}^{[\bl]}(q_3^{-1}z):\vphi_\o^-(z)\Phi_{\bl}^\ast:,
\end{split}
\end{align}
and the zero-modes part,
\begin{align}
\begin{split}
&P_\o(z)t_\bl=\mr{f}_\o^{[\bl]}(z):P_\o(z)t_\bl:,\quad [Q_\o(z),t_\bl]=0,\\
&P_\o(z)t_\bl^\ast=\mr{f}_\bo^{[\bl]}(q_3^{-1}z)^{-1}:P_\o(z)t_\bl^\ast:,\quad t_\bl^\ast Q_\o(z)=\mr{f}_\bo^{[\bl]}(q_3^{-1}z)^{-1}:t_\bl^\ast Q_\o(z):.
\end{split}
\end{align}
From these relations, we to deduce the normal ordering relations for the currents $x_\o^\pm$ and $\psi_\o^\pm$. Then, the relation \ref{intw_I} and \ref{intw_II} for the Cartan currents $\psi_\o^\pm$ follow directly, provided that the weights and levels satisfy the relation \ref{rel_u_n}. This condition is related to the difference between the functions $f\oY$ and $\mr{f}\oY$,
\begin{equation}
\dfrac{u'_\o z^{-n'_\o}}{u_\o z^{-n_\o}}=\dfrac{\mr{f}\boY(q_3^{-1}z)}{f\boY(q_3^{-1}z)}.
\end{equation} 
The relations \ref{intw_I} and \ref{intw_II} involving the currents $x_\o^\pm$ are harder to prove. This is done by decomposition of the functions $\CYY(z)$ as sum over poles. We refer the reader to \cite{Bourgine2017b} for a more detailed explanation.

\section{Example of $qq$-characters}\label{AppG}
We provide here the expansion of the fundamental $qq$-characters for several simple theories at the first few orders in the exponentiated gauge couplings $\qf_\o$. For simplicity, all the theories considered here are pure $U(m)$ gauge theories without Chern-Simons terms. This expansion of $qq$-characters has been done using a short Python script.% \textbf{We might want to remove this appendix in the final version.}

\subsection{$U(1)$ gauge group}
In this case, we can set $v_0=1$, and choose, for definiteness, $c_0=0$.

\paragraph{$\mZ_2$ orbifold (surface defect $\nu_1=1$, $\nu_2=0$)}
\begin{align}
\begin{split}
\Zinst&=1+\dfrac{\qf_0}{1-q_2}+\dfrac{\qf_0^2}{(1-q_2)(1-q_2^2)}+\dfrac{\qf_0^3}{(1-q_2)(1-q_2^2)(1-q_2^3)}+\dfrac{\qf_0\qf_1}{(1-q_2)(1-q_1^2)}+\dfrac{\qf_0^2\qf_1}{(1-q_1^2)(1-q_2)^2}+O(\qf^4)\\
\chi_0(z)&=1-z-\qf_0\dfrac{z}{q_2}-\qf_0^2\dfrac{z}{q_2^2}-\qf_0^3\dfrac{z}{q_2^3}+\qf_1-\qf_0\qf_1+O(\qf^4),\quad \chi_1(z)=1+O(\qf^4)\\
\chi_0^\ast(z)&=1-\dfrac1z+\dfrac{\qf_0}{z}-\dfrac{\qf_0^2}{z}-\dfrac{\qf_1}{q_2z}+\dfrac{\qf_0\qf_1(1+q_2)}{q_2z}+O(\qf^4),\quad \chi_1^\ast(z)=1+\dfrac{\qf_0}{q_2}+\dfrac{\qf_0^2}{q_2^2}+\dfrac{\qf_0^3}{q_2^3}+O(\qf^4).
\end{split}
\end{align}

\paragraph{$\mZ_2$ orbifold (ALE $\nu_1=-\nu_2=1$)}
\begin{align}
\begin{split}
\Zinst&=1+\qf_0+\dfrac{\qf_0\qf_1(1+q_1q_2)}{(1-q_1^2)(1-q_2^2)}+O(\qf^3)\\
\chi_0(z)&=1-z+\qf_0z-\qf_0^2(z-1-q_3^{-2}z^{-1})+\qf_0\qf_1q_3z+O(\qf^3),\\
\chi_1(z)&=1+\qf_0(1-\qf_0)q_3z(z-q_1-q_2)+\qf_1(1-\qf_0)+O(\qf^3)\\
\chi_0^\ast(z)&=1-z^{-1}+\qf_0z^{-1}-\qf_0^2(z-q_3^{-2}+z^{-1})+\qf_0\qf_1z^{-1}+O(\qf^3),\\
\chi_1^\ast(z)&=1+\qf_0(1-\qf_0)z^{-1}((q_3z)^{-1}-q_1-q_2)+\qf_1(1-\qf_0)q_3^{-1}+O(\qf^3).
\end{split}
\end{align}

\paragraph{$\mZ_3$ orbifold (surface defect $\nu_1=1$, $\nu_2=0$)}
\begin{align}
\begin{split}
\Zinst&=1+\dfrac{\qf_0}{1-q_2}+\dfrac{\qf_0^2}{(1-q_2)(1-q_2^2)}+\dfrac{\qf_0\qf_1}{(1-q_2)}+O(\qf^3)\\
\chi_0(z)&=1-z-\qf_0q_2^{-1}z+\qf_2-\qf_0\qf_1q_2^{-1}z-\qf_0^2q_2^{-2}z+O(\qf^3),\quad \chi_1(z)=1+O(\qf^3),\quad \chi_2(z)=1-\qf_0+\qf_1+O(\qf^3)\\
\chi_0^\ast(z)&=1-z^{-1}+\dfrac{\qf_0}{z}-\dfrac{\qf_2}{q_2z}+\dfrac{\qf_0\qf_1}{z}+\dfrac{\qf_0\qf_2}{q_2z}+O(\qf^3),\quad \chi_1^\ast(z)=1+\dfrac{\qf_0}{q_2}+\dfrac{\qf_0^2}{q_2^2}+O(\qf^3),\quad \chi_2^\ast(z)=1+\dfrac{\qf_1}{q_2}+O(\qf^3).
\end{split}
\end{align}
Taking instead $\nu_1=2$ simply exchanges $\qf_1\leftrightarrow\qf_2$ and $\chi_{1,2}\leftrightarrow\chi_{2,1}$.

\paragraph{$\mZ_3$ orbifold (ALE $\nu_1=-\nu_2=1$)}
\begin{align}
\begin{split}
\Zinst&=1+\qf_0+\qf_0\qf_1+\qf_0\qf_2+O(\qf^3)\\
\chi_0(z)&=1-z+\qf_0z-\qf_0^2(z-1-q_3^{-2}z^{-1})+\qf_0\qf_1z+\qf_0\qf_2z+O(\qf^3),\\
\chi_1(z)&=1-\qf_0q_1^{-1}z+\qf_1+\qf_0^2q_1^{-1}z-\qf_0\qf_1+\qf_0\qf_2q_1^{-1}q_2^{-2}z(z-q_1-q_2^2)+O(\qf^3),\\
\chi_2(z)&=1-\qf_0q_2^{-1}z+\qf_2+\qf_0^2q_2^{-1}z-\qf_0\qf_2+\qf_0\qf_1q_1^{-2}q_2^{-1}z(z-q_1^2-q_2)+O(\qf^3),\\
\chi_0^\ast(z)&=1-z^{-1}+\qf_0z^{-1}-\qf_0^2(z-1-q_3^{-2}z^{-1})+\qf_0\qf_1z^{-1}+\qf_0\qf_2z^{-1}+O(\qf^3),\\
\chi_1^\ast(z)&=1-\qf_0q_1z^{-1}+\qf_1q_3^{-1}+\qf_0^2q_1z^{-1}-\qf_0\qf_1q_3^{-1}-\qf_0\qf_2(q_1+q_2^2-q_1q_2^2z^{-1})z^{-1}+O(\qf^3),\\
\chi_2^\ast(z)&=1-\qf_0q_2z^{-1}+\qf_2q_3^{-1}+\qf_0^2q_2z^{-1}-\qf_0\qf_2q_3^{-1}-\qf_0\qf_1(q_1^2+q_2-q_1^2q_2z^{-1})z^{-1}+O(\qf^3).
\end{split}
\end{align}

\paragraph{$\mZ_3$ orbifold (case $\nu_1=\nu_2=1$)}
\begin{align}
\begin{split}
\Zinst&=1+\qf_0+\qf_0\qf_1+O(\qf^3)\\
\chi_0(z)&=1-z+\qf_0z+\qf_1-\qf_0^2z-\qf_0\qf_1(1-z)+O(\qf^3),\\
\chi_1(z)&=1+\qf_0(1-\qf_0)q_3z(z-q_1-q_2)+\qf_2-\qf_0\qf_1q_3(q_1+q_2)z-\qf_0\qf_2q_3^{-2}z^{-1}+O(\qf^3),\\
\chi_2(z)&=1+\qf_0q_3z+\qf_0^2+O(\qf^3),\\
\chi_0^\ast(z)&=1-z^{-1}+(\qf_0+\qf_1-\qf_0^2)z^{-1}-\qf_0\qf_1+O(\qf^3),\\
\chi_1^\ast(z)&=1-\qf_0(1-\qf_0)(q_1+q_2-q_3^{-1}z^{-1})z^{-1}-\qf_2+\qf_0\qf_2(1-q_3^{-1}z^{-2}+q_3^{-3}z^{-3})+O(\qf^3),\\
\chi_2^\ast(z)&=1-\qf_0+\qf_0^2(q_3z+1)-\qf_0\qf_1(q_1^2+q_1q_2+q_2^2)z^{-1}+O(\qf^3).
\end{split}
\end{align}

\subsection{$U(2)$ gauge group}
For a gauge group $U(2)$, and a $\mZ_p$-orbifold, one of the important examples is the case of surface defects. When the surface defect lies on the $z_1$-plane, this is realized by choosing $(\nu_1,\nu_2) = (0,1)$. Then, the qq-characters take the following form
\begin{align}
\begin{split}
&\mathcal{X}_\o^{[\bl]} (z)=\CY\oY(z)+\qf_{\o-1}  \dfrac{(q_3z)^{\k_{\o-1}}}{q_1 f\oY(z)\CY_{\o-1}^{[\bl]}(q_3z)},\\
&\mathcal{X}_\o^{[\bl]\ast} (z)=\CY_\o^{[\bl]\ast}(z)+\qf_{\o-1} \dfrac{(q_3z)^{\k_{\o-1}}}{\CY_{\o-1}^{[\bl]}(q_3z)},
\end{split}
\end{align}
with $\o \in\Zp$. Their expansion at first orders in the instanton counting parameters can be obtained with the same Python sketch. For instance, in the case of $p=3$ with colors $(c_1,c_2)=(0,1)$, taking the Chern-Simons levels $\k_\o=0$, and setting $(v_1,v_2)=(1,v)$ by a suitable rescaling of the weight and spectral parameter $z$, we find that the partition function expands in the instanton counting parameters as
\begin{align}
\begin{split}
\Zinst&=1+\dfrac{\qf_0v}{(1-q_2)(v-q_1q_2)}+\dfrac{\qf_1}{1-q_1q_2v}+\dfrac{\qf_0^2v^2}{(1-q_2)(1-q_2^2)(v-q_1q_2)(v-q_1q_2^2)}\\
&+\dfrac{\qf_1^2}{(1-q_2)(1-q_2^2)}+\dfrac{\qf_0\qf_1v}{(1-q_2)^2(v-q_1q_2)}+\dfrac{\qf_1\qf_2}{(1-q_2)(1-vq_1^2q_2)}+O(\qf^3),
\end{split}
\end{align}
and the qq-characters expansion reads
\begin{align}
\begin{split}
\chi_0(z)&=1-z-\dfrac{\qf_0vz}{q_2(v-q_1q_2)}+\dfrac{\qf_1(z-1+vq_1^2q_2)}{1-vq_1^2q_2}+\qf_2-\qf_0^2\dfrac{zv^2(v-q_1q_2-q_1q_2^2)}{q_2^2(v-q_1q_2)^2(v-q_1q_2^2)}+O(\qf^3)\\
\chi_1(z)&=1-\dfrac{z}{v}+\dfrac{\qf_0z}{v-q_1q_2}-\dfrac{\qf_1z}{q_2v}-\dfrac{\qf_0^2zvq_1q_2}{(v-q_1q_2)^2(v-q_1q_2^2)}-\dfrac{\qf_1^2z}{q_2^2v}+\dfrac{\qf_0\qf_1z}{q_2(v-q_1q_2)}-\dfrac{\qf_1\qf_2z}{vq_2(1-vq_1^2q_2)}+O(\qf^3)\\
\chi_2(z)&=1+O(\qf^3)\\
\chi_0^\ast(z)&=1-z^{-1}+\dfrac{\qf_0v}{z(v-q_1q_2)}-\dfrac{\qf_1vq_1^2}{z(1-vq_1^2q_2)}-\dfrac{\qf_2}{zq_2}-\dfrac{\qf_0^2v^2q_1q_2}{z(v-q_1q_2)^2(v-q_1q_2^2)}+\dfrac{\qf_0\qf_2v}{zq_2(v-q_1q_2)}+O(\qf^3),\\
\chi_1^\ast(z)&=1-\dfrac{v}{z}-\dfrac{\qf_0v^2}{zq_2(v-q_1q_2)}+\dfrac{\qf_1v}{z}-\dfrac{\qf_0^2v^3(v-q_1q_2-q_1q_2^2)}{zq_2^2(v-q_1q_2)^2(v-q_1q_2^2)}+\dfrac{\qf_0\qf_1v^2}{zq_2(v-q_1q_2)}+\dfrac{\qf_1\qf_2v}{z(1-vq_1^2q_2)}+O(\qf^3),\\
\chi_2^\ast(z)&=1+\dfrac{\qf_1}{q_2}+\dfrac{\qf_1^2}{q_2^2}+O(\qf^3).
\end{split}
\end{align}
Similar expansions can be obtained for other values of the parameters. These results, or the sketch itself, can be made available upon request.

\bibliographystyle{utphys}
\providecommand{\href}[2]{#2}\begingroup\raggedright\endgroup
\end{document}